\documentclass[aps,prb,twocolumn,showpacs,superscriptaddress,longbibliography,floatfix]{revtex4-2}

\usepackage{amsmath, amsfonts, amssymb, bm}
\usepackage{mathtools}
\usepackage[inline,shortlabels]{enumitem}
\usepackage{graphicx}
\usepackage{comment}
\usepackage{xcolor}
\usepackage[breaklinks,hypertexnames=false]{hyperref}
\hypersetup{colorlinks,linkcolor={magenta},citecolor={blue},urlcolor={blue}} 
\usepackage{lipsum}
\usepackage[capitalize]{cleveref}
\usepackage{multirow}
\usepackage{url}
\usepackage{glossaries}

\glsdisablehyper
\newacronym{tmd}{TMD}{transition metal dichalcogenide}
\newacronym{dos}{LDOS}{local density of states}
\newacronym{gf}{GF}{Green's function}

\newcommand{\e}{\mathrm{e}}

\usepackage{sidecap,tikz}
\definecolor{lime}{HTML}{A6CE39}
\DeclareRobustCommand{\orcidicon}{\hspace{-1.5mm}
	\begin{tikzpicture}
		\draw[lime, fill=lime] (0,0) 
		circle [radius=0.16] 
		node[white] {{\fontfamily{qag}\selectfont \tiny \,ID}};
		\draw[white, fill=white] (-0.0525,0.095) 
		circle [radius=0.007];
	\end{tikzpicture}
	\hspace{-3mm}
}
\foreach \x in {A, ..., Z}{\expandafter\xdef\csname orcid\x\endcsname{\noexpand\href{https://orcid.org/\csname orcidauthor\x\endcsname}
		{\noexpand\orcidicon}}
}

\begin{document}
	
\title{Microscopic Green's function approach for generalized Dirac Hamiltonians}

\author{Jeyson Támara-Isaza\orcidA{}
}
\affiliation{Departamento de F\'{\i}sica, Universidad Nacional de Colombia, 110911 Bogot\'a, Colombia}
\affiliation{Quantum Technology Center, University of Maryland, College Park, Maryland, 20742, USA}

\author{Pablo Burset\orcidB{}}
\affiliation{Department of Theoretical Condensed Matter Physics, Condensed Matter Physics Center (IFIMAC) and Instituto Nicol\'as Cabrera, Universidad Aut\'onoma de Madrid, 28049 Madrid, Spain}

\author{William J. Herrera\orcidC{}}
\affiliation{Departamento de F\'{\i}sica, Universidad Nacional de Colombia, 110911 Bogot\'a, Colombia}
	
\date{\today}
	
\begin{abstract}
The rising interest on Dirac materials, condensed matter systems where low-energy electronic excitations are described by the relativistic Dirac Hamiltonian, entails a need for microscopic effective models to analytically describe their transport properties. 
Specifically, for the study of quantum transport these effective models must take into account the effect of atomic scale interfaces and the presence of well-defined edges, while reproducing the correct band structure. 
We develop a general method to analytically compute the microscopic Green's function of Dirac materials valid for infinite, semi-infinite, and finite two-dimensional layers with zigzag or armchair edge orientations. We test our method computing the density of states and scattering probabilities of germanene and some transition metal dichalcogenides, obtaining simple analytical formulas. 
Our results provide a useful analytical tool for the interpretation of transport experiments on Dirac materials and could be extended to describe additional degrees of freedom like extra layers, superconductivity, etc. 
\end{abstract}

\maketitle
	

\section{Introduction}


Two dimensional (2D) materials have become an excellent playground to engineer quantum devices with exotic properties~\cite{RecentAdvancesinUltrathin2DNanomaterials_Tan2017May,Varghese2015Sep,Liu2020Jul,Schwierz2015Apr,Akinwande2019Sep,Fiori2014Oct}. Examples include one- or few-atom thick materials like graphene, silicene, stanene, germanene, and \glspl{tmd}~\cite{GermaneneReview_Acun_2015,Arsene2Dhoneycombsystem_kamal_2015,ReviewElectronicMagneticOpticalPropertiesSilicene_Chowdhury_2016,2DTMDs_Manzeli_2017}. 
Despite their differences, many of these materials share low-energy characteristics like the presence of nodal points or a linear band dispersion, tunable carrier density, and high mobility due to suppressed backscattering~\cite{CastroNeto2009Jan,GrapheneLike2DMaterials_Xu2013May,rare2DmaterialsDiracCones_Wang_2015}. 
All these properties make them very promising candidates for quantum transport applications~\cite{EmergingDeviceApplications2DTMDs_Jariwala2014Feb,Schaibley2016Aug,Feng2017Sep,Zhu2017Mar,Akinwande2019Sep,Ahn2020Jun}. 
This class of condensed matter systems is known as Dirac materials since, in the infrared regime, charge carriers follow a relativistic Dirac equation yielding the same universal behavior for, e.g., the optical conductivity or the specific heat~\cite{DiracMaterials_Oliver_2014,Cayssol2013,GrapheneLike2DMaterials_Xu2013May,rare2DmaterialsDiracCones_Wang_2015}. 
The origin of the Dirac-like behavior varies with the material, but in all cases some specific symmetries protect the formation of Dirac nodes in the spectrum~\cite{DiracMaterials_Oliver_2014}. 
For example, in the recently discovered quantum spin Hall insulator~\cite{Konig2007,Sullivan2011}, time-reversal symmetry promotes and protects the formation of Dirac-like metallic one-dimensional (1D) edge states on an otherwise 2D insulator. Similarly, three-dimensional (3D) topological insulators feature 2D edge states described by a single (or an odd number of) Dirac cones~\cite{Hasan2010,Kou2017Apr,Xia2009Jun,TIQSHP_Hsieh2008Apr,Chen2009Jul}. 

Dirac materials are characterized by their band structure~ \cite{GermaneneReview_Acun_2015,Arsene2Dhoneycombsystem_kamal_2015,ReviewElectronicMagneticOpticalPropertiesSilicene_Chowdhury_2016,2DTMDs_Manzeli_2017}, and new materials can be predicted from numerical methods, like density functional theory~\cite{Bradlyn2017}. 
Once identified, their microscopic properties are more easily accessed by numerical lattice calculations~\cite{QSHEinSiliceneAnd2DGermaniun_Cheng_2011,TopologicalInsulatorHelicalModeSiliceneInhoumogeneusElectricField_Ezawa_2012,Valley-PolarizedMetalandQAHEinSilicene_Ezawa_2012,Lewenkopf2013Jun,Thorgilsson2014Mar,Ezawa2015Oct,Gerivani2022}, which can include finite-size effects and the presence of edges~\cite{Marmolejo-Tejada2016Feb,Wang2021Dec}. For the study of quantum transport, it is also important to include the effect of atomic scale interfaces, or the presence of well-defined edges in layers of 2D materials; for example, the electronic spectrum of a graphene nanoribbon strongly depends on the edge orientation~\cite{ElectronicStateGrapheneNanoDiracEquation_Brey_2006,EdgeEffectsGrapheneNanostructures_Jurgen_2011,Aidelsburger2018Sep}. 
In a complementary approach, low-energy, effective microscopic Hamiltonians offer a way to include all these effects (finite-size, edges and interfaces, etc.), while, at the same time, providing analytical results for the study of transport~\cite{ElectronicStateGrapheneNanoDiracEquation_Brey_2006}. 
In the absence of Coulomb interactions, a scattering theory can be derived from the Dirac Hamiltonian including interesting effects like edge orientation, spin-orbit coupling, magnetization, or even superconductivity~\cite{AndreevLevelsGrapheneSuperSurface_Manjarres_2009,GreenFunctionApproachWellDefinedEdges_Herrera_2010,ProximityInterfaceBoundStateSuperGraphJunction_burset_2009,DiracPointAndreevTunnelingSuperlatticeGraphene-SuperconductorJunctions_Paez_2019,GreenFunctionTIjunctionwithMagneticandSuperRegion_Casas_2020,Andelkovic2023,Chaves2023}. 
A particularly interesting generalization of effective Hamiltonians are the \gls{gf} methods combined with Dyson's equation~\cite{GreenFunctionApproachWellDefinedEdges_Herrera_2010,EdgeEffectsGrapheneNanostructures_Jurgen_2011}. The \gls{gf} facilitates the study of correlations (interactions, superconductivity, etc) while naturally allowing the computation of observables like the electric current and the density of states. 
The Hamiltonian approach to \gls{gf} techniques is perfectly suited to consider edge and interface effects in 2D systems and, in many cases, provides simple, analytical results~\cite{GreenFunctionApproachWellDefinedEdges_Herrera_2010, StudyGreenFunctionTISurface_Lu_2018, Burset2008, GreenFunctionTIjunctionwithMagneticandSuperRegion_Casas_2020}. 

In this work, we develop a systematic and general method to analytically compute the microscopic \gls{gf} of systems with a general Dirac Hamiltonian and apply this method to 2D honeycomb structure. The resulting \gls{gf} accounts for the presence of well-defined edges or interfaces at the atomic scale. We then obtain the \gls{gf} of relevant examples of Dirac materials, like germanene and \glspl{tmd}, and obtain transport properties (density of states and scattering probabilities) of infinite, semi-infinite, and finite layers. 
Our general method can distinguish specific edge orientations like zigzag, which only involves one Dirac node or valley, and armchair that combines two valleys. In all cases, our method provides simple analytical formulas. 


The rest of the paper is organized as follows. In \cref{sec:GF-bulk}, we present the general Dirac Hamiltonian and calculate the related \gls{gf}. Then, in \cref{sec:GF-semi,sec:GF-fini}, we develop the recipe for the \gls{gf} of a semi-infinite and a finite system, respectively. In \cref{sec:GF-ZZ,sec:GF-AC}, we apply our method to zigzag and armchair edge orientations and include specific applications for \glspl{tmd} and germanene. We then explore the interesting behavior of germanene under an out-of-plane electric field in \cref{sec:Germ}. Finally, we present some concluding remarks in \cref{sec:conc}. An appendix has also been added to show details of the calculations. 

\begin{figure}[!t]
    \centering
\includegraphics[width=0.45\textwidth]{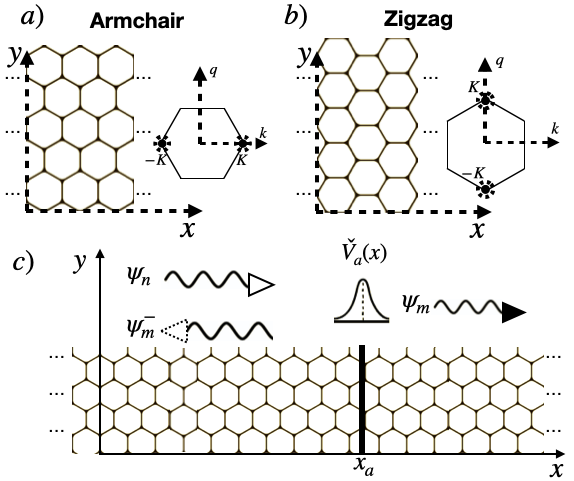}
    \caption{Real-space atomic a) armchair and b) zigzag terminations, with their corresponding reciprocal unit cell showing $K$ and $K'=-K$ valleys. c) Perturbation potential forming two semi-infinite layers at $x=x_a$. The arrows correspond to the incident ($\psi_n$) and scattered wave functions ($\psi_m$ and $\psi_{m}^{-}$). }
    \label{fig:sketch}
\end{figure}

\section{Generalized Dirac Hamiltonian\label{sec:GF-bulk}}

We consider the following general Dirac Hamiltonian, acting on a two-dimensional space
\begin{equation}\label{eq:Dirac-general}
	\check{H}(x,y)= -i\hbar v_{F} (\check{\alpha}_x \partial_x + \check{\alpha}_y \partial_y )+ \check{V}(x,y),
\end{equation}
with $v_{F}$ the Fermi velocity and $\check{V}(x,y)$ an electrostatic potential. 
Here, $\check{\alpha}_{x,y}$ are $2N\times2N$ matrices acting on the $SU(2)$ degree of freedom that defines the Dirac Hamiltonian and the $N$-dimensional space containing the rest of degrees of freedom. 
In the following, we assume translation invariance along the $y$-axis, with $k_y\equiv q$ a conserved quantity, and consider inhomogeneous systems along the $x$ direction. Then, \cref{eq:Dirac-general} reduces to $\check{H}_{q}=-i\hbar \mathbf{\partial }_{x}\check{\alpha}_{x} + q\check{\alpha}_{y} + \check{V}(x)$, which obeys the generalized Dirac equation 
\begin{equation}\label{eq:Dirac-x}
\check{H}_{q}\psi_q \left( x,y\right) =E_{q}\psi _{q}\left( x,y\right) ,
\end{equation}
with solutions of the form
\begin{equation}\label{eq:sols-H}
	\psi _{q}\left( x,y\right) = \e^{iqy} \left[ \psi _{n}^{+} \e^{ik_{n}^{+}x} + \psi_{n}^{-} \e^{ik_{n}^{-}x} \right] ,
\end{equation}
where $\psi_n=(\psi_{n}^+,\psi_{n}^-)^T$ are $2N$-dimensional spinors in the $SU(2)$ space spawned by $\check{\alpha}_x$, with $n$ labeling the other quantum numbers. 
The states $\psi _{n}^{\pm} \e^{ik_{n}^{\pm}x}$, where, usually, $k_{n}^{-}=-k_{n}^{+}$, represent right ($\psi _{n}^{+}$) and left ($\psi _{n}^{-}$) moving solutions along the $x$ direction, with probability flux current given by $J_{n,x}^{\pm}= v_{F} \psi_{n}^{\pm\dagger }\check{\alpha}_{x}\psi _{n}^{\pm}= \pm v_{F}$. 
These states are not, however, orthogonal. To obtain an orthogonality relation, we must define the states
\begin{equation}\label{eq:orthogonal-states}
	\tilde{\psi}_{n}^{\pm} = \pm \check{\alpha}_{x} ^ \dagger \psi _{n} ^{\pm}, \quad(\tilde{\psi}_{n}^{\pm}) ^\dagger = \pm (\psi _{n} ^{\pm})^{\dagger }\check{\alpha}_{x},
\end{equation}
which fulfill 
\begin{equation}
	(\tilde{\psi}_{n}^{\varepsilon } )^ \dagger \psi _{m}^{\varepsilon ^{\prime }} = \delta_{nm}\delta _{\varepsilon \varepsilon ^{\prime }},
\end{equation}
with $\varepsilon=+,-$. Combining $\psi_{n}^{\pm}$ and $\tilde{\psi}_{n}^{\pm}$ states we then find the completeness relation  
\begin{equation}\label{eq:Dirac-complete}
	\sum\limits_{\varepsilon,n} \psi^{\varepsilon}_{n} (\tilde{\psi}^{\varepsilon}_{n})^{\dagger} = \check{1} ,
\end{equation}
with $\check{1}$ being the $2N\times2N$ unit matrix. 

We can now define the \gls{gf} associated to the Dirac Hamiltonian in \cref{eq:Dirac-x} as
\begin{equation}
	\left( E \check{1} -\check{H}_{q}\right) \check{G}_{q}\left( x, x^{\prime }\right) =E \delta \left(x-x^{\prime }\right) \check{1}. 
\end{equation}
When $k_{n}^{-}=-k_{n}^{+}=-k_{n}$, with $k_{n}\geq0$, the Green's functions of the unbounded (bulk) system become 
\begin{subequations}\label{eq:GF-bulk}
\begin{align}
	\check{g}^{<}\left( x<x', x^{\prime }\right) ={}&
	\frac{-i}{2\hbar v_{F}} \sum\limits_{n}f_{n}^{-} \left( x- x^{\prime }\right) 
	\psi^{-}_{n} (\bar{\psi}_{n}^{-})^{T} ,
	\\
	\check{g}^{>}\left( x>x', x^{\prime }\right) ={}& 
	\frac{-i}{2\hbar v_{F}} \sum\limits_{n}f_{n}^{+} \left( x- x^{\prime }\right) 
	\psi^+_{n} (\bar{\psi}^{+}_{n})^{T} .
\end{align}
\end{subequations}
Here, the $x$-dependence is encoded in the functions $f^{\varepsilon}_{n}(x)= \e^{i\varepsilon k_n x }$, and the states $\bar{\psi}^{\varepsilon}_{n}$ are solutions to the transposed Hamiltonian in \cref{eq:Dirac-general}. Transposing a Dirac Hamiltonian results in the exchange $\mathbf{k}\rightarrow-\mathbf{k}$, so the transposed states can be related to the left and right moving states in \cref{eq:sols-H} as
\begin{equation} \label{eq:transposed-states}
	\bar{\psi}_{n}^{+T}=\left( \check{\gamma} \psi _{n}^{-} \right) ^{T} ,\quad 
	\bar{\psi}_{n}^{-T}=\left( \check{\gamma} \psi _{n}^{+} \right) ^{T} ,
\end{equation}
with $\check{\gamma}$ a matrix such that the scalar product $\psi^{\dagger} \check{\gamma} \psi$ is invariant under Lorentz transformations and spatial inversion~\cite{GreenFunctionApproachWellDefinedEdges_Herrera_2010}. 


\section{Semi-infinite systems\label{sec:GF-semi}}

We can define a sharp edge localized at $x=x_a$ by means of the perturbation potential $\check{V}\left( x\right) =U_a\check{\tau}_a \delta \left(x_{a}-x\right)$, with $U_a$ the potential strength that takes the limit $U_a\rightarrow\infty$ when we consider the edge of a semi-infinite layer. Here, $\check{\tau}_a$ is a matrix that encodes the specific boundary conditions at the edge. 
The \gls{gf} perturbed by this potential is given by Dyson's equation as
\begin{align}
	\check{G}_a \left( x,x^{\prime }\right) ={}& \check{g}\left( x,x^{\prime }\right) + \int
	\mathrm{d}x_{1}\check{g}\left( x,x_{1}\right) \check{V}\left( x_{1}\right) \check{G}_a \left(x_{1},x^{\prime }\right) \notag \\
	={}& \check{g}\left( x,x^{\prime }\right) +\check{g} \left( x,x_{a}\right) U_a \check{\tau}_a\check{G}_a \left( x_{a},x^{\prime }\right) .
	\label{eq:Dyson}
\end{align}
The solution of this equation when both $x$ and $x'$ are on the right (left) of $x_a$ takes the form
\begin{equation}\label{eq:Gsemi-sol}
	\check{G}_a^{RR(LL)}\left( x,x^{\prime }\right) = \check{g}\left( x,x^{\prime}\right) + \check{Q}^{>(<)}\left( x\right) \check{g}^{<(>)} \left( x_{a},x^{\prime}\right) ,  
\end{equation}
with the superindex RR (LL) indicating that $x,x'>x_a$ ($x,x'<x_a$), and where we have defined
\begin{align}
	\check{Q}^{>(<)}\left( x\right) ={}&\check{g}^{>(<)} \left( x,x_{a}\right) U_a \check{\tau}_a D^{>(<)}, 
	\label{eq:Gsemi-R}\\
	\check{D}_{a}^{>(<)} ={}&\left( \check{1} -\check{g}^{>(<)}\left( x_a,x_a\right) U_a \check{\tau}_a\right) ^{-1}. 
	\label{eq:Gsemi-S}
\end{align}

An iterative method allows us to find an expression for \cref{eq:Gsemi-R} as (see \cref{sec:appendix-iterative} for details)
\begin{equation}\label{eq:Rm_pert}
	\check{Q}^{>}\left( x\right) =U_a \check{g}^{>}\left( x,x_{a}\right) \check{\tau}_a \sum\limits_{p=0}^{\infty} \left(U_a \check{g}^{>}\left( x_a,x_a\right) \check{\tau}_a\right) ^{p}.
\end{equation}
Using the completeness relation, \cref{eq:Dirac-complete}, we find
\begin{align}
	\check{g}^{>}\left( x_a, x_a\right) \check{\tau}_a ={}& \frac{-i}{2\hbar v_{F}} \sum\limits_{n} \psi _{n}^{+} (\bar{\psi}_{n}^{+})^T \check{\tau}_a \sum\limits_{m,\varepsilon} \psi _{m}^{\varepsilon } (\tilde{\psi}_{m}^{\varepsilon})^{\dagger}
	\nonumber \\
	=& \frac{-i}{2\hbar v_{F}}\sum\limits_{n,m,\varepsilon } \tau_{a,nm}^{+\varepsilon } \check{P}_{nm}^{+\varepsilon } ,
\end{align}
where we have defined the projector operators  
\begin{equation}\label{eq:projector}
	\check{P}_{nm}^{\varepsilon\varepsilon'}= \psi _{n}^{\varepsilon} (\tilde{\psi}_{m}^{\varepsilon'} )^{\dagger} ,
\end{equation}
and the matrix representation of $\check{\tau}_a$ as
\begin{equation}\label{eq:hopping-LR-space}
\tau_{a,nm}^{\varepsilon\varepsilon'} = (\bar{\psi}_{n}^{\varepsilon})^{T} \check{\tau}_a \psi _{m}^{\varepsilon'} .
\end{equation}
In what follows, it is more insightful to write explicitly the subspace spawned by the left and right moving solutions of \cref{eq:Dirac-x}, that is, the $N\times N$ matrices $\hat{\tau}^{\varepsilon\varepsilon'}$, so that 
\begin{equation}
	\check{\tau} = \begin{pmatrix}
		\hat{\tau}^{++} & \hat{\tau}^{+-} \\ \hat{\tau}^{-+} & \hat{\tau}^{--}
	\end{pmatrix}. 
\end{equation}
Henceforth, we use the symbol $\hat{\dots}$ to denote the $2\times2$ matrices in the left and right mover space. 
As a result, \cref{eq:Gsemi-R} becomes
\begin{equation}\label{eq:Rmay_sol}
	\check{Q}^{>}\left( x\right) =
	\sum\limits_{n,m,\varepsilon} f^{+}_{n} \left( x- x_{a}\right) (\hat{r}_{a}^{+\varepsilon})_{nm} \check{P}_{nm}^{+\varepsilon}, 
\end{equation}
with $\hat{r}_{a}^{+\varepsilon}= \hat{D}_a^{+}\hat{\tau}_a^{+\varepsilon}$, and
\begin{equation}\label{eq:Deno}
	\hat{D}_a^{\varepsilon}= \frac{-iU_a}{2\hbar v_{F}}\left( 1+\frac{iU_a}{2\hbar v_{F}}\hat{\tau}_a^{\varepsilon\varepsilon} \right) ^{-1}. 
\end{equation}
Analogously, 
\begin{equation}
	\check{Q}^{<}\left( x\right) =
\sum\limits_{n,m,\varepsilon} f^{-}_{n} \left( x- x_{a}\right) (\hat{r}_{a}^{-\varepsilon})_{nm} \check{P}_{nm}^{-\varepsilon}, 
\end{equation}%
with $\hat{r}^{-\varepsilon}_a=\hat{D}_a^{-}\hat{\tau}_a^{-\varepsilon}$. 

Consequently, the local \glspl{gf} at each side of the perturbation adopt the form
\begin{gather}\label{eq:GFsemi_RR}
	\check{G}_a^{RR}\left( x,x^{\prime }\right) = \check{g}\left( x,x^{\prime }\right) 
	\\ + \sum\limits_{nm} f_{n}^{+}(x-x_a) (\hat{r}^{+-}_a)_{nm} f^{-}_{m} (x_a-x') \psi_{n}^+ (\bar{\psi}_{m}^-)^T, \notag
\end{gather}
and 
\begin{gather}\label{eq:GFsemi_LL}
	\check{G}_a^{LL}\left( x,x^{\prime }\right) = \check{g}\left( x,x^{\prime }\right) 
	\\ + \sum\limits_{nm} f_{n}^{-}(x-x_a) (\hat{r}^{-+}_a)_{nm} f^+_{m}(x_a-x') \psi_{n}^- (\bar{\psi}_{m}^+)^T. \notag
\end{gather}
Here, $\check{r}_a$ corresponds to the scattering matrix of reflection amplitudes. For a potential acting on $x_a$, we can define the transmission amplitudes as $\hat{t}_a^{\varepsilon\varepsilon} = \hat{1} + \hat{r}_a^{\varepsilon\varepsilon}$, and the scattering matrix results in 
\begin{equation}\label{eq:scat-mat}
	\check{S}_a = \check{1}+\check{r}_a = 
	\begin{pmatrix}
		\hat{t}^{++}_a & \hat{r}^{+-}_a \\ 
		\hat{r}^{-+}_a & \hat{t}^{--}_a
	\end{pmatrix} ,
\end{equation}
which fulfills the unitarity condition $\check{S}_a\check{S}_a^\dagger = \check{S}_a^\dagger \check{S}_a = \check{1}$; for more details we refer the reader to \cref{sec:appendix-scattering}. 

The semi-infinite system requires a hard edge at $x=x_a$, which we obtain taking the limit $U_a\rightarrow \infty$. Consequently, the denominators in \cref{eq:Deno} become $\hat{D}_a^\varepsilon \rightarrow -(\hat{\tau}_a^{\varepsilon\varepsilon})^{-1}$. The scattering matrix for the semi-infinite system thus reduces to 
\begin{equation}
\label{eq:scat-mat-potential-infinite}
	\check{S}_a=
	\begin{pmatrix}
		0 & ( \hat{\tau}_a^{^{++}} ) ^{-1}\hat{\tau}_a^{^{+-}} \\ 
		( \hat{\tau}_a^{^{--}} ) ^{-1}\hat{\tau}_a^{^{-+}} & 0
	\end{pmatrix} ,
\end{equation}
where the transmission amplitudes are zero at the edge. As a result, the matrix of reflection amplitudes is unitary, $\left(\hat{r}_a^{+-}\right) ^{\dagger } = \left( \hat{r}_a^{+-}\right) ^{-1}$, and, therefore, $[ ( \hat{\tau}_a^{^{++}} ) ^{-1} ]^{\dagger } ( \hat{\tau}_a^{+-} )^{\dagger }= ( \hat{\tau}_a^{+-} ) ^{-1}\hat{\tau}_a^{++}$. 


\section{Nanoribbon\label{sec:GF-fini}}

We now extend the previous results to include a second edge potential barrier as $\check{V}(x)= \sum_{j=a}^{b} U_{j} \check{\tau}_{j} \delta(x_{j} - x)$. We approach this problem sequentially by first considering only one edge potential. For example, by taking $U_b=0$, we find, from \cref{eq:Dyson}, 
\begin{equation}\label{eq:GFsemi}
	\check{G}_{a} \left( x,x^{\prime }\right) =\check{g}\left( x,x^{\prime }\right) + \check{g}\left( x,x_{a}\right) U_{a}\check{\tau}_a \check{G}\left( x_{a},x^{\prime} \right) ,
\end{equation}
Similarly, by taking $U_a=0$ and $U_b\neq0$, we obtain $\check{G}_{b} ( x,x^{\prime } )$. Restricting ourselves to the region where $x_{a}\leq x,x^{\prime }\leq x_{b}$, the \gls{gf} adopts the form given in \cref{eq:GFsemi_RR} for barrier potential $U_a\check{\tau}_a$, while it follows \cref{eq:GFsemi_LL} for barrier potential $U_b\check{\tau}_b$. 
We apply again Dyson's equation to \cref{eq:GFsemi} to introduce the edge potential $U_b$ and reach
\begin{align}\label{eq:GF-finito}
	\check{G}_{ab}\left( x,x^{\prime }\right) = {}& \check{G}_{a} \left( x,x^{\prime}\right) + \check{G}_{a}^< \left(x,x_{b}\right) U_{b}\check{\tau}_b \\
	\times &\left( 1- \check{G}_{a}^< \left( x_{b},x_{b}\right) U_{b}\check{\tau}_b \right) ^{-1} \check{G}_{a}^> \left(x_{b},x^{\prime}\right) , \nonumber
\end{align}
with $\check{G}^\lessgtr=\check{G}(x\lessgtr x',x')$, cf. \cref{eq:GFsemi_RR}. 
The solution to \cref{eq:GF-finito} reads as
\begin{equation}\label{eq:GF-finito_vf}
	\check{G}_{ab}\left( x,x^{\prime }\right) =\check{G_{a}} \left( x,x^{\prime }\right)
	+ \check{Q}^{<}\left( x\right) \check{G}_{a}^{>}\left( x_{b},x^{\prime }\right) ,
\end{equation}
with
\begin{equation}
	\check{Q}^{<}\left( x\right) =\sum\limits_{n,m, \varepsilon,\varepsilon'}  f_{n}^{\varepsilon} \left(x- x_{b}\right) (\hat{r}^{\varepsilon\varepsilon'}_b)_{nm} \check{P}_{nm}^{\varepsilon\varepsilon'} ,
\end{equation}
and
\begin{equation}
	\check{r}_{b}=-i\frac{U_{b}\check{\tau}_{b}}{2\hbar v_{F}}\left( \check{1}+i
	\frac{U_{b}\check{\tau}_{b}}{2\hbar v_{F}}\right) ^{-1} .
\end{equation}

Using the explicit expression for the semi-infinite \gls{gf}, \cref{eq:GFsemi_RR}, we write the \gls{gf} with $x>x'$ for the central region as 
\begin{gather}\label{eq:GF-finite}
	\check{G}_{ab}^{<}\left( x,x^{\prime }\right) = \frac{-i}{2\hbar v_{F}} 
	\\ \times 
	\sum\limits_{n,m, \varepsilon,\varepsilon'} f_{n}^{\varepsilon}( x- x_{b} ) (\check{w}^{>})_{nm}^{\varepsilon \varepsilon'} f_{m}^{\varepsilon'} (x_{b}- x^{\prime}) \psi_n^{\varepsilon} (\bar{\psi}_{m}^{\varepsilon'})^T , \notag
\end{gather}
where we have defined the matrices
\begin{equation}\label{eq:pert-finite}
	\check{w}^{>} = \begin{pmatrix}
				\hat{1}+\hat{w}_b^{++} & \left( \hat{1}+\hat{w}_b^{++}\right) \hat{r}_a^{+-}\left( W\right)  \\ 
		\hat{w}_b^{-+} & \hat{w}_b^{-+}\hat{r}_a^{+-}\left( W\right) 
	\end{pmatrix} ,
\end{equation}
with $W=x_b-x_a>0$ the width of the finite region. 
Here, $\hat{w}_b^{\varepsilon\varepsilon' }$ are the submatrix elements of
\begin{equation}
	\check{w}_{b}=-i\frac{U_{b}}{2\hbar v_{F}}\check{\tau}\left( \check{1} + i \frac{U_{b}}{2\hbar v_{F}}\check{\tau}\right) ^{-1},
\end{equation}
with 
\begin{equation}
\check{\tau}=
\begin{pmatrix}
	\hat{r}_{a}^{+-}\left( W\right) \hat{\tau}_{b}^{-+} & \hat{r}_{a}^{+-}\left(
	W\right) \hat{\tau}_{b}^{--} \\ 
	\hat{\tau}_{b}^{-+} & \hat{\tau}_{b}^{--}
\end{pmatrix} ,
\end{equation}
and $\hat{r}_a^{+-}(W)$ is defined from the reflection matrix at the left interface, $\hat{r}_a^{+-}$, as $\hat{r}_a^{+-}(W)= \hat{f}^{++}(-W) \hat{r}_a^{+-} \hat{f}^{--}(W)$, with the diagonal matrices $(\hat{f}^{\varepsilon\varepsilon})_{nm} (x) = \delta_{nm} f_{m}^{\varepsilon}(x)$. 
Analogously, we can also define
\begin{equation}\label{eq:pert-finite2}
	\check{w}^{<} = \begin{pmatrix}
				\ \hat{w}_b^{++} & \left( \hat{1}+\hat{w}_b^{++}\right) \hat{r}_a^{+-}\left( W\right)  \\ 
		\hat{w}_b^{-+} & \hat{1}+\hat{w}_b^{-+}\hat{r}_a^{+-}\left( W\right) 
	\end{pmatrix} ,
\end{equation}
where we have used the projector operators, \cref{eq:projector}. This result is equivalent to the reflection matrix for a single barrier at $x=x_{b}$, $\hat{r}^{++}$, changing $\check{w}_{b}$ by $\hat{\tau}^{++}$. 

To obtain the \gls{gf} of a nanoribbon we must take the limits $U_{a,b} \rightarrow \infty$, obtaining
\begin{equation}\label{eq:finite-GF}
	\check{G}_{ab}^{\lessgtr}\left( x,x^{\prime }\right) = \frac{-i}{2\hbar v_{F}} \sum\limits_{\substack{n,m\\ \varepsilon,\varepsilon'}} f_{n}^{\varepsilon} \left( x\right)( \check{w}^{\lessgtr} )_{nm}^{\varepsilon\varepsilon'} f_{m}^{\varepsilon'} \left(-x'\right) \psi_n^{\varepsilon} (\bar{\psi}_{m}^{\varepsilon'})^T  ,
\end{equation}
where \cref{eq:pert-finite,eq:pert-finite2} reduce to
\begin{subequations}\label{eq:finite-r-matrix}
	\begin{align}
		\check{w}^{>}= {}& 
			\begin{pmatrix}
				\hat{D}^{++} & \hat{D}^{++}\hat{r}_a^{+-}\left( x_{a}\right) \\ 
				\hat{r}_b^{-+}\left( x_{b}\right) \hat{D}^{++} & \hat{r}_b^{-+} \left(x_{b}\right) \hat{D}^{++} \hat{r}_a^{+-} \left( x_{a}\right)	
			\end{pmatrix} ,
								\\
		\check{w}^{<}= {}& 
			\begin{pmatrix}
				\hat{D}^{++}\hat{r}_a^{+-}\left( x_{a}\right) \hat{r}_b^{-+}\left(x_{b}\right) & \hat{D}^{++}\hat{r}_a^{+-}\left( x_{a}\right) \\ 
				\hat{r}_b^{-+}\left( x_{b}\right) \hat{D}^{++} & \hat{D}^{--}
			\end{pmatrix} ,
	\end{align}
\end{subequations}
with
\begin{subequations}\label{eq:finite-D-matrix}
\begin{align}
	\hat{r}^{+-}_a(x_a)={}&  \hat{f}^{++}(-x_a) \hat{r}_a^{+-} \hat{f}^{--} (x_a) , 
	\\
	\hat{r}^{-+}_b(x_b)={}& \hat{f}^{--}(x_b) \hat{r}_b^{-+} \hat{f}^{++} (-x_b) ,
	\\
	\hat{D}^{++} ={}&\left[ \hat{1}-\hat{r}_a^{+-}\left( x_{a}\right) \hat{r}_b^{-+}\left( x_{b}\right) \right] ^{-1} , 
	\\
	\hat{D}^{--} ={}&\left[ \hat{1}-\hat{r}_b^{-+}\left( x_{b}\right) \hat{r}_a^{+-}\left( x_{a}\right) \right] ^{-1} .
\end{align}
\end{subequations}
The later equations allow us to obtain the bound states of the finite region by taking the condition $\hat{D}^{\pm\pm}=0$, or, analogously, $\hat{r}_a^{+-} (x_{a} ) \hat{r}_b^{-+} ( x_{b} ) = \hat{1}$. Consequently, the bound states of the finite region are tied to the reflection matrices at each independent edge.


\section{Dirac system with zigzag edges\label{sec:GF-ZZ}}
Having established the general method for the computation of the \gls{gf} of a Dirac system, we now present some examples showcasing specific edge orientations. 
We start with layers ending in zigzag edges, see \cref{fig:sketch}, and modify the Hamiltonian of an infinite system, \cref{eq:Dirac-general}, so that it describes the low-energy physics of graphene-like materials like germanene, silicene, and \glspl{tmd}. 
We thus get the Hamiltonian
\begin{equation} 
\label{eq:hamiltonian 2D-zz}
\check{H}_{s\eta } \left( \mathbf{k}\right) = \mu _{s\eta }\hat{\sigma} _{0}+\hbar
v_{F} \left(k\hat{\sigma}_{x}+\eta q\hat{\sigma}_{y}\right)+m_{s\eta }\hat{\sigma}_{z} 
,
\end{equation}
with
\begin{align}\label{eq:fermi energy}
\mu _{s \eta }={}&-E_{F}+\eta s\lambda _{SO} , 
\\
\label{eq:mass}
m_{s\eta }={}&\lambda _{z}-\eta s\lambda _{SO}+s\lambda _{AF} + \frac{\Delta}{2}.
\end{align}
Here, the Pauli matrices $\hat{\sigma}_{0,x,y,z}$ act in sublattice space denoted by A- and B-type atoms, $\psi=(\psi_A,\psi_B)^T$, and the parameters $\eta=\pm$ and $s=\pm$ are the valley index and the spin $S_z$ component, respectively. The Fermi velocity is written in terms of the lattice constant $a$ and the hopping parameter $t$ as $v_F=\sqrt{3}at/(2\hbar)$. 
The parameter $m_{s\eta}$ describes a generic mass term given by the intrinsic gap $\Delta$, and extra terms that depend on the specific material. 
For example, $\lambda_{SO}$ represents the spin-orbit coupling, $\lambda_{AF}$ a magnetic gap, and $\lambda_z$ the energy due to an external electrical field perpendicular to the monolayer. 
Indeed, $\lambda_{SO}$ is strong in systems such as germanene, silicene, and \glspl{tmd}, but negligible in graphene. A magnetic material in close contact to the monolayer can induce a magnetization $\lambda_{AF}$ by proximity effect. 
Previous works have analyzed the effect on the electronic state of the system caused by an electric field applied perpendicularly to monolayer of \glspl{tmd}, germanene, or silicene~\cite{Qian2014Dec,Ezawa2015Oct,Xiao2012May}. 
Here, we represent this effect by the parameter $\lambda_z$. 
Finally, the doping level $\mu _{s\eta }$ is determined by the Fermi energy $E_F$ and, if present, the spin-orbit coupling $\lambda_{SO}$. 

The eigenstates of \cref{eq:hamiltonian 2D-zz} can be found in \cref{sec:app-applications}, with transposed states defined using \cref{eq:transposed-states} and $\check{\gamma} = \hat{\sigma}_z$. 
With these states, the \gls{gf} of the bulk system is given by \cref{eq:GF-bulk}. 
The resulting wavevectors are  $k_{s\eta}^{\pm}=\pm k_{s\eta}$, where
\begin{equation}
k_{s \eta }=\sqrt{(E -\mu _{s \eta })^2 - m_{s\eta }^{2} - (\hbar v_F q)^2}/(\hbar v_F) .
\end{equation}
For a given excitation energy $E$, the conserved momentum parallel to the interface $\hbar q$ can be parameterized by the angle $\alpha_{s \eta }$ defined as
\begin{equation}\label{eq:alpha}
\e^{\pm i\alpha_{s \eta }} =\hbar v_{F}\frac{ k_{s \eta }  \pm i q}{\sqrt{%
\left( \mu _{s \eta }-E\right) ^{2}-m_{s\eta }^{2}}} .
\end{equation}
We can now define the spectral density of states from the retarded \gls{gf} as
\begin{equation}\label{eq:LDOS}
    \rho(E,q,x)= \frac{1}{\pi} \operatorname{Im}\left\{ \operatorname{Tr}\left[ \check{G}\left(E;q;x,x\right)  \right]\right\},
\end{equation}
and the \gls{dos} is then
\begin{equation}\label{eq:DOS}
    \rho_{T}(E,x)= \int \rho\left(E;q;x\right) \mathrm{d}q .
\end{equation}
The bulk spectral density for a given spin-valley configuration adopts the simple form $\rho_{T,s\eta}=1/\left(\pi \hbar v_F  k_{s\eta} \right)$. 
Henceforth, we normalize the spectral density of states by $\rho_0 = t a$, which is equivalent to measuring energies and distances in units of the hopping $t$ and the lattice constant $a$, respectively.  
We now proceed to define and apply the edge potentials to obtain the \gls{gf} of a semi-infinite and a finite layer, or nanoribbon, with zigzag and armchair edges. 

\subsection{Zigzag semi-infinite layer.} 
For zigzag edge orientation, the matrices that encode the boundary conditions for border of A and B atoms are, respectively, 
\begin{equation}\label{eq:boundary-condition-matrix}
	\check{\tau}_A=
	\begin{pmatrix}
		1 & 0 \\ 
		0 & 0
	\end{pmatrix} , \quad 
 \check{\tau}_B=
	\begin{pmatrix}
		0 & 0 \\ 
		0 & 1
	\end{pmatrix}. 
\end{equation} 
For the semi-infinite layer, we can use either atom type to define the perturbation potential $\check{V}(x_a)$, at position $x_a$. Following the method described above, we reach \cref{eq:GFsemi_RR,eq:GFsemi_LL} at each side of $x_a$. 

To fully characterize the semi-infinite \gls{gf}, we must define the scattering amplitudes, see \cref{eq:scat-mat,eq:scat-mat-potential-infinite}, which yield
$\hat{r}^{+-}_a=( \hat{\tau}_a^{^{++}} ) ^{-1}\hat{\tau}_a^{^{+-}}$. For instance, $\hat{\tau}_a^{^{++}}$ corresponds to the projection of $\hat{\tau}$ onto the right-propagating states and their transposed counterparts. 
As a result, for border type A (B) we get $\hat{r}^{+-}_{a,s\eta}=\hat{r}^{+-}_{s \eta}(x_a)=-(+)\e^{\pm i\alpha_{s \eta}} \e^{-2ik_{s \eta} x_a}$. 
The phase factor $\e^{-2ik_{s \eta} x_a}$ is irrelevant for the semi-infinite case, but very important for the finite layer. 
Plugging $\hat{r}^{+-}_{s \eta}(x_a)$ into \cref{eq:GFsemi_RR}, we find $\rho_{s \eta}= \operatorname{Im}(iN^2_{s \eta}\e^{i\alpha_{s \eta}})/(\pi\hbar v_F)$ for A-type termination, and $\rho_{s \eta}= \operatorname{Im}(i\e^{i\alpha_{s \eta}}/N^2_{s \eta})/(\pi\hbar v_F)$ for border B. 
In both cases, 
\begin{equation}
\label{eq:N}
N_{s \eta}^2 =\frac{\sqrt{E-\mu _{s \eta }-m_{s\eta }}}{\sqrt{E-\mu _{s \eta }+m_{s\eta }}} , 
\end{equation}
see more details in \cref{sec:app-applications}. 

\begin{figure}[t!]
\centering
\includegraphics[scale=0.36]{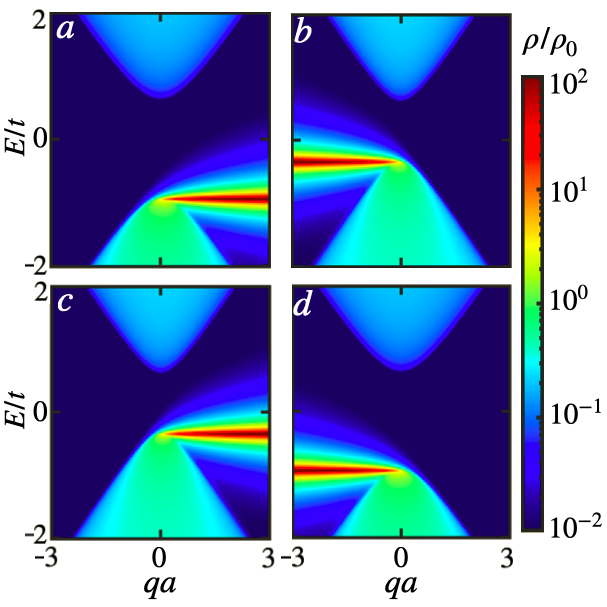}
\caption{Spectral density of states at the edge of a semi-infinite monolayer of \gls{tmd}, where $\Delta$=$1.28t$ and $\lambda_{SO}$=$0.15t$. 
Panels (a,b) represent spin up valleys K' (a) and K (b), while (c,d) correspond to spin down K' (c) and K (d).}
\label{fig: zz_semi_TMDC}
\end{figure}

\begin{figure}[t!]
\centering
\includegraphics[scale=0.36]{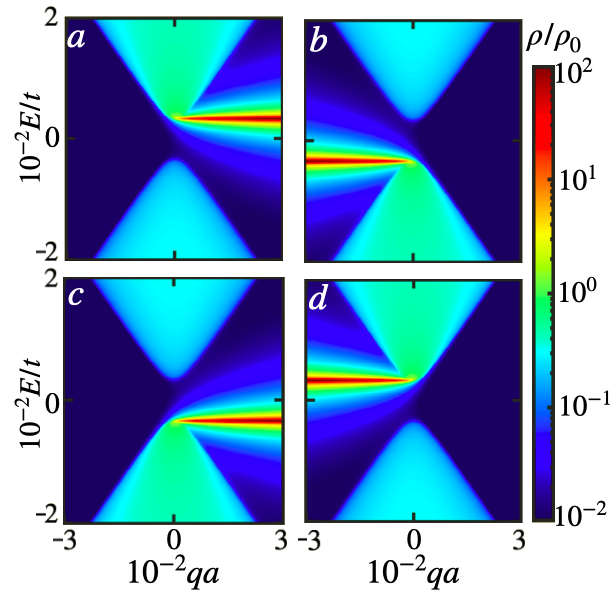} 
\caption{Spectral density of states at the edge of a semi-infinite zigzag germanene monolayer. 
Panels (a,b) represent spin up valleys $K'$ (a) and $K$ (b), while (c,d) correspond to spin down $K'$ (c) and $K$ (d).  Where $\Delta$=$0$ and $\lambda_{SO}$=$0.0033t$.}  
\label{fig: zz_semi_germanene}
\end{figure}

We now illustrate different applications of the semi-infinite \gls{gf} associated to \cref{eq:hamiltonian 2D-zz}. 
By setting $\Delta$=$1.28t$ and $\lambda_{SO}$=$0.15t$, the Hamiltonian of \cref{eq:hamiltonian 2D-zz} describes the low-energy electronic excitations of a semiconductor \gls{tmd} like, e.g., WSe$_2$ (usually, $t\sim1.4$ eV~\cite{Fang2015Nov,Xiao2012May,Kormanyos2015Apr,Xiao2012May}). 
We plot in \cref{fig: zz_semi_TMDC} the spectral density of states, \cref{eq:LDOS}, at the zigzag edge, for each valley and spin configuration. 
Similarly, choosing $\lambda_{so}=0.0033t$ and $\Delta=0$, we obtain the edge dispersion of a germanene layer with one zigzag edge, see \cref{fig: zz_semi_germanene}. 
While the \gls{tmd} features a large, semiconducting gap of the order of $t$, the germanene gap is much smaller, two orders of magnitude, as it corresponds to a semimetallic material. Moreover, a finite intrinsic gap $\Delta$ combined with the spin-orbit gap $\lambda_{SO}$ in the \gls{tmd}, yields a strong asymmetry in the resulting band gap for the different spin-valley configurations. 
By contrast, the band gap of the germanene zigzag layer, with $\Delta=0$, is only due to the spin-orbit coupling and is thus smaller and symmetric. 

\begin{figure}[t!]
\centering
\includegraphics[scale=0.35]{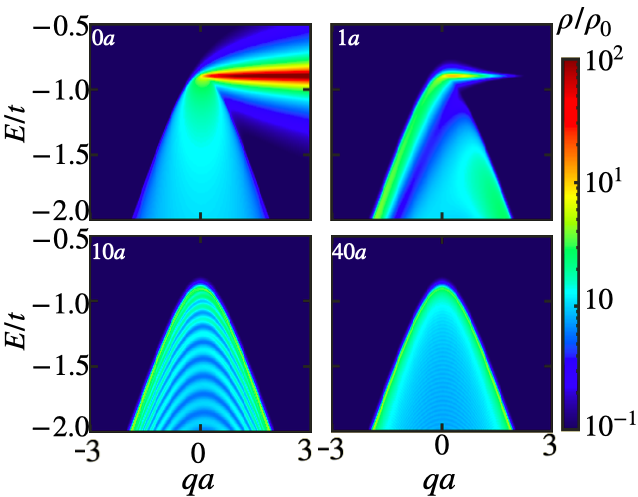}
\caption{Spectral density of states for spin-up, $K'$ valley valence band of a semi-infinite zigzag \gls{tmd} monolayer, at different distances from the edge. The parameters in this case are the same as in figure \cref{fig: zz_semi_TMDC}.}
\label{fig:edge_to_bulk_tmdc}
\end{figure}


The zigzag termination features an edge state in both materials \cite{Ma2013Jan}. However, while the edge state trivially connects the different valence bands for the \gls{tmd}, it couples conduction and valence bands for the germanene case. As a result, the semi-infinite zigzag germanene layer displays two edge states that cross the Fermi energy with opposite velocities, meaning that germanene is a topological insulator \cite{TopologicalInsulatorHelicalModeSiliceneInhoumogeneusElectricField_Ezawa_2012,Si2014Mar,Ezawa2012Mar,Molle2017Feb}(\cref{fig: zz_semi_germanene}). 
We explore in \cref{fig:edge_to_bulk_tmdc} the decay of the edge state inside the semi-infinite layer by computing the spectral density of states at different positions away from the zigzag edge. The edge state has almost completely disappeared a few unit cells away from the edge. As we continue inside the layer, more bands appear in the spectrum, until it is almost a continuum like in the bulk case. 


\subsection{Zigzag nanoribbon.} 
The \gls{gf} for the zigzag nanoribbon, or finite layer with two zigzag edges, is obtained from \cref{eq:finite-GF}, see \cref{sec:appendix-nanoribbon} for a detailed calculation. We now highlight the main steps in its derivation for the Hamiltonian in \cref{eq:hamiltonian 2D-zz}. 
Importantly, the geometry of the zigzag nanoribbon is such that the two edges at $x_a$ and $x_b$ must belong to different A-B atoms, see \cref{fig:sketch}. For simplicity, we now assume that $x_a<x_b$ is made of A atoms. 
To fully characterize the nanoribbon \gls{gf}, we must first obtain the scattering amplitudes forming \cref{eq:finite-r-matrix,eq:finite-D-matrix}. With our choice of atomic terminations for the zigzag nanoribbon, we have $\hat{r}^{+-}_{s\eta}(x_{a(b)})=-(+)\e^{+(-)i\alpha_{s \eta}} \e^{-2ik_{s \eta} x_{a(b)}}$ from the semi-infinite case. 
Analogously, we get $\hat{r}^{-+}_{s\eta}(x_{a(b)})= -(+) \e^{-(+)i\alpha_{s \eta} +2ik_{s \eta}x_{a(b)}}$. As a result, we obtain the local \gls{gf} at the A-type zigzag edge at $x_a$ to be
\begin{equation}\label{eq:GF-zz-nano}
\check{G}_{ab}^{>}\left( x_{a},x_{a}\right)  \\
=
\begin{pmatrix}
0 &  -i\frac{1+\e^{2i(\alpha_{s \eta}+k_{s \eta}W)} }{1+\e^{2i\left(\alpha_{s \eta}+k_{s \eta}W\right)}} \\
0 & -iN_{s \eta}^{2}\e^{i\alpha_{s \eta}}\frac{ 1-\e^{i2k_{s \eta}W}}{1+\e^{2i\left(\alpha_{s \eta}+k_{s \eta}W\right)}}
\end{pmatrix} ,
\end{equation}
with the nanoribbon width being $W=x_b-x_a$. 

The local spectral density for that edge is
\begin{equation}\label{eq:LDOS-zz-nano}
\rho\left( x_{a}\right)=\frac{1}{\pi \hbar v_{F}} \operatorname{Re}\left( \frac{ N_{s \eta}^2\e^{i\alpha_{s \eta} }\left(
1-\e^{i2k_{s \eta}W}\right)} { 1+\e^{2i(\alpha_{s \eta}+k_{s \eta}W) }}\right) .
\end{equation}
When $ W\rightarrow \infty$, the result in \cref{eq:LDOS-zz-nano} is equal to the semi-infinite case. 

\begin{figure}[t!]
\centering
\label{fig:Edge-To-Bulk-TMDC}
\hspace*{-0.4cm}  
\includegraphics[scale=0.38]{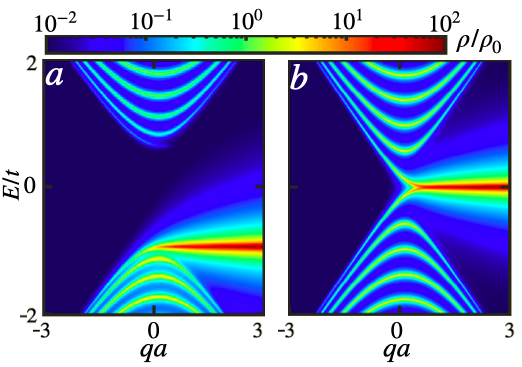}
\caption{Edge spectral density of a zigzag nanoribbon of \gls{tmd} (a) and germanene (b). In both cases, $W/a=4$, and the density is computed for spin-up electrons at valley $K'$. The parameters, in this case, are the same as in \cref{fig: zz_semi_TMDC,fig: zz_semi_germanene}.}
\label{fig:nano_zz_tmdc_germa}
\end{figure}

\begin{figure}[b!]
\centering
\includegraphics[scale=0.36]{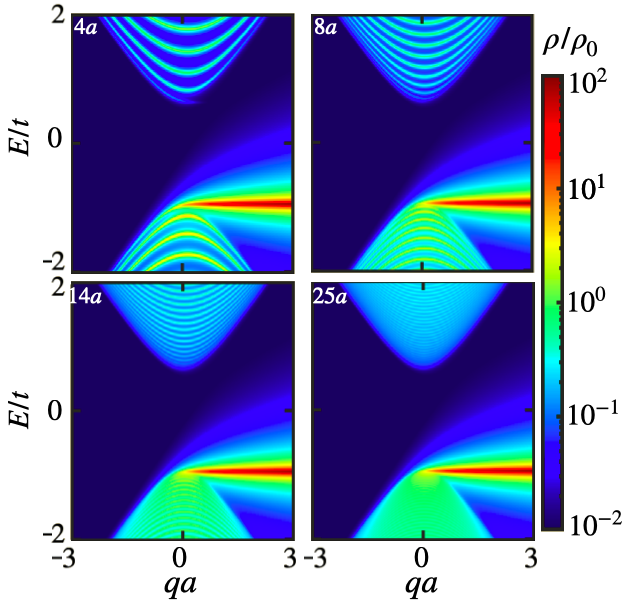}
\caption{Evolution of the edge spectral density of a \gls{tmd} zigzag nanoribbon with the width $W$. All cases correspond to spin-up electrons at valley $K'$. The parameters, in this case, are the same as in figure \cref{fig: zz_semi_TMDC}.}
\label{fig: zz_nano_tmdc_from_w_to_infty}
\end{figure}

In \cref{fig:nano_zz_tmdc_germa}, we compare the edge spectral density for a \gls{tmd} [\cref{fig:nano_zz_tmdc_germa}(a)] and a germanene [\cref{fig:nano_zz_tmdc_germa}(b)] zigzag nanoribbon. The finite size effect is manifested in the appearance of periodic subbands, with periodicity determined by the nanoriboon width $W$. 
The effect of the finite width on the subband periodicity is clearly shown in \cref{fig: zz_nano_tmdc_from_w_to_infty}, where we plot the edge spectral density for a \gls{tmd} zigzag nanoribbon with different widths. As the width $W$ increases, there are more subbands per energy unit, until an almost continuum is recovered for large $W$. Notice that the plots show roughly the same number of subbands as unit cells form the nanoribbon width, i.e., $4a$, $8a$, etc. 
%

\begin{figure}[t!]
\centering
\hspace*{-0.4cm}  
\includegraphics[scale=0.29]{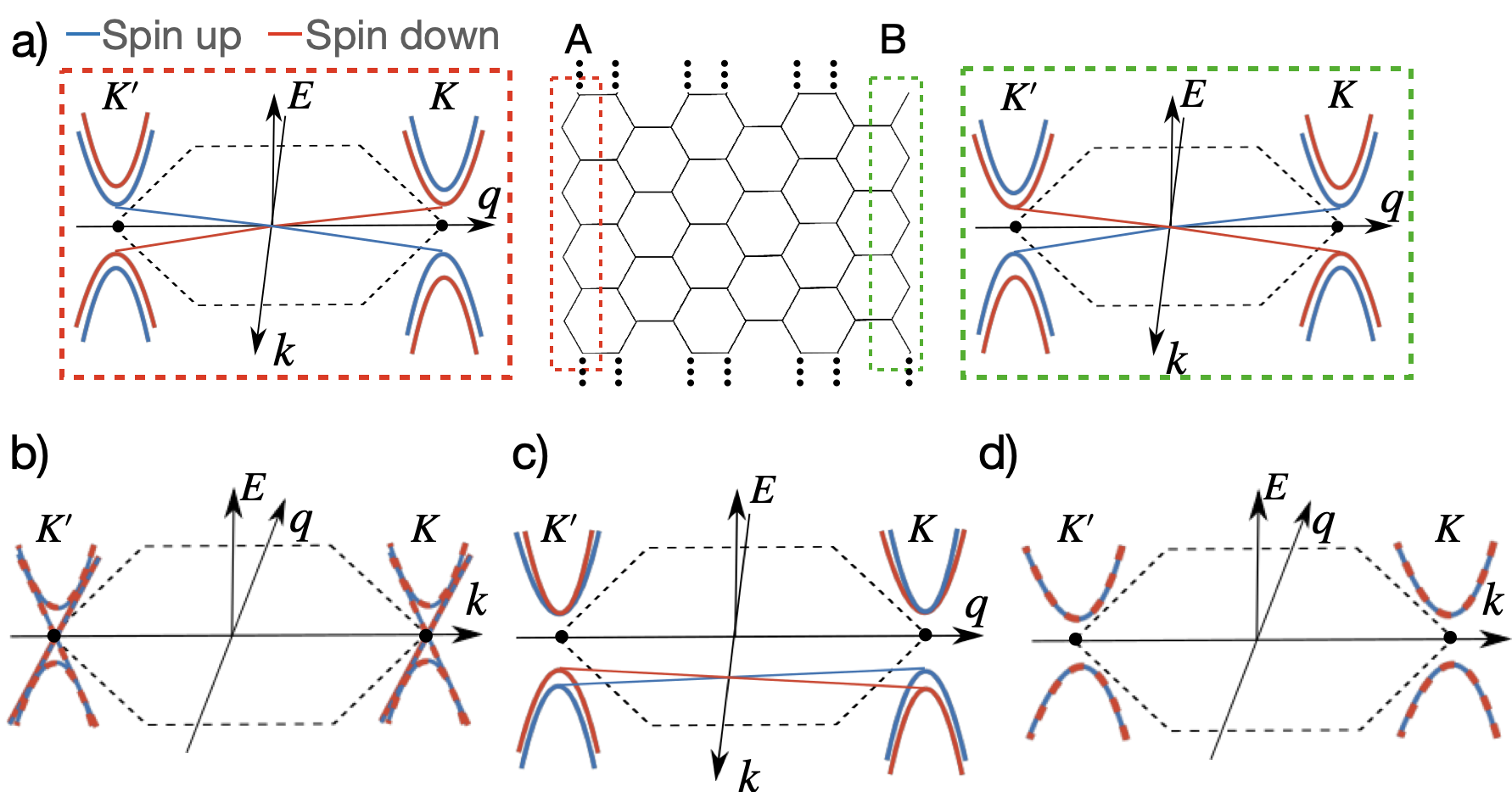}
\caption{Summary of the different band dispersions for \gls{tmd} and germanene nanoribbons. 
a) Sketch of the honeycomb lattice of a zigzag nanoribbon (middle) and band dispersion for the A (left in red) and B (right in green) type of terminations, for germanene nanoribbons. 
b) Edge electronic band structure for armchair germanene. c,d) Edge electronic bands of a \gls{tmd} nanoribbon with zigzag (c) or armchair (d) edges. 
}
\label{fig: qualitative_result_band_for_Edge}
\end{figure}

\Cref{fig: qualitative_result_band_for_Edge} summarizes the obtained spectral properties of zigzag germanene, \cref{fig: qualitative_result_band_for_Edge}(a), and \glspl{tmd}, \cref{fig: qualitative_result_band_for_Edge}(c), showing the lowest energy spin-valley bands and the dispersion of the edge states (red and blue lines). 
Notice the important inversion of the bands when the nanoribbon edge type is changed from A to B in \cref{fig: qualitative_result_band_for_Edge}(a), that basically cames from the \cref{eq:LDOS-zz-nano} by changing $N^2_{s \eta} \rightarrow \frac{1}{N^2_{s \eta}} $. 
The effect of the termination on the band shape is encoded in the parameter $N_{s\eta}$, \cref{eq:N}, which is inverted when the termination changes. 
We also sketch the lowest band dispersion for armchair germanene, \cref{fig: qualitative_result_band_for_Edge}(b), and \gls{tmd}, \cref{fig: qualitative_result_band_for_Edge}(d). We explore this edge termination in more detail below. 

\section{Dirac system with armchair edges\label{sec:GF-AC}}
In contrast to the zigzag termination that does not mix valleys, the armchair edge mixes the valley degree of freedom and thus requires a description that explicitly takes this into account. 
As we did before, we start with a general Dirac Hamiltonian, valid for different graphene-like materials like germanene and \glspl{tmd}, 
now explicitly expressed in the valley subspace as 
\begin{equation}\label{eq:hamil-arm}
    \check{H}_s= \begin{pmatrix}
    \check{H}_{sK} & 0 \\ 0 & \check{H}_{sK'}
\end{pmatrix} ,
\end{equation}
with $s=\pm$ the (degenerate) spin index and
\begin{equation} \label{eq:hamiltonian 2D-Armchair}
\check{H}_{s\eta }\left( \mathbf{k}\right)= \mu_{s\eta }\hat{\sigma} _{0}+\hbar v_{F}\left(
q\hat{\sigma}_{y}-\ \eta k\hat{\sigma}_{x}\right)+m_{s\eta }\hat{
\sigma}_{z} ,
\end{equation}
with the same parameters as \cref{eq:hamiltonian 2D-zz}. The solutions of \cref{eq:hamil-arm} are now spinors in the valley subspace, $\Psi_s=(\psi_{sK},\psi_{sK'})^T$, with each $\psi_{s\eta}$ an eigenstate of \cref{eq:hamiltonian 2D-Armchair}, see more details in \cref{sec:app-applications}. 
%
The orthogonal and transposed states are again calculated following \cref{eq:orthogonal-states,eq:transposed-states} with 
\begin{equation}\label{eq:orthogonal-states-armchair}
	\check{\alpha}_x=
	\begin{pmatrix}
		\hat{\sigma}_x & 0 \\ 
		0 & -\hat{\sigma}_x
	\end{pmatrix} 
 , \quad 
 	\check{\gamma}=
	\begin{pmatrix}
		\hat{\sigma}_z & 0 \\ 
		0 & -\hat{\sigma}_z
	\end{pmatrix} . 
\end{equation}

\subsection{Armchair semi-infinite layer.} 
The armchair edge contains atoms of both sublattices, A and B, see \cref{fig:sketch}. It is thus not necessary to specify the edge type as for zigzag terminations. 
However, the boundary conditions for an armchair edge mix the valley $K$ and $K'$ and, thus, must be defined in the valley subspace as
\begin{equation}\label{eq:orthogonal-states-armchair2}
	\check{\tau}=
	\begin{pmatrix}
		\hat{\sigma}_x & \hat{\sigma}_x \\ 
		\hat{\sigma}_x  & \hat{\sigma}_x
	\end{pmatrix} . 
\end{equation}
As before, the edge \gls{gf} for a semi-infinite layer is given by \cref{eq:GFsemi_RR}, which, when evaluated at the edge ($x=x_0$), adopts the form
\begin{gather}\label{eq:GF-Semi_Arm}
\check{G}_{RR}^{<}\left( x_{0},x_{0}\right) = \\
\begin{pmatrix}
\hat{M}_{KK}^{--}+r_{KK}^{+-}\hat{M}_{KK}^{+-} & r_{KK^{\prime }}^{+-}\hat{M}%
_{KK^{\prime }}^{+-} \\ 
r_{K^{\prime }K}^{+-}\hat{M}_{K^{\prime }K}^{+-} & \hat{M}_{K
^{\prime }K^{\prime }}+r_{K^{\prime }K^{\prime }}^{+-}\hat{M}%
_{K^{\prime }K^{\prime }}^{+-}%
\end{pmatrix} , \notag 
\end{gather}
with 
\begin{equation}\label{eq:M-matrix-arm}
    \hat{M}_{nm}^{\epsilon \epsilon'}=\psi_{n}^{\epsilon}(\bar{\psi}_{m}^{\epsilon'})^{T} ,
\end{equation}
$n,m=K,K'$ labelling the valleys, and $\epsilon,\epsilon'=+,-$ for, respectively, left and right propagating states, see \cref{eq:sols-H} and \cref{fig:sketch}. 

The key ingredients are again the scattering amplitudes $\hat{r}^{+-}=( \hat{\tau}^{^{++}} ) ^{-1}\hat{\tau}^{^{+-}}$ and $\hat{r}^{-+}=( \hat{\tau}^{^{--}} ) ^{-1}\hat{\tau}^{^{-+}}$. 
Importantly, these matrices are now defined in the valley subspace, and, within that subspace, every element corresponds to a projection into a specific valley index. For instance, the element $\hat{r}^{+-}_{KK'}$ of $\hat{r}^{+-}$ corresponds to the reflection amplitude for a state incident from the right ($x<x_0$) on valley $K$ that backscatters to the left into valley $K'$. 
Defining the scalars $h_{nm}^{\epsilon\epsilon'}=\tilde{\varphi}_n^{\epsilon \dagger}\varphi_m^{\epsilon'}$, we find that $r_{KK}^{+-}= h_{KK^{\prime }}^{++} / h_{KK^{\prime }}^{-+}$. The other elements are computed in a similar way, as shown in \cref{sec:app-applications}.  
%

\begin{figure}[t!]
\centering
\includegraphics[scale=0.36]{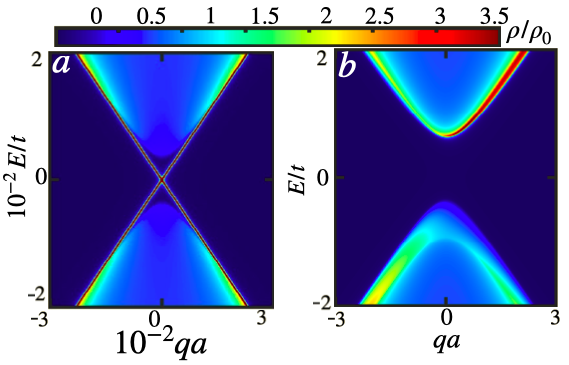}
\caption{Edge spectral density of states for a semi-infinite armchair layer of (a) germanene and (b) and \gls{tmd}. The parameters, in this case, are the same as in \cref{fig: zz_semi_TMDC,fig: zz_semi_germanene}.}
\label{fig:LDOS_Germ+TMDs_semi_arm}
\end{figure}

With these simplifications, the spectral density of states reads as (see more details in \cref{sec:app-applications})
\begin{equation}
\label{eq:arm_semi_tmdc}
\rho_s (E,q)= \frac{4}{\pi\hbar v_F}\operatorname{Re}\left( \frac{N^2N^{\prime 2}(1+\e^{i( \alpha_s
-\alpha_s^{\prime })}) }{ N^2 \e^{i\alpha_s }+N^{\prime 2} \e^{-i\alpha_s ^{\prime }} } 
 \right) .
\end{equation}
Here, we defined $N\!\equiv\! N_{sK}$ and $N'\!\equiv\! N_{sK'}$, following \cref{eq:N} evaluated for valley $K$ and $K'$, respectively. Analogously for $\alpha_s$ and $\alpha_s^\prime$ with \cref{eq:alpha}. 

We plot the edge spectral density in \cref{fig:LDOS_Germ+TMDs_semi_arm} for a germanene (a) and a \gls{tmd} (b) semi-infinite armchair layer. The topology of germanene is now manifested by a gapless state closing the, otherwise, small spin-orbit gap. The \gls{tmd}, by contrast, presents a large semiconducting gap, like for the zigzag case. The main difference for the armchair termination is that the bands are spin symmetric, but display a small asymmetry with the conserved momentum $q$. Both effects stem from the symmetrization imposed by the armchair edge after combining the different spin-valley bands. 


\begin{figure}[t!]
\centering
\hspace*{-0.4cm}  
\includegraphics[width=0.98\columnwidth]{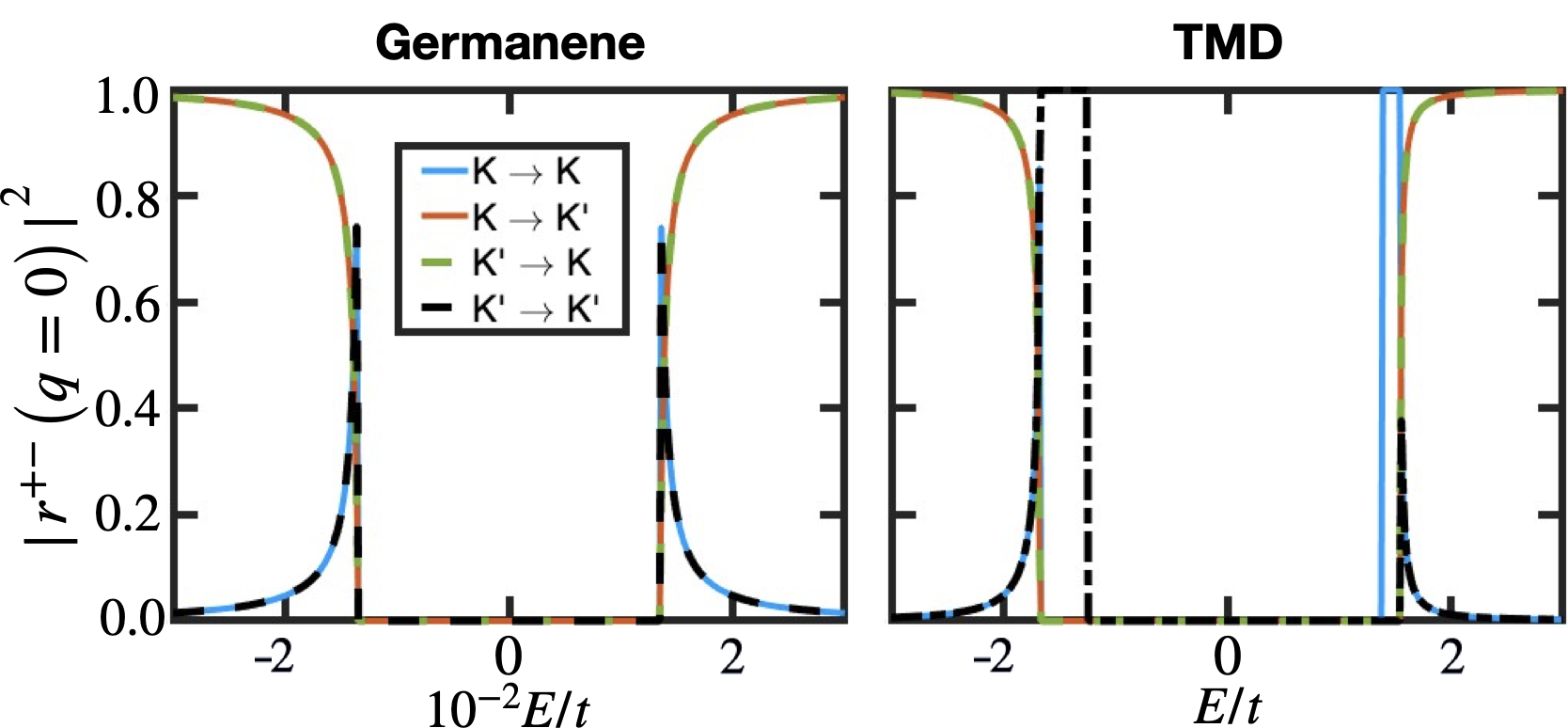}
\caption{Electron reflection probabilities at normal incidence ($q=0$) in the armchair edge of a semi-infinite layer of germanene and \gls{tmd}, showing intravalley (blue and black) and intervalley (red and green) scattering probabilities. The parameters, in this case, are the same as in \cref{fig: zz_semi_TMDC,fig: zz_semi_germanene}.}
\label{fig:reflection_coefficients}
\end{figure}

The main feature about the armchair termination, that different valleys are mixed on scattering, can be better understood by checking the specific inter- and intravalley scattering amplitudes. We show in \cref{fig:reflection_coefficients} the scattering coefficient for an incident electron from valley $K$ (blue and red solid lines) and $K'$ (green and black dashed lines) into the same or the opposite valley. 
For simplicity, we only show the normal incident case ($q=0$), but similar results are obtained for any incident angle. 
For germanene (left panel), intervalley scattering is dominant (red and green lines), except for energies very close to the gap edge. This behavior has been previously reported for graphene \cite{Rutter2007Jul,CastroNeto2009Jan}. Both inter- and intravalley coefficients are symmetric with respect to the energy, as the spin-valley bands are also symmetric, see \cref{fig: qualitative_result_band_for_Edge}. 
The strong asymmetry in the spin-valley bands for the \glspl{tmd} (right panel) results in an intervalley scattering that is very asymmetric with the energy, with small energy windows close to the edge of the valence and conduction bands where intravalley scattering is perfect since the intervalley one is forbidden. 


\subsection{Armchair nanoribbon.} 

The nanoribbon \gls{gf} associated to the Hamiltonian in \cref{eq:hamil-arm} spawns in a combined spin-valley-sublattice space. Being spin-degenerate, we can focus on its valley-sublattice structure, which adopts the form
\begin{equation}\label{eq:gf-arm-finite}
G_{nm}^{>} (x,x')= \sum\limits_{ \varepsilon,\varepsilon^\prime=\pm }
f_{n}^{\varepsilon}(x)f_{m}^{\epsilon^\prime}(x')
\hat{r}_{nm}^{>,\varepsilon\varepsilon^{\prime}} \hat{M}_{nm}^{\varepsilon\varepsilon^{\prime}} .
\end{equation}
Here, $f_{n}^{\varepsilon}(x)=\e^{i(K_n+\varepsilon k_n)x}$, with $K_n=K,K'$ and $k_n=k_s,k_s'$. 
The matrices $\hat{M}_{nm}^{\varepsilon\varepsilon^{\prime}}$ are defined in \cref{eq:M-matrix-arm} and the reflection matrices according to \cref{eq:finite-D-matrix}. See more details in \cref{sec:app-applications}. 


\begin{figure}[t!]
\centering
\hspace*{-0.4cm}  
\includegraphics[scale=0.38]{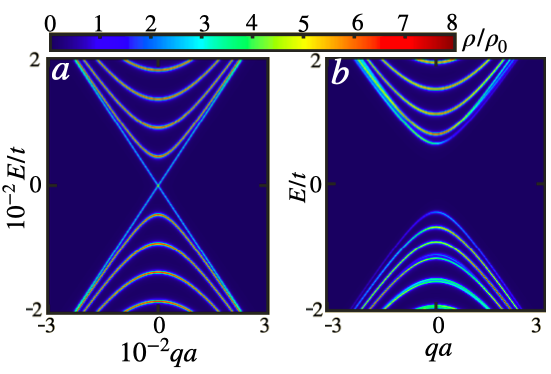}
\caption{Edge spectral density of states for a finite armchair nanoribbon of (a) germanene or (b) \gls{tmd}. For both cases the width is 6 units cells. The parameters, in this case, are the same as in \cref{fig: zz_semi_TMDC,fig: zz_semi_germanene}.}
\label{fig: arm_nano_tmdc_germanene}
\end{figure}

\begin{figure}[t!]
\centering
\hspace*{-0.4cm}  
\includegraphics[scale=0.38]{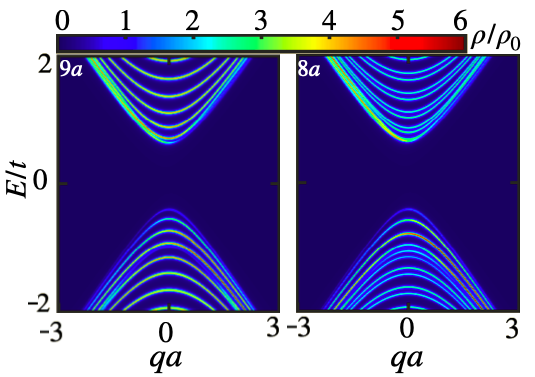}
\caption{Spectral density of states at the armchair edge of \glspl{tmd} nanoribbons for $W/a=8$ (right) and $W/a=9$ (left). The parameters, in this case, are the same as in \cref{fig: zz_semi_TMDC}.}
\label{fig: Multiplicity_TMDC_ARM}
\end{figure}
\begin{figure}[t!]
\centering
\hspace*{-0.4cm}  
\includegraphics[scale=0.38]{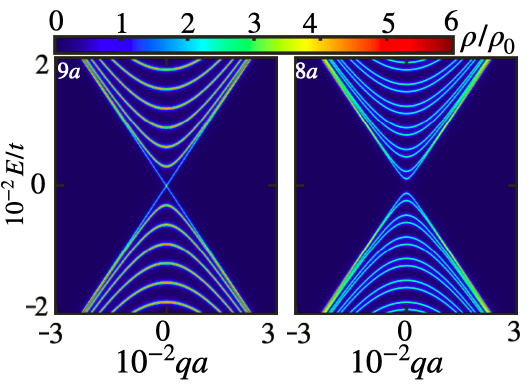}
\caption{Spectral density of states at the armchair edge of germanene nanoribbons for $W/a=8$ (right) and $W/a=9$ (left). The parameters, in this case, are the same as in \cref{fig: zz_semi_germanene}.}
\label{fig: Multiplicity_GERM_ARM}
\end{figure}
We plot the spectral density of states for an armchair nanoribbon of germanene and \gls{tmd} in \cref{fig: arm_nano_tmdc_germanene}(a) and \cref{fig: arm_nano_tmdc_germanene}(b), respectively, for a width of six unit cells. As expected, we observe the band discretization in both cases and the topological edge state for the germanene nanoribbon. 
We note that the nature of the edge state in germanene stems from a topological phase, while a similar, trivial edge state in graphene nanoribbons arises due to interference processes~\cite{ElectronicStateGrapheneNanoDiracEquation_Brey_2006,Rosales2013Jan}.  
We find that the edge state only appears when the nanoribbon width is a multiple of three, see \cref{fig: Multiplicity_TMDC_ARM,fig: Multiplicity_GERM_ARM}, and only for the germanene case, as the semiconducting gap of the \gls{tmd} completely suppresses this effect. 


\section{Germanene gap manipulation\label{sec:Germ}}
Using the \gls{gf} formalism introduced in \cref{sec:GF-bulk,sec:GF-semi,sec:GF-fini}, we have successfully described the electronic band structure of bulk, semi-infinite, and finite layers of graphene-like materials. These \gls{gf} methods give us access to the spectral density of states and can easily take into account extra parameters in the Hamiltonian. 
In this section, we showcase this by studying the effect of an external perpendicular electric field on the electronic and topological properties of a germanene layer. For simplicity, we only consider a semi-infinite amrchair layer. 
As introduced in \cref{eq:hamiltonian 2D-zz,eq:hamiltonian 2D-Armchair}, the parameter $\lambda_z$ represents the effect of such an external field, perpendicular to the layer. 



\begin{figure}[t!]
\centering
\hspace*{-0.4cm}  
\includegraphics[scale=0.36]{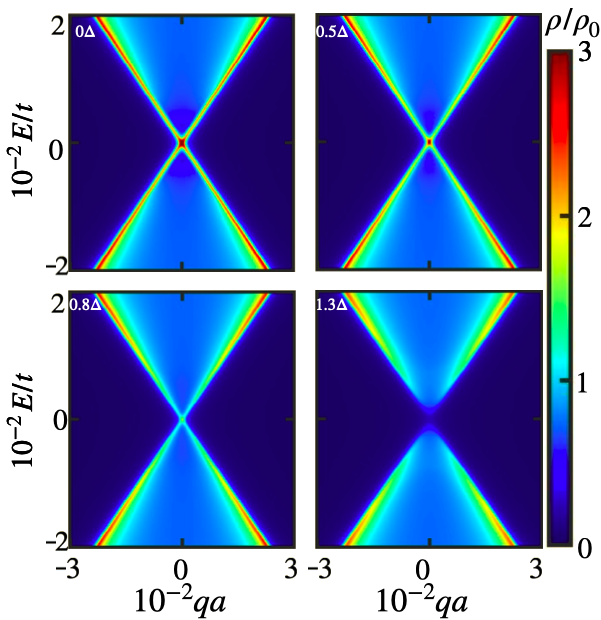}
\caption{Spectral density of states at the armchair edge of a semi-infinite germanene layer, for $\lambda_z/\Delta=0,0.5,0.8$, and $1.3$. The parameters, in this case, are the same as in \cref{fig: zz_semi_germanene}.}
\label{fig:semi-germanene-arm-voltages}
\end{figure}

\begin{figure}[t!]
\centering
\hspace*{-1cm}  
\includegraphics[scale=0.4]{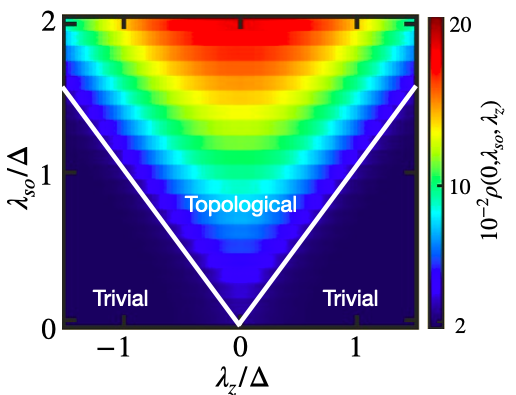}
\caption{Zero-energy density of states of a semi-infinite armchair germanene layer as a function of an external electric field and the spin-orbit coupling. The parameters, in this case, are the same as in \cref{fig: zz_semi_germanene}.}
\label{fig:dos-semi-germanene-arm-voltages-spin}
\end{figure}

The semi-infinite armchair germanene layer features a topological edge state crossing the Fermi energy, see \cref{fig:LDOS_Germ+TMDs_semi_arm}(a). For comparison, we reproduce this result in the top-left panel of \cref{fig:semi-germanene-arm-voltages} ($\lambda_z=0$) and proceed to study the evolution of the edge state for finite $\lambda_z$. As the external electric field increases, the gap closes, removing the edge state ($\lambda_z/\Delta\simeq0.8$), and the layer becomes semimetallic. By further increasing the field, the gap reopens into a trivial, semiconducting phase, i.e., without edge state. 

The gap induced by the external electric field is thus competing with the topological gap from the spin-orbit coupling, $\lambda_{SO}$ in \cref{eq:hamiltonian 2D-Armchair}. We explore such interplay in \cref{fig:dos-semi-germanene-arm-voltages-spin} by computing the zero-energy \gls{dos} as a function of both $\lambda_z$ and $\lambda_{SO}$. The semi-infinite armchair germanene layer is topologically nontrivial (trivial) when $\lambda_z<\lambda_{SO}$ ($\lambda_z>\lambda_{SO}$), with critical lines at $\lambda_z=\lambda_{SO}$. 
The spectral properties of germanene and other graphene-like materials are of great interest after the recent development of gap engineering methods~\cite{Raja2017May,Chaves2020Aug}. The \gls{gf} methods developed here provide a useful approach for the study of their spectral and topological properties.

%



\section{Conclusions\label{sec:conc}}
We have developed a general and analytic method for computing the microscopic Green's functions of a two-dimensional Dirac Hamiltonian with atomic-scale boundary conditions. Our approach is thus valid for honeycomb lattices with atomically well-defined zigzag and armchair edges, and contains all the relevant information about the scattering processes that take place at these edges, including spin, sublattice (from the honeycomb structure), and valley degrees of freedom. 

We tested our Green's function formalism by computing the spectral and topological properties of several well-known Dirac materials, like semimetal germanene and semiconducting transition metal dichalcogenides. Our approach allowed us to characterize bulk (infinite two-dimensional layer), semi-infinite, and finite layers. The zigzag termination resulted in the emergence of localized edge states, which have a topological nature in germanene samples \cite{Ezawa2015Oct}, and are affected, in nanoribbons (finite layers), by the spectral quantization resulting from the finite ribbon length. 

We go beyond previous works~\cite{Valley-PolarizedMetalandQAHEinSilicene_Ezawa_2012} showing that the topological properties of germanene are maintained for armchair terminations. We also showed how the topological gap for this material can be manipulated by an external electric field, which could be helpful for field effect transistor using germanene or silicene. This effect in silicene has been recently measured~\cite{Tao2015Mar}, and its potential use as a topological field effect transistor has also been theoretically explored~\cite{Ezawa2015Oct}. 

The analytical formalism presented here aligns with current efforts to explore and engineer novel two-dimensional junctions with interesting quantum applications \cite{Roy2014Jun}, as it paves the way for the exploration of exotic electronic phases in nanoscale junctions based on Dirac materials, including important effects like edge terminations, valley-dependent scattering, and finite size effects. 
In general, our approach provides an analytical alternative, with low computational cost, to simulate the low-energy electronic properties of graphene-like materials at the atomic level. It is straightforward to generalize it to study quantum transport in junctions with atomic-scale electric contacts~\cite{Guimaraes2016Jun, Leonard2011Dec, Allain2015Dec, OneDimensionalElectricalContactto2DMaterial_wang_2013}, or to include magnetic or superconducting orders~\cite{Milosevic2021}. 
In addition, the analytical expressions computed here offer great flexibility to include several effects, such as band gap engineering, external fields, and even light-matter interactions~\cite{Raja2017May,Chaves2020Aug, AntiferroSuperJunctionCAR_WeiTao_2021,Milosevic2022,Kheirabadi2022}. 

\section{Acknowledgments}

J.T. and W.J.H. acknowledge support from the Universidad 
Nacional de Colombia, project No.~57522. 
P. B. acknowledges support from the Spanish CM ``Talento Program'' project No.~2019-T1/IND-14088 and the Agencia Estatal de Investigaci\'on project No.~PID2020-117992GA-I00 and No.~CNS2022-135950. 

\appendix

\section{Iterative method}\label{sec:appendix-iterative}
In this section, we obtain $\check{Q}^{>}\left( x\right) $ given in \cref{eq:Rm_pert} following an iterative method. First, we start using \cref{eq:Gsemi-R} with $x<x^{\prime }$ in the definition of the \gls{gf}, 
\begin{equation}\label{eq:app1}
	\check{G}_{a}^{<}\left( x,x^{\prime }\right) = 
	\check{g}^{<}\left(x,x^{\prime }\right) + \check{g}^{>}\left( x,x_{a}\right) U_{a}\check{\tau}_{a} \check{G}_{RR}^{<}\left( x_{a},x^{\prime }\right) .
\end{equation}

Here, $\check{G}_{RR}^{<}\left( x_{a},x^{\prime }\right) $ is given by Dyson's equation, \cref{eq:Gsemi-sol}, as
\begin{equation}
	\check{G}_{RR}^{<}\left( x_{a},x^{\prime }\right) =\left[ 1+\check{Q}^{>}\left( x_{a}\right) \right] \check{g}^{<}\left( x_{a},x^{\prime }\right) .
\end{equation}
Substituting into \cref{eq:app1} we get
\begin{align}\label{eq:app2}
	\check{G}_{RR}^{<}\left( x,x^{\prime }\right) =&\check{g}^{<}\left(x,x^{\prime }\right) + \check{g}^{>}\left( x,x_{a}\right) \\ & \times U_{a}\check{\tau}_{a}\left( 1+\check{Q}^{>}\left( x_{a}\right) \right) \check{g}^{<}\left(x_{a},x^{\prime }\right) . \notag
\end{align}
Comparing \cref{eq:app2} with \cref{eq:Gsemi-sol}, namely,
\begin{equation*}
	\check{G}_{a}^{RR,<}\left( x,x^{\prime }\right) =\check{g}^{<}\left(
	x,x^{\prime }\right) + \check{Q}^{>}\left(x\right) \check{g}^{<}\left( x_{a},x^{\prime }\right) ,
\end{equation*}

we obtain the following equation for $\check{Q}^{>}\left( x\right) $, 
\begin{equation}\label{eq:app3}
	\check{Q}^{>}\left( x\right) =\check{g}^{>}\left( x,x_{a}\right) U_{a}\check{%
		\tau}_{a}+\check{g}^{>}\left( x,x_{a}\right) U_{a}\check{\tau}_{a}\check{Q}^{>}\left( x_{a}\right) . 
\end{equation}
We then solve \cref{eq:app3} recursively to find 
\begin{equation}
	\check{Q}^{>}\left( x\right) =U_{a}\check{g}^{>}\left( x,x_{a}\right) \check{%
		\tau}_{a}\sum\limits_{p=0}U_{a}^{p}\left( \check{g}^{>}\left( 0\right) 
	\check{\tau}_{a}\right) ^{p}.  \label{Eq_Rm_pert}
\end{equation}

\section{Scattering matrix}\label{sec:appendix-scattering}
We now derive the expression and general properties of the scattering matrix in \cref{eq:scat-mat}. 
To define the scattering problem, we consider a potential barrier at $x=x_a$. Incoming states from the left (right) of the barrier are $\psi_{m}^{+}\left( x\right)$ with amplitudes $a_{m}^{+}$ ($\psi _{m}^{-}\left( x\right) $ with $a_{m}^{-}$), which are solutions of \cref{eq:Dirac-x}. In the absence of a potential barrier, a general unperturbed scattering state reads
\begin{equation}
	\psi _{0}\left( x\right) =\sum\limits_{m,\varepsilon }a_{m}^{\varepsilon
	}\psi _{m}^{\varepsilon }\left( x\right) .
\end{equation}

In the presence of the barrier potential, the perturbed state to the left or right of the barrier is obtained using Dyson's equation as
\begin{equation}
	\psi _{L(R)}\left( x\right) =\psi _{0}\left( x\right) + \check{Q}^{<(>)}\left(x\right) \psi _{0}\left( x_{a}\right) ,
\end{equation}
with $\psi _{m}^{\varepsilon }\left( x\right) = \psi _{m}^{\varepsilon}f_{m}^{\varepsilon }\left( x\right) $ and $f^{\varepsilon}_{m}(x)= \e^{i\varepsilon k_m x }$. 

Owing to the translational invariance along the $x$-direction, we proceed taking $x_a=0$ without loss of generality. 
For simplicity, we also define $\psi_{R}\left( 0\right) =\psi _{R}$ and $\psi _{0}\left( 0\right) =\psi _{0}$ and obtain
\begin{align}
	\psi _{R} ={}&\left[ 1+\check{Q}^{>}\left( 0\right) \right] \psi _{0} 
	\nonumber \\
	={}&(1+\sum\limits_{n,m,\varepsilon }\left( \hat{r}_{a}^{+\varepsilon }\right)
	_{nm}\hat{P}_{nm}^{+\varepsilon })\sum\limits_{m^{\prime },\varepsilon
		^{\prime }}a_{m^{\prime }}^{\varepsilon ^{\prime }}\psi _{m^{\prime
	}}^{\varepsilon ^{\prime }}  \nonumber \\
	={}&\sum\limits_{m,\varepsilon }a_{m}^{\varepsilon }\psi _{m}^{\varepsilon
	}+\sum\limits_{n,m,\varepsilon }\left( \hat{r}_{a}^{+\varepsilon }\right)
	_{nm}a_{m}^{\varepsilon }\psi _{m}^{+},
\end{align}
where we have used that $\hat{P}_{nm}^{+\varepsilon }\psi _{m^{\prime
}}^{\varepsilon ^{\prime }}=\delta _{\varepsilon ,\varepsilon ^{\prime
}}\delta _{m,m^{\prime }}\psi _{m^{\prime }}^{\varepsilon ^{\prime }}$. 
Summing over $\varepsilon $ we get
\begin{equation}
	\psi _{R}=\sum\limits_{m}a_{m}^{-}\psi _{m}^{-}+\sum\limits_{n}b_{n}^{+}\psi_{n}^{+} ,
\end{equation}
with $b_{n}^{+}$ being the outgoing scattering amplitudes, 
\begin{equation}\label{eq:appb1}
	b_{n}^{+}=\sum\limits_{m}\left( \hat{r}_{a}^{+-}\right)
	_{nm}a_{m}^{-}+\sum\limits_{m}\left( \hat{t}_{a}^{++}\right) _{nm}a_{m}^{+} .
\end{equation}
Here, we have defined the transmission amplitudes as the elements of the following matrix
\begin{equation}
	\hat{t}_{a}^{++}=\hat{1}+\hat{r}_{a}^{++}, 
\end{equation}
which allows us to rewrite \cref{eq:appb1} as
\begin{equation}
\mathbf{b}^{+}=\hat{r}_{a}^{+-}\mathbf{a}^{-}+\hat{t}_{a}^{++}\mathbf{a}^{+},
\end{equation}
with 
\begin{align}
	\mathbf{b}^{+T} ={}&\left( b_{1},b_{2},...,b_{N}\right) ^{T}, 
	\label{Eq_def_t++} \\
	\mathbf{a}^{+T} ={}&\left( a_{1},a_{2},...,a_{N}\right) ^{T}. 
\end{align}

The perturbed scattering state to the left of the barrier is obtained analogously as
\begin{equation}
	\psi _{L}\left( 0\right) =\sum\limits_{m}a_{m}^{+}\psi_{m}^{+}+\sum\limits_{n}b_{n}^{-}\psi _{n}^{-},
\end{equation}
with
\begin{equation}
\mathbf{b}^{-}=\hat{r}_{a}^{-+}\mathbf{a}^{+}+\hat{t}_{a}^{--}\mathbf{a}^{-}, 
\end{equation}
and 
\begin{equation}\label{Eq_def_t--} 
	\hat{t}_{a}^{--}=\hat{1}+\hat{r}_{a}^{--} .
\end{equation}

Combining the previous results, we define the scattering matrix $\hat{S}$ as 
\begin{equation}
	\mathbf{b}=\check{S}\mathbf{a} ,
\end{equation}
with
\begin{equation}
	\mathbf{b} = 
	\begin{pmatrix}
		\mathbf{b}^{+} \\ 
		\mathbf{b}^{-}%
	\end{pmatrix}
	 , \quad \mathbf{a}=
	\begin{pmatrix}
		\mathbf{a}^{+} \\ 
		\mathbf{a}^{-}%
	\end{pmatrix} .
\end{equation}
Note that in the usual definition of the scattering matrix the reflection amplitudes are in the diagonal, which results from taking $\mathbf{b}^{T}=(\mathbf{b}^{-T}, \mathbf{b}^{+T})$. 
In what follows, however, we use
\begin{align}
	\check{S}_{a} ={}&
	\begin{pmatrix}
		\hat{t}_{a}^{++} & \hat{r}_{a}^{+-} \\ 
		\hat{r}_{a}^{-+} & \hat{t}_{a}^{--}%
	\end{pmatrix} \label{Eq_def_s} \\
	={}& 
	\begin{pmatrix}
		\hat{1}+\hat{r}_{a}^{++} & \hat{r}_{a}^{+-} \\ 
		\hat{r}_{a}^{-+} & \hat{1}+\hat{r}_{a}^{--}%
	\end{pmatrix} \notag \\
	={}& \check{1}+\check{r}_{a} . \notag
\end{align}

We can now set the potential barrier at an arbitrary position $x_{a}\neq 0$ changing the scattering matrix as 
\begin{equation}
	\check{S}\left( x_{a}\right) =\check{f}\left( -x_{a}\right) \check{S}\check{f}\left( x_{a}\right) ,
\end{equation}
with
\begin{equation}
\check{f}\left( x_{a}\right) =
\begin{pmatrix}
	\check{f}^{++}\left( x_{a}\right)  & 0 \\ 
	0 & \check{f}^{--}\left( x_{a}\right) 
\end{pmatrix} ,
\end{equation}
and 
\begin{equation}
\left( \hat{f}^{\varepsilon \varepsilon }\right) _{nm}\left( x\right) =\delta _{nm}\hat{f}_{m}^{\varepsilon }\left( x\right) . 
\end{equation}
Here, $\hat{S}\left( x_{a}\right) $ and $\hat{S}$ are related by an unitary transformation since $\hat{f}\left( -x_{0}\right) \hat{f} \left( x_{0}\right) =\hat{1}$. 

As it is usually interpreted, $\mathbf{b}$ represents the outgoing flux and $\mathbf{a}$ the incoming one. Consequently, the probability flux to the right and left of the barrier reads
\begin{align}
	J_{R} ={}& v_{F}\sum\limits_{n}\left( a_{n}^{-\ast }\psi _{n}^{-\dagger
	}+b_{n}^{+\ast }\psi _{n}^{+\dagger }\right) \hat{\alpha}_{x}\sum\limits_{n}%
	\left( a_{n}^{-}\psi _{n}^{-}+b_{n}^{+}\psi _{n}^{+}\right)  \nonumber \\
	={}&v_{F}\sum\limits_{n}\left( \left\vert b_{n}^{+}\right\vert
	^{2}-\left\vert a_{n}^{-}\right\vert ^{2}\right) , \\
	J_{L} ={}&v_{F}\sum\limits_{n}\left( a_{n}^{+\ast }\psi _{n}^{+\dagger
	}+a_{n}^{-\ast }\psi _{n}^{-\dagger }\right) \hat{\alpha}_{x}\sum\limits_{n}%
	\left( a_{n}^{+}\psi _{n}^{+}+a_{n}^{-}\psi _{n}^{-}\right)  \nonumber \\
	={}&v_{F}\sum\limits_{n}\left( \left\vert a_{n}^{+}\right\vert
	^{2}-\left\vert b_{n}^{-}\right\vert ^{2}\right) .
\end{align}
Conservation of the probability flux requires that $J_{L}=J_{R}$; therefore, 
\begin{align}
	\sum\limits_{n}\left( \left\vert a_{n}^{+}\right\vert ^{2}-\left\vert
	b_{n}^{-}\right\vert ^{2}\right) =&\sum\limits_{n}\left( \left\vert
	b_{n}^{+}\right\vert ^{2}-\left\vert a_{n}^{-}\right\vert ^{2}\right) , \\
	\sum\limits_{n}\left( \left\vert a_{n}^{+}\right\vert ^{2}+\left\vert
	a_{n}^{-}\right\vert ^{2}\right) =&\sum\limits_{n}\left( \left\vert
	b_{n}^{+}\right\vert ^{2}+\left\vert b_{n}^{-}\right\vert ^{2}\right) ,
\end{align}
which we can recast in vector form as $|\mathbf{b|}^{2}=|\mathbf{a|}^{2}$. 
As a result, we find that 
\begin{equation}
	\mathbf{a}^{\dagger }\check{S}^{\dagger }\check{S}\mathbf{a}\mathbf{=}|\mathbf{a|}^{2},
\end{equation}
and thus, 
\begin{equation}\label{eq:appb2}
	\mathbf{a}^{\dagger }\left( \check{S}^{\dagger }\check{S}-1\right) \mathbf{a} = 0 . 
\end{equation}
\Cref{eq:appb2} proves that the scattering matrix is unitary, that is, 
\begin{equation}
\check{S}^{\dagger }\check{S}=\check{S}\check{S}^{\dagger }=\check{1} ,
\end{equation}
which we can recast using the reflection matrix $\check{r}$ as
\begin{equation}
	\left( 1+\check{r}_{a}\right) \left( 1+\check{r}_{a}^{\dagger }\right)
	=\left( 1+\check{r}_{a}^{\dagger }\right) \left( 1+\check{r}_{a}\right) = \check{1} .
\end{equation}

\section{Nanoribbon Green's function and bound states}\label{sec:appendix-nanoribbon}

In this section we provide more details on the derivation of the nanoribbon \gls{gf}, \cref{eq:finite-GF}, and the associated bound states. 
For the case of two potential barriers placed at $x=x_a$ and $x=x_b>x_a$, the perturbed \gls{gf} for the region between the barriers is 
\begin{equation}\label{eq:appc1}
	\check{G}_{ab}\left( x,x^{\prime }\right) =\check{G}_{a}\left(x,x^{\prime }\right) + \check{Q}^{<}\left( x\right) \check{G}_{ab}^{>}\left( x_{b},x^{\prime }\right) ,
\end{equation}
for $x>x^{\prime }$, where 
\begin{gather}\label{eq:GF-finite2}
	\check{G}_{ab}^{>}\left( x,x^{\prime }\right) = \frac{-i}{2\hbar v_{F}} 
	\\ \times 
	\sum\limits_{n,m, \varepsilon,\varepsilon'} f_{n}^{\varepsilon}( x- x_{b} ) (\check{w}^{>})_{nm}^{\varepsilon \varepsilon'} f_{m}^{\varepsilon'} (x_{b}- x^{\prime}) \psi_n^{\varepsilon} (\bar{\psi}_{m}^{\varepsilon'})^T , \notag
\end{gather}
with $\check{w}^{>}$ given by \cref{eq:pert-finite}, see also \cref{eq:GF-finite}. 

\Cref{eq:appc1} can be solved after obtaining $\check{G}_{ab}^{>}$. However, to do so, one can not take the limit $U_{b}\rightarrow \infty $ and invert $\check{w}$, because the matrix $\check{\tau}$ has no inverse ($\mathrm{det}[\check{\tau}]=0$). 
To circumvent this problem, we define the matrix
\begin {equation}
	\check{N}=\hat{1}-AU_{b}\check{\tau}, 
\end{equation}
with $A=-i/(2\hbar v_{F})$, and
\begin{equation}
	\check{N}=  
	\begin{pmatrix}
		\hat{N}_{1} & \hat{N}_{2} \\ 
		\hat{N}_{3} & \hat{N}_{4}%
	\end{pmatrix} ,
\end{equation}
where
\begin{align*}
	\hat{N}_{1} ={}&\hat{1}-AU_{b}\hat{r}_{a}^{+-}\left( W\right) \hat{\tau}_{b}^{-+} , \, &
	\hat{N}_{2} ={}&-AU_{b}\hat{r}_{a}^{+-}\left( W\right) \hat{\tau}_{b}^{--} , \\
	\hat{N}_{3} ={}&-AU_{b}\hat{\tau}_{b}^{-+} , \, &
	\hat{N}_{4} ={}&AU_{b}\hat{\tau}_{b}^{-+}\left( \hat{r}_{b}^{-+}\right) ^{-1}.
\end{align*}
We can now compute the inverse of $\hat{N}$ and then take the limit $U_{b}\rightarrow \infty $. To do so, we use Schur complement, which, for example, is defined for submatrix $\hat{N}_{4}$ as
\begin{equation*}
\hat{C}=\hat{N}_{1}-\hat{N}_{2}\hat{N}_{4}{}^{-1}\hat{N}_{3} .
\end{equation*}
Therefore, defining
\begin{equation*}
	\hat{r}_{b}^{-+}=AU_{b}\left( 1-AU_{b}^{--}\hat{\tau}_{b}\right) ^{-1}\hat{\tau}_{b}^{^{-+}},
\end{equation*}
we get 
\begin{equation}
	\hat{C}=\hat{1}-\hat{r}_{a}^{+-}\left( W\right) \hat{r}_{b}^{-+} ,
\end{equation}
with $W=x_b-x_a$. 

\begin{widetext}
The inverse matrix of $\hat{N}$ is thus
\begin{equation}
	\check{N}^{-1}= 
	\begin{pmatrix}
		\hat{1} & 0 \\ 
		-\hat{N}_{4}^{-1}\hat{N}_{3} & \hat{1}%
	\end{pmatrix}
	\begin{pmatrix}
		\hat{C}^{-1} & 0 \\ 
		0 & \hat{N}_{4}^{-1}%
	\end{pmatrix}
	\begin{pmatrix}
		\hat{1} & -\hat{N}_{2}\hat{N}_{4}^{-1} \\ 
		0 & \hat{1}%
	\end{pmatrix} ,
\end{equation}
which simplifies to 
\begin{equation}
	\check{N}^{-1}= 
	\begin{pmatrix}
		\hat{C}^{-1} & \hat{C}^{-1}\hat{r}_{a}^{+-}\left( W\right) \hat{\tau}%
		_{b}^{--}\hat{r}_{b}^{-+}\left( \hat{\tau}_{b}^{-+}\right) ^{-1} \\ 
		\hat{r}_{b}^{-+}\hat{C}^{-1} & U_{b}^{-1}\hat{r}_{b}^{-+}\hat{C}^{-1}\left( 
		\hat{1}-U_{b}\hat{r}_{a}^{+-}\left( W\right) \hat{\tau}_{b}^{-+}\right)
		\left( \hat{\tau}_{b}^{-+}\right) ^{-1}%
	\end{pmatrix} .
\end{equation}

\end{widetext}

We can now take the limit $U_{b}\rightarrow \infty $, resulting in
\begin{equation}
	\check{w}^{<}= 
	\begin{pmatrix}
		\hat{C}^{-1}\hat{r}_{a}^{+-}\left( W\right) \hat{r}_{b}^{-+} & -\hat{C}^{-1}%
		\hat{r}_{a}^{+-}\left( W\right)  \\ 
		\hat{r}_{b}^{-+}\hat{C}^{-1} & -\hat{r}_{b}^{-+}\hat{C}^{-1}\left( \hat{r}%
		_{b}^{-+}\right) ^{-1}%
	\end{pmatrix} .
\end{equation}
Analogously, we get 
\begin{equation}
	\check{w}^{>}= 
	\begin{pmatrix}
		\hat{D}^{++} & \hat{D}^{++}\hat{r}_{a}^{+-}\left( x_{a}\right)  \\ 
		\hat{r}_{b}^{-+}\left( x_{b}\right) \hat{D}^{++} & \hat{r}_{b}^{-+}\left(
		x_{b}\right) \hat{D}^{++}\hat{r}_{a}^{+-}\left( x_{a}\right) 
	\end{pmatrix} ,
\end{equation}
with 
\begin{align}
	\hat{D}^{++} ={}&\left[ \hat{1}-\hat{r}_{a}^{+-}\left( x_{a}\right) \hat{r}_{b}^{-+}\left( x_{b}\right) \right] ^{-1} , \label{Eq_Dmm_pelicula} \\
	\hat{D}^{--} ={}&\left[ \hat{1}-\hat{r}_{b}^{-+}\left( x_{b}\right) \hat{r}_{a}^{+-}\left( x_{a}\right) \right] ^{-1} . 
\end{align}
As a result, we obtain the final form of the nanoribbon \gls{gf} as 
\begin{equation}\label{eq:finite-GF-app}
	\check{G}_{ab}^{\lessgtr}\left( x,x^{\prime }\right) = \frac{-i}{2\hbar v_{F}} \sum\limits_{\substack{n,m\\ \varepsilon,\varepsilon'}} f_{n}^{\varepsilon} \left( x\right)( \check{w}^{\lessgtr} )_{nm}^{\varepsilon\varepsilon'} f_{m}^{\varepsilon'} \left(-x'\right) \psi_n^{\varepsilon} (\bar{\psi}_{m}^{\varepsilon'})^T  .
\end{equation}

The nanoribbon's bound states are obtained from the denominator of the \gls{gf}, that is, setting the inverse of $\hat{D}^{++}$ or, equivalently, $\hat{D}^{--}$, to zero. 
From \cref{Eq_Dmm_pelicula}, this condition reduces to
\begin{equation}
	\hat{r}_{a}^{+-}\left( x_{a}\right) \hat{r}_{b}^{-+}\left( x_{b}\right) = \hat{1}.  \label{Eq_estados}
\end{equation}
We can interpret this condition as follows: inside the nanoribbon, the scattering state is a superposition of left and right movers, namely,
\begin{equation}
	\psi =\sum\limits_{n}\left( c_{n}^{+}\psi _{n}^{+}+c_{n}^{-}\psi_{n}^{-}\right) .
\end{equation}
At each potential barrier, $x=x_{a,b}$, the amplitudes for left and right movers, $c_{n}^{-}$ and $c_{n}^{+}$, respectively, are related by
\begin{align}
	c_{n}^{-}={}&\sum\limits_{m}\hat{r}_{nm}^{-+}\left( x_{b}\right) c_{m}^{+} , \\
	c_{n}^{+}={}&\sum\limits_{m}\hat{r}_{nm}^{+-}\left( x_{a}\right) c_{m}^{-}. 
\end{align}
In matrix form we have
\begin{align}
	\mathbf{c}^{-} ={}& \hat{r}^{-+}\left( x_{b}\right) \mathbf{c}^{+} , \\
	\mathbf{c}^{+} ={}& \hat{r}^{+-}\left( x_{a}\right) \mathbf{c}^{-} ,
\end{align}
which form the closed cycle
\begin{equation}
	\mathbf{c}^{+}=\hat{r}^{+-}\left( x_{a}\right) \hat{r}^{-+}\left(x_{b}\right) \mathbf{c}^{+} .
\end{equation}
It is thus straightforward to get
\begin{equation}
	\mathbf{c}^{+}\left( \hat{1}-\hat{r}^{+-}\left( x_{a}\right) \hat{r}^{-+}\left( x_{b}\right) \right)  =\mathbf{0} ,
\end{equation}
which corresponds to the zeroes of the \gls{gf}, as shown in \cref{Eq_estados}. 
This result shows that the nanoribbon's bound states are determined by the reflection matrices at each edge of the finite region, and can thus be obtained by the zeroes of the inverse of \cref{Eq_Dmm_pelicula}.

\section{Green's functions for Dirac systems with well-defined edges}\label{sec:app-applications}
This appendix contains the necessary definitions and supplementary calculations to derive the \gls{gf} of the zigzag Hamiltonian in \cref{eq:hamiltonian 2D-zz} and the armchair one in \cref{eq:hamil-arm}. 

\subsection{Zigzag Hamiltonian. }
The normalized eigenstates associated to \cref{eq:hamiltonian 2D-zz} are
\begin{subequations}\label{eq:zz-eigen}
\begin{align}
\psi_{s \eta}^{+} ={}&\frac{1}{\sqrt{2\cos \alpha_{s\eta} }}\left( 
\begin{array}{c}
N_{s \eta}^{-1}\e^{-i\frac{\alpha_{s\eta} }{2}} \\ 
N_{s \eta}\e^{i\frac{\alpha_{s\eta} }{2}}%
\end{array}%
\right) ,
\\
\psi_{s \eta} ^{-}={}&\frac{1}{\sqrt{2\cos
\alpha_{s\eta} }}\left( 
\begin{array}{c}
N_{s \eta}^{-1}\e^{i\frac{\alpha_{s\eta} }{2}} \\ 
-N_{s \eta}\e^{-i\frac{\alpha_{s\eta} }{2}}%
\end{array}%
\right) ,
\\
\tilde{\psi}_{s \eta}^{+} ={}&\frac{1}{\sqrt{%
2\cos \alpha_{s\eta} }}\left( 
\begin{array}{c}
N_{s \eta}\e^{i\frac{\alpha_{s\eta} }{2}} \\ 
N_{s \eta}^{-1}\e^{-i\frac{\alpha_{s\eta} }{2}}%
\end{array}
\right) ,
\\
\tilde{\psi}_{s \eta}^{-}={}&\frac{1}{%
\sqrt{2\cos \alpha_{s\eta} }}\left( 
\begin{array}{c}
N_{s \eta}\e^{i\frac{-\alpha_{s\eta} }{2}} \\ 
-N_{s \eta}^{-1}\e^{i\frac{\alpha_{s\eta} }{2}}%
\end{array}%
\right) .
\end{align}
\end{subequations}
The corresponding transposed states are
\begin{align}
\bar{\psi}_{s \eta}^{+} ={}&\frac{1}{\sqrt{%
2\cos \alpha_{s\eta} }}\left( 
\begin{array}{c}
N_{s \eta}^{-1}\e^{i\frac{\alpha_{s\eta} }{2}} \\ 
N_{s \eta}\e^{-i\frac{\alpha_{s\eta} }{2}}%
\end{array}%
\right) , \\
\bar{\psi}_{s \eta}^{-} ={}&\frac{1}{\sqrt{%
2\cos \alpha_{s\eta} }}\left( 
\begin{array}{c}
N_{s \eta}^{-1}\e^{-i\frac{\alpha_{s\eta} }{2}} \\ 
-N_{s \eta}\e^{i\frac{\alpha_{s\eta} }{2}}%
\end{array}%
\right).
\end{align}

Next, the bulk \glspl{gf} reads
\begin{align}
g_{s\eta}^{<}\left( x,x^{\prime }\right) ={}&\frac{\e^{ik_{s\eta}\left( x^{\prime }-x\right) }
}{2\cos \alpha_{s\eta} } 
\begin{pmatrix}
N_{s \eta}^{-2} & -\e^{i\alpha_{s\eta} } \\ 
-\e^{-i\alpha_{s\eta} } & N_{s \eta}^2%
\end{pmatrix} ,%
\\
g_{s\eta}^{>}\left( x,x^{\prime }\right) = {}& \frac{\e^{ik_{s\eta}\left( x-x^{\prime }\right) }
}{2\cos \alpha_{s\eta} }
\begin{pmatrix}
N_{s \eta}^{-2} & \e^{-i\alpha_{s\eta} } \\ 
\e^{i\alpha_{s\eta} } & N_{s \eta}^{2}%
\end{pmatrix} .%
\end{align}

For a semi-infinite zigzag layer with A-type border, the projections of $\hat{\tau}$ into the eigenstates are
\begin{subequations}\label{eq:s11}
\begin{align}
\hat{S}_{s \eta}^{++} ={}&\hat{S}_{s \eta}^{--}=\frac{1}{2N_{s \eta}^{2}\cos \alpha_{s\eta} } ,
\\
\hat{S}_{s \eta}^{+-} ={}&\frac{\e^{i\alpha_{s\eta} }}{2N_{s \eta}^2\cos \alpha_{s\eta} } ,
\\
\hat{S}_{s \eta}^{-+} ={}&\frac{\e^{-i\alpha_{s\eta} }}{2N_{s \eta}^2\cos \alpha_{s\eta} } .
\end{align}
\end{subequations}
Similar expressions are obtained for border B. 

The semi-infinite zigzag \gls{gf} can thus be compactly written as
\begin{equation}
\hat{G}_{RR,s\eta}\left( x,x^{\prime }\right) =\hat{g}_{s\eta}\left( x,x^{\prime
}\right) -
\e^{i\alpha_{s\eta} }\e^{ik_{s\eta}\left( x+x^{\prime }-2x_{0}\right) }\hat{M}^{+-}_{s \eta} ,
\end{equation}
where, 
\begin{equation}
    \hat{M}_{s \eta}^{\epsilon \epsilon'}=\psi_{s \eta}^{\epsilon}(\bar{\psi}_{s \eta}^{\epsilon'})^{T} .
\end{equation}
Substituting the eigenstates, \cref{eq:zz-eigen}, we get
\begin{gather}
\hat{G}_{RR,s\eta}^{>}\left( x,x^{\prime }\right)
= A \left[ \frac{\e^{ik_{s\eta}\left( x-x^{\prime }\right)
}}{2\cos \alpha_{s\eta} } 
\begin{pmatrix}
N_{s \eta}^{-2} & \e^{-i\alpha_{s\eta} } \\ 
\e^{i\alpha_{s\eta} } & N_{s \eta}^{2}%
\end{pmatrix} - \right. \nonumber
\\ %
\left. \frac{\e^{i\alpha_{s\eta}}\e^{ik_{s\eta}\left( x+x^{\prime }-2x_{a}\right)}}{2\cos \alpha_{s\eta} } 
\begin{pmatrix}
N_{s \eta}^{-2}\e^{-i\alpha_{s\eta} } & -1 \\ 
1 & -N_{s \eta}^{2}\e^{i\alpha_{s\eta}}  
\end{pmatrix} \right],
\end{gather}
with $A=-i/(2\hbar v_{F})$. 

\begin{figure}[t!]
\centering
\hspace*{-1cm}  
\includegraphics[scale=0.4]{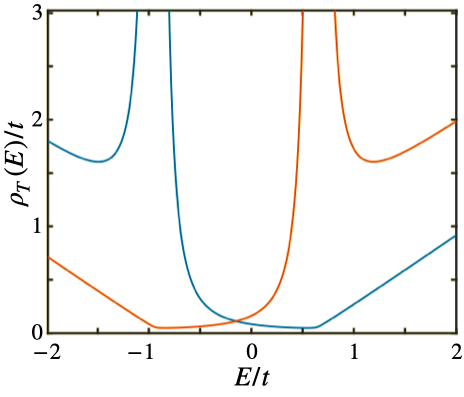}
\caption{Density of states at the zigzag edge of a semi-infinite \gls{tmd} layer with A- (blue) and B-type (red) termination. All parameters are the same as in \cref{fig: zz_semi_TMDC} for valley $K$ and spin up electrons. }
\label{fig:border-DOS}
\end{figure}

The \gls{gf} evaluated at the zigzag edge with A-atom termination reads as
\begin{equation}
\label{eq:GF-Semi-zigzag}
\hat{G}_{RR,s\eta}^{>}\left( x_{a},x_{a}\right) =-\frac{i}{\hbar v_{F}}
\begin{pmatrix}
0 & 1 \\ 
0 & N_{s \eta}^2\e^{i\alpha_{s\eta} }%
\end{pmatrix} .
\end{equation}
Analogously, when the zigzag edge is terminated in B-type atoms, the \gls{gf} adopts the same for as in \cref{eq:GF-Semi-zigzag} with the change $ N_{s \eta}^2 \rightarrow 1/N_{s \eta}^2$, cf. \cref{eq:N}. This simple inversion plays an important role in the spectral properties of the layer, as schematically described in \cref{fig: qualitative_result_band_for_Edge}. 
To show this effect more clearly, we plot in \cref{fig:border-DOS} the density of states for each edge termination, A and B, for a semiconducting \gls{tmd}, showcasing the band inversion. 

Finally, a compact expression for the zigzag nanoribbon \gls{gf} is
\begin{gather}
\hat{G}_{ab,s\eta}^{>}\left( x,x^{\prime }\right) 
=AD_{s \eta}^{++}\left( \e^{ik_{s\eta}x}\psi_{s \eta} ^{+}+r^{-+}_{s \eta}\e^{-ik_{s\eta}\left( x-2x_{b}\right) }\psi_{s \eta}
^{-}\right) \nonumber
\\ \times 
\left( \e^{-ik_{s\eta}x^{\prime }}\psi_{s \eta} ^{+\dagger }+r^{+-}_{s \eta} \e^{ik_{s\eta}\left(
x^{\prime }-2x_{a}\right) }\psi_{s \eta} ^{-\dagger }\right) ,
\end{gather}
where 
\begin{equation}
D^{++}_{s \eta}=\left[ 1-r^{+-}_{s \eta}r^{-+}_{s \eta} \e^{2ik_{s\eta}W }\right] ^{-1} .
\end{equation}
It is thus straightforward to get the edge \gls{gf} as
\begin{equation}
\label{eq:GF-Nano-ZigZag}
\hat{G}_{ab,s\eta}^{>}\left( x_{a},x_{a}\right) = \frac{-i}{\hbar v_{F}}
\begin{pmatrix}
0 & \frac{ 1+\e^{2i\alpha_{s\eta} }\e^{i2k_{s\eta}W} }{ 1+\e^{+2i\alpha_{s\eta} }\e^{2ik_{s\eta}W}}\\ 
0 & \frac{N_{s \eta}^2\e^{i\alpha_{s\eta} }\left( 1-\e^{i2k_{s\eta}W}\right)}{1+\e^{+2i\alpha_{s\eta} }\e^{2ik_{s\eta}W} }%
\end{pmatrix} ,
\end{equation}
valid when the left edge ($x_a$) has A-type termination. As explained above, for terminations on B atoms, \cref{eq:GF-Nano-ZigZag} is changed as $ N_{s \eta}^2 \rightarrow N_{s \eta}^{-2}$. 

\subsection{Armchair Hamiltonian. }
Being block-diagonal, the solutions of \cref{eq:hamil-arm} are projected into each valley, 
with normalized eigenstates 
\begin{subequations}\label{eq:armchair}
\begin{align}
\varphi _{K}^{+} ={}&\frac{1}{\sqrt{2\cos \alpha_s }}
\begin{pmatrix}
N^{-1}\e^{-i\frac{\alpha_s }{2}} \\ 
N\e^{i\frac{\alpha_s }{2}}%
\end{pmatrix} ,
\\
\varphi _{K}^{-}={}&\frac{1}{\sqrt{2\cos \alpha_s }}
\begin{pmatrix}
N^{-1}\e^{i\frac{\alpha_s }{2}} \\ 
-N\e^{-i\frac{\alpha_s }{2}}
\end{pmatrix} ,
\\
\varphi _{K^{\prime }}^{+} ={}& \frac{1}{\sqrt{2\cos \alpha_s ^{\prime }}}
\begin{pmatrix}
N^{\prime -1}\e^{i\frac{\alpha_s ^{\prime }}{2}} \\ 
-N^{\prime }\e^{-i\frac{\alpha_s ^{\prime }}{2}}%
\end{pmatrix} ,
\\
\varphi _{K^{\prime }}^{-}={}&\frac{1}{\sqrt{2\cos \alpha_s ^{\prime }}}
\begin{pmatrix}
N^{\prime -1}\e^{-i\frac{\alpha_s ^{\prime }}{2}} \\ 
N^{\prime }\e^{i\frac{\alpha_s ^{\prime }}{2}}%
\end{pmatrix} , 
\end{align}
\end{subequations}
and transposed states
\begin{equation}
\label{eq:armchair-5}
\tilde{\varphi}_{K^{\prime }}^{-}=\frac{1}{\sqrt{2\cos \alpha_s ^{\prime }}}\left( 
\begin{array}{c}
N^{\prime }\e^{i\frac{\alpha_s ^{\prime }}{2}} \\ 
N^{\prime -1 }\e^{-i\frac{\alpha_s ^{\prime }}{2}}
\end{array}\right)
\end{equation}
and $\tilde{\varphi}_{n}^{\epsilon}=\hat{\sigma}_{x}\varphi_{n}^{\epsilon}$. 
We define $N\!\equiv\! N_{sK}$ and $N'\!\equiv\! N_{sK'}$, see \cref{eq:N}, for valley $K$ and $K'$, respectively. Similarly, primed quantities ($\alpha_s',k_s',\dots$) refer to valley $K'$. 

The semi-infinite armchair \gls{gf} at the edge is 
\begin{gather}
\hat{G}_{RR}^{<}\left( x_{0},x_{0}\right)
= \left( \psi _{K}^{-}\allowbreak +r_{KK}^{+-}\psi _{K}^{+}+r_{K^{\prime}K}^{+-}\psi _{K^{\prime }}^{+}\right) \psi _{K}^{-\dagger } 
\nonumber \\
+ \left( \psi
_{K^{\prime }}^{-}\allowbreak +r_{K^{\prime }K^{\prime }}^{+-}\psi
_{K^{\prime }}^{+}+r_{KK^{\prime }}^{+-}\psi _{K}^{+}\right) \psi
_{K^{\prime }}^{-\dagger } .
\end{gather}

The valley-projected scattering amplitudes, such as $r_{K^{\prime }K^{\prime }}^{+-} $, can be recast in terms of the auxiliary scalar quantities $h_{nm}^{\epsilon\epsilon'}=\tilde{\varphi}_n^{\epsilon \dagger}\varphi_m^{\epsilon'}$. 
For instance, $r_{K^{\prime }K^{\prime }}^{+-} = h_{K^{\prime }K}^{++} / h_{K^{\prime}K}^{-+}$. 
Using \cref{eq:armchair}, we get 
\begin{subequations}
\begin{gather}
h_{KK^{\prime }}^{++} =\tilde{\varphi}_{K}^{+\dagger}\varphi _{K^{\prime }}^{+}
\\ \nonumber
=\frac{\e^{i\frac{(\alpha_s ^{\prime }-\alpha_s )%
}{2}}}{2\sqrt{\cos \alpha_s \cos
\alpha_s ^{\prime } }} \left( \frac{N}{N^{\prime}}-\frac{N^{\prime }}{N}\e^{-i(\alpha_s
^{\prime }-\alpha_s )}\right) ,
\end{gather}
\begin{gather}
h_{K^{\prime }K}^{-+} = h_{KK^{\prime }}^{-+} = \tilde{\varphi}_{K}^{-\dagger}\varphi _{K^{\prime }}^{+}
\\ \nonumber
=\frac{\e^{i\frac{\left( \alpha_s ^{\prime }+\alpha_s \right) }{2}}}{2\sqrt{\cos \alpha_s ^{\prime }\cos
\alpha_s }}\left( \frac{N}{N^{\prime }}+%
\frac{N^{\prime }}{N}\e^{-i\left( \alpha_s ^{\prime }+\alpha_s \right) }\right) ,
\end{gather}
\begin{gather}
h_{K^{\prime }K}^{++} = 
\tilde{\varphi}_{K^{\prime }}^{+\dagger}\varphi _{K}^{+}
\\ \nonumber
=\frac{\e^{i\frac{(\alpha_s ^{\prime }-\alpha_s )}{2}}}{%
2\sqrt{\cos \alpha_s \cos \alpha_s ^{\prime }}}\left( \frac{N'}{N}-\frac{N}{N'}\e^{-i(\alpha_s ^{\prime }-\alpha_s )}\right) .
\end{gather}
\end{subequations}

The resulting scattering amplitudes are
\begin{subequations}
\begin{align}
r_{KK}^{+-}
={}& \e^{i\alpha_s }\frac{N^{2}\e^{-i\alpha_s }-N^{\prime 2 }\e^{-i\alpha_s ^{\prime }}}{%
N^{2}\e^{i\alpha_s }+N^{\prime 2}\e^{-i\alpha_s ^{\prime }}} ,
\\
r_{K^{\prime }K}^{+-} ={}& -2 \frac{NN^{\prime }\sqrt{\cos \alpha_s \cos \alpha_s ^{\prime }}}{N^2\e^{i\alpha_s }+N^{\prime
2}\e^{-i\alpha_s ^{\prime }}} \e^{i\frac{\alpha_s -\alpha_s
^{\prime }}{2}} ,
\\
r_{KK^{\prime }}^{+-} 
={}& -2 \frac{NN^{\prime } \sqrt{\cos \alpha_s ^{\prime }\cos \alpha_s } }{N^{\prime 2
}\e^{-i\alpha_s ^{\prime }}+N^{2}\e^{i\alpha_s }} \e^{i\frac{\alpha_s -\alpha_s
^{\prime }}{2}} ,
\\
r_{K^{\prime }K^{\prime }}^{+-} ={}& \e^{i\alpha_s ^{\prime }}\frac{N^{\prime 2}-N^{2}\e^{-i(\alpha _s^{\prime
}-\alpha_s )}}{N^{\prime 2}+N^{2}\e^{i\left( \alpha_s ^{\prime }+\alpha_s
\right) }} .
\end{align}
\end{subequations}

Finally, the armchair nanoribbon \gls{gf} contains elements in all valley projections. The projection into valleys $n=K,K'$ and $m=K,K'$, $G_{nm}^{>} (x,x')$, is shown in the main text, \cref{eq:gf-arm-finite}, and repeated here for completeness:
\begin{equation}
G_{nm}^{>} (x,x')= \sum\limits_{ \varepsilon,\varepsilon^{\prime}=\pm  }
\hat{f}_{nm}^{\varepsilon\varepsilon^{\prime}}\left(
x,x^{\prime }\right) \hat{r}_{nm}^{>,\varepsilon\varepsilon^{\prime}} \hat{M}_{nm}^{\varepsilon\varepsilon^{\prime}} .
\end{equation}

Here, $f_{nm}^{\varepsilon\varepsilon^{\prime}}(x,x^{\prime})=f_{n}^{\varepsilon}(x)f_{m}^{\epsilon^\prime}(x')$, where $f_{n}^{\varepsilon}(x)=\e^{i(K_n+\varepsilon k_n)x}$, $K_n=K,K'$, and $k_n=k_s,k_s'$. 
Moreover, the matrices of reflection coefficients projected into each valley subspace, $r^{>,\varepsilon\varepsilon^{\prime}}_{nm}$, with $\varepsilon,\varepsilon^{\prime}=\pm$, are obtained from 
\begin{subequations}
\begin{align}
\hat{r}_{a}^{+-}\left( x_{a}\right) = {}& 
\begin{pmatrix}
\frac{h_{KK^{\prime }}}{h_{K K^{\prime }}}\e^{-2ik_s x_{a}} & \frac{-1}{h_{K^{\prime }K}}\e^{-i\left( 2K+k_s+k_s^{\prime }\right) x_{a}} \\ 
\frac{-1}{h_{KK^{\prime }}}\e^{i\left( 2K-k_s-k_s^{\prime }\right)
x_{a}} & \frac{h_{K^{\prime }K}}{h_{K^{\prime }K}}\e^{-2ik_s^{\prime
}x_{a}}%
\end{pmatrix} ,
\\
\hat{r}_{b}^{-+}\left( x_{b}\right) = {}&
\begin{pmatrix}
\frac{h_{KK^{\prime }}^{--}}{h_{KK^{\prime }}^{+-}}\e^{2ik_s x_{b}} &
\frac{-1}{h_{K^{\prime }K}^{+-}}\e^{i\left( -2K+k_s+k_s^{\prime }\right) x_{b}} \\ 
\frac{-1}{h_{KK^{\prime }}^{+-}}\e^{i\left( 2K+k_s^{\prime }+k_s \right)
x_{b}} & \frac{h_{K^{\prime }K}^{--}}{h_{K^{\prime }K}^{+-}}\e^{2ik_s^{\prime} x_{b}}
\end{pmatrix} .
\end{align}
\end{subequations}

\bibliography{Dirac_Gen_GF}

\begin{thebibliography}{71}%
\makeatletter
\providecommand \@ifxundefined [1]{%
 \@ifx{#1\undefined}
}%
\providecommand \@ifnum [1]{%
 \ifnum #1\expandafter \@firstoftwo
 \else \expandafter \@secondoftwo
 \fi
}%
\providecommand \@ifx [1]{%
 \ifx #1\expandafter \@firstoftwo
 \else \expandafter \@secondoftwo
 \fi
}%
\providecommand \natexlab [1]{#1}%
\providecommand \enquote  [1]{``#1''}%
\providecommand \bibnamefont  [1]{#1}%
\providecommand \bibfnamefont [1]{#1}%
\providecommand \citenamefont [1]{#1}%
\providecommand \href@noop [0]{\@secondoftwo}%
\providecommand \href [0]{\begingroup \@sanitize@url \@href}%
\providecommand \@href[1]{\@@startlink{#1}\@@href}%
\providecommand \@@href[1]{\endgroup#1\@@endlink}%
\providecommand \@sanitize@url [0]{\catcode `\\12\catcode `\$12\catcode
  `\&12\catcode `\#12\catcode `\^12\catcode `\_12\catcode `\%12\relax}%
\providecommand \@@startlink[1]{}%
\providecommand \@@endlink[0]{}%
\providecommand \url  [0]{\begingroup\@sanitize@url \@url }%
\providecommand \@url [1]{\endgroup\@href {#1}{\urlprefix }}%
\providecommand \urlprefix  [0]{URL }%
\providecommand \Eprint [0]{\href }%
\providecommand \doibase [0]{https://doi.org/}%
\providecommand \selectlanguage [0]{\@gobble}%
\providecommand \bibinfo  [0]{\@secondoftwo}%
\providecommand \bibfield  [0]{\@secondoftwo}%
\providecommand \translation [1]{[#1]}%
\providecommand \BibitemOpen [0]{}%
\providecommand \bibitemStop [0]{}%
\providecommand \bibitemNoStop [0]{.\EOS\space}%
\providecommand \EOS [0]{\spacefactor3000\relax}%
\providecommand \BibitemShut  [1]{\csname bibitem#1\endcsname}%
\let\auto@bib@innerbib\@empty
\bibitem [{\citenamefont {Tan}\ \emph {et~al.}(2017)\citenamefont {Tan},
  \citenamefont {Cao}, \citenamefont {Wu}, \citenamefont {He}, \citenamefont
  {Yang}, \citenamefont {Zhang}, \citenamefont {Chen}, \citenamefont {Zhao},
  \citenamefont {Han}, \citenamefont {Nam}, \citenamefont {Sindoro},\ and\
  \citenamefont {Zhang}}]{RecentAdvancesinUltrathin2DNanomaterials_Tan2017May}%
  \BibitemOpen
  \bibfield  {author} {\bibinfo {author} {\bibfnamefont {C.}~\bibnamefont
  {Tan}}, \bibinfo {author} {\bibfnamefont {X.}~\bibnamefont {Cao}}, \bibinfo
  {author} {\bibfnamefont {X.-J.}\ \bibnamefont {Wu}}, \bibinfo {author}
  {\bibfnamefont {Q.}~\bibnamefont {He}}, \bibinfo {author} {\bibfnamefont
  {J.}~\bibnamefont {Yang}}, \bibinfo {author} {\bibfnamefont {X.}~\bibnamefont
  {Zhang}}, \bibinfo {author} {\bibfnamefont {J.}~\bibnamefont {Chen}},
  \bibinfo {author} {\bibfnamefont {W.}~\bibnamefont {Zhao}}, \bibinfo {author}
  {\bibfnamefont {S.}~\bibnamefont {Han}}, \bibinfo {author} {\bibfnamefont
  {G.-H.}\ \bibnamefont {Nam}}, \bibinfo {author} {\bibfnamefont
  {M.}~\bibnamefont {Sindoro}},\ and\ \bibinfo {author} {\bibfnamefont
  {H.}~\bibnamefont {Zhang}},\ }\bibfield  {title} {\bibinfo {title} {Recent
  advances in ultrathin two-dimensional nanomaterials},\ }\href
  {https://doi.org/10.1021/acs.chemrev.6b00558} {\bibfield  {journal} {\bibinfo
   {journal} {Chem. Rev.}\ }\textbf {\bibinfo {volume} {117}},\ \bibinfo
  {pages} {6225} (\bibinfo {year} {2017})}\BibitemShut {NoStop}%
\bibitem [{\citenamefont {Varghese}\ \emph {et~al.}(2015)\citenamefont
  {Varghese}, \citenamefont {Varghese}, \citenamefont {Swaminathan},
  \citenamefont {Singh},\ and\ \citenamefont {Mittal}}]{Varghese2015Sep}%
  \BibitemOpen
  \bibfield  {author} {\bibinfo {author} {\bibfnamefont {S.~S.}\ \bibnamefont
  {Varghese}}, \bibinfo {author} {\bibfnamefont {S.~H.}\ \bibnamefont
  {Varghese}}, \bibinfo {author} {\bibfnamefont {S.}~\bibnamefont
  {Swaminathan}}, \bibinfo {author} {\bibfnamefont {K.~K.}\ \bibnamefont
  {Singh}},\ and\ \bibinfo {author} {\bibfnamefont {V.}~\bibnamefont
  {Mittal}},\ }\bibfield  {title} {\bibinfo {title} {Two-dimensional materials
  for sensing: Graphene and beyond},\ }\href
  {https://doi.org/10.3390/electronics4030651} {\bibfield  {journal} {\bibinfo
  {journal} {Electronics}\ }\textbf {\bibinfo {volume} {4}},\ \bibinfo {pages}
  {651} (\bibinfo {year} {2015})}\BibitemShut {NoStop}%
\bibitem [{\citenamefont {Liu}\ \emph {et~al.}(2020)\citenamefont {Liu},
  \citenamefont {Chen}, \citenamefont {Wang}, \citenamefont {Liu},
  \citenamefont {Jiang}, \citenamefont {Zhang}, \citenamefont {Liu},\ and\
  \citenamefont {Zhou}}]{Liu2020Jul}%
  \BibitemOpen
  \bibfield  {author} {\bibinfo {author} {\bibfnamefont {C.}~\bibnamefont
  {Liu}}, \bibinfo {author} {\bibfnamefont {H.}~\bibnamefont {Chen}}, \bibinfo
  {author} {\bibfnamefont {S.}~\bibnamefont {Wang}}, \bibinfo {author}
  {\bibfnamefont {Q.}~\bibnamefont {Liu}}, \bibinfo {author} {\bibfnamefont
  {Y.-G.}\ \bibnamefont {Jiang}}, \bibinfo {author} {\bibfnamefont {D.~W.}\
  \bibnamefont {Zhang}}, \bibinfo {author} {\bibfnamefont {M.}~\bibnamefont
  {Liu}},\ and\ \bibinfo {author} {\bibfnamefont {P.}~\bibnamefont {Zhou}},\
  }\bibfield  {title} {\bibinfo {title} {{Two-dimensional materials for
  next-generation computing technologies}},\ }\href
  {https://doi.org/10.1038/s41565-020-0724-3} {\bibfield  {journal} {\bibinfo
  {journal} {Nat. Nanotechnol.}\ }\textbf {\bibinfo {volume} {15}},\ \bibinfo
  {pages} {545} (\bibinfo {year} {2020})}\BibitemShut {NoStop}%
\bibitem [{\citenamefont {Schwierz}\ \emph {et~al.}(2015)\citenamefont
  {Schwierz}, \citenamefont {Pezoldt},\ and\ \citenamefont
  {Granzner}}]{Schwierz2015Apr}%
  \BibitemOpen
  \bibfield  {author} {\bibinfo {author} {\bibfnamefont {F.}~\bibnamefont
  {Schwierz}}, \bibinfo {author} {\bibfnamefont {J.}~\bibnamefont {Pezoldt}},\
  and\ \bibinfo {author} {\bibfnamefont {R.}~\bibnamefont {Granzner}},\
  }\bibfield  {title} {\bibinfo {title} {{Two-dimensional materials and their
  prospects in transistor electronics}},\ }\href
  {https://doi.org/10.1039/C5NR01052G} {\bibfield  {journal} {\bibinfo
  {journal} {Nanoscale}\ }\textbf {\bibinfo {volume} {7}},\ \bibinfo {pages}
  {8261} (\bibinfo {year} {2015})}\BibitemShut {NoStop}%
\bibitem [{\citenamefont {Akinwande}\ \emph {et~al.}(2019)\citenamefont
  {Akinwande}, \citenamefont {Huyghebaert}, \citenamefont {Wang}, \citenamefont
  {Serna}, \citenamefont {Goossens}, \citenamefont {Li}, \citenamefont {Wong},\
  and\ \citenamefont {Koppens}}]{Akinwande2019Sep}%
  \BibitemOpen
  \bibfield  {author} {\bibinfo {author} {\bibfnamefont {D.}~\bibnamefont
  {Akinwande}}, \bibinfo {author} {\bibfnamefont {C.}~\bibnamefont
  {Huyghebaert}}, \bibinfo {author} {\bibfnamefont {C.-H.}\ \bibnamefont
  {Wang}}, \bibinfo {author} {\bibfnamefont {M.~I.}\ \bibnamefont {Serna}},
  \bibinfo {author} {\bibfnamefont {S.}~\bibnamefont {Goossens}}, \bibinfo
  {author} {\bibfnamefont {L.-J.}\ \bibnamefont {Li}}, \bibinfo {author}
  {\bibfnamefont {H.-S.~P.}\ \bibnamefont {Wong}},\ and\ \bibinfo {author}
  {\bibfnamefont {F.~H.~L.}\ \bibnamefont {Koppens}},\ }\bibfield  {title}
  {\bibinfo {title} {{Graphene and two-dimensional materials for silicon
  technology}},\ }\href {https://doi.org/10.1038/s41586-019-1573-9} {\bibfield
  {journal} {\bibinfo  {journal} {Nature}\ }\textbf {\bibinfo {volume} {573}},\
  \bibinfo {pages} {507} (\bibinfo {year} {2019})}\BibitemShut {NoStop}%
\bibitem [{\citenamefont {Fiori}\ \emph {et~al.}(2014)\citenamefont {Fiori},
  \citenamefont {Bonaccorso}, \citenamefont {Iannaccone}, \citenamefont
  {Palacios}, \citenamefont {Neumaier}, \citenamefont {Seabaugh}, \citenamefont
  {Banerjee},\ and\ \citenamefont {Colombo}}]{Fiori2014Oct}%
  \BibitemOpen
  \bibfield  {author} {\bibinfo {author} {\bibfnamefont {G.}~\bibnamefont
  {Fiori}}, \bibinfo {author} {\bibfnamefont {F.}~\bibnamefont {Bonaccorso}},
  \bibinfo {author} {\bibfnamefont {G.}~\bibnamefont {Iannaccone}}, \bibinfo
  {author} {\bibfnamefont {T.}~\bibnamefont {Palacios}}, \bibinfo {author}
  {\bibfnamefont {D.}~\bibnamefont {Neumaier}}, \bibinfo {author}
  {\bibfnamefont {A.}~\bibnamefont {Seabaugh}}, \bibinfo {author}
  {\bibfnamefont {S.~K.}\ \bibnamefont {Banerjee}},\ and\ \bibinfo {author}
  {\bibfnamefont {L.}~\bibnamefont {Colombo}},\ }\bibfield  {title} {\bibinfo
  {title} {{Electronics based on two-dimensional materials}},\ }\href
  {https://doi.org/10.1038/nnano.2014.207} {\bibfield  {journal} {\bibinfo
  {journal} {Nat. Nanotechnol.}\ }\textbf {\bibinfo {volume} {9}},\ \bibinfo
  {pages} {768} (\bibinfo {year} {2014})}\BibitemShut {NoStop}%
\bibitem [{\citenamefont {Acun}\ \emph {et~al.}(2015)\citenamefont {Acun},
  \citenamefont {Zhang}, \citenamefont {Bampoulis}, \citenamefont {Farmanbar},
  \citenamefont {van Houselt}, \citenamefont {Rudenko}, \citenamefont
  {Lingenfelder}, \citenamefont {Brocks}, \citenamefont {Poelsema},
  \citenamefont {Katsnelson},\ and\ \citenamefont
  {Zandvliet}}]{GermaneneReview_Acun_2015}%
  \BibitemOpen
  \bibfield  {author} {\bibinfo {author} {\bibfnamefont {A.}~\bibnamefont
  {Acun}}, \bibinfo {author} {\bibfnamefont {L.}~\bibnamefont {Zhang}},
  \bibinfo {author} {\bibfnamefont {P.}~\bibnamefont {Bampoulis}}, \bibinfo
  {author} {\bibfnamefont {M.}~\bibnamefont {Farmanbar}}, \bibinfo {author}
  {\bibfnamefont {A.}~\bibnamefont {van Houselt}}, \bibinfo {author}
  {\bibfnamefont {A.~N.}\ \bibnamefont {Rudenko}}, \bibinfo {author}
  {\bibfnamefont {M.}~\bibnamefont {Lingenfelder}}, \bibinfo {author}
  {\bibfnamefont {G.}~\bibnamefont {Brocks}}, \bibinfo {author} {\bibfnamefont
  {B.}~\bibnamefont {Poelsema}}, \bibinfo {author} {\bibfnamefont {M.~I.}\
  \bibnamefont {Katsnelson}},\ and\ \bibinfo {author} {\bibfnamefont
  {H.~J.~W.}\ \bibnamefont {Zandvliet}},\ }\bibfield  {title} {\bibinfo {title}
  {Germanene: the germanium analogue of graphene},\ }\href
  {https://doi.org/10.1088/0953-8984/27/44/443002} {\bibfield  {journal}
  {\bibinfo  {journal} {Journal of Physics: Condensed Matter}\ }\textbf
  {\bibinfo {volume} {27}},\ \bibinfo {pages} {443002} (\bibinfo {year}
  {2015})}\BibitemShut {NoStop}%
\bibitem [{\citenamefont {Kamal}\ and\ \citenamefont
  {Ezawa}(2015)}]{Arsene2Dhoneycombsystem_kamal_2015}%
  \BibitemOpen
  \bibfield  {author} {\bibinfo {author} {\bibfnamefont {C.}~\bibnamefont
  {Kamal}}\ and\ \bibinfo {author} {\bibfnamefont {M.}~\bibnamefont {Ezawa}},\
  }\bibfield  {title} {\bibinfo {title} {Arsenene: Two-dimensional buckled and
  puckered honeycomb arsenic systems},\ }\href
  {https://doi.org/10.1103/PhysRevB.91.085423} {\bibfield  {journal} {\bibinfo
  {journal} {Phys. Rev. B}\ }\textbf {\bibinfo {volume} {91}},\ \bibinfo
  {pages} {085423} (\bibinfo {year} {2015})}\BibitemShut {NoStop}%
\bibitem [{\citenamefont {Chowdhury}\ and\ \citenamefont
  {Jana}(2016)}]{ReviewElectronicMagneticOpticalPropertiesSilicene_Chowdhury_2016}%
  \BibitemOpen
  \bibfield  {author} {\bibinfo {author} {\bibfnamefont {S.}~\bibnamefont
  {Chowdhury}}\ and\ \bibinfo {author} {\bibfnamefont {D.}~\bibnamefont
  {Jana}},\ }\bibfield  {title} {\bibinfo {title} {A theoretical review on
  electronic, magnetic and optical properties of silicene},\ }\href
  {https://doi.org/10.1088/0034-4885/79/12/126501} {\bibfield  {journal}
  {\bibinfo  {journal} {Reports on Progress in Physics}\ }\textbf {\bibinfo
  {volume} {79}},\ \bibinfo {pages} {126501} (\bibinfo {year}
  {2016})}\BibitemShut {NoStop}%
\bibitem [{\citenamefont {{Manzeli}}\ \emph {et~al.}(2017)\citenamefont
  {{Manzeli}}, \citenamefont {{Ovchinnikov}}, \citenamefont {{Pasquier}},
  \citenamefont {{Yazyev}},\ and\ \citenamefont {{Kis}}}]{2DTMDs_Manzeli_2017}%
  \BibitemOpen
  \bibfield  {author} {\bibinfo {author} {\bibfnamefont {S.}~\bibnamefont
  {{Manzeli}}}, \bibinfo {author} {\bibfnamefont {D.}~\bibnamefont
  {{Ovchinnikov}}}, \bibinfo {author} {\bibfnamefont {D.}~\bibnamefont
  {{Pasquier}}}, \bibinfo {author} {\bibfnamefont {O.~V.}\ \bibnamefont
  {{Yazyev}}},\ and\ \bibinfo {author} {\bibfnamefont {A.}~\bibnamefont
  {{Kis}}},\ }\bibfield  {title} {\bibinfo {title} {{{2D} transition metal
  dichalcogenides}},\ }\href {https://doi.org/10.1038/natrevmats.2017.33}
  {\bibfield  {journal} {\bibinfo  {journal} {Nature Reviews Materials}\
  }\textbf {\bibinfo {volume} {2}},\ \bibinfo {pages} {17033} (\bibinfo {year}
  {2017})}\BibitemShut {NoStop}%
\bibitem [{\citenamefont {Castro~Neto}\ \emph {et~al.}(2009)\citenamefont
  {Castro~Neto}, \citenamefont {Guinea}, \citenamefont {Peres}, \citenamefont
  {Novoselov},\ and\ \citenamefont {Geim}}]{CastroNeto2009Jan}%
  \BibitemOpen
  \bibfield  {author} {\bibinfo {author} {\bibfnamefont {A.~H.}\ \bibnamefont
  {Castro~Neto}}, \bibinfo {author} {\bibfnamefont {F.}~\bibnamefont {Guinea}},
  \bibinfo {author} {\bibfnamefont {N.~M.~R.}\ \bibnamefont {Peres}}, \bibinfo
  {author} {\bibfnamefont {K.~S.}\ \bibnamefont {Novoselov}},\ and\ \bibinfo
  {author} {\bibfnamefont {A.~K.}\ \bibnamefont {Geim}},\ }\bibfield  {title}
  {\bibinfo {title} {{The electronic properties of graphene}},\ }\href
  {https://doi.org/10.1103/RevModPhys.81.109} {\bibfield  {journal} {\bibinfo
  {journal} {Rev. Mod. Phys.}\ }\textbf {\bibinfo {volume} {81}},\ \bibinfo
  {pages} {109} (\bibinfo {year} {2009})}\BibitemShut {NoStop}%
\bibitem [{\citenamefont {Xu}\ \emph {et~al.}(2013)\citenamefont {Xu},
  \citenamefont {Liang}, \citenamefont {Shi},\ and\ \citenamefont
  {Chen}}]{GrapheneLike2DMaterials_Xu2013May}%
  \BibitemOpen
  \bibfield  {author} {\bibinfo {author} {\bibfnamefont {M.}~\bibnamefont
  {Xu}}, \bibinfo {author} {\bibfnamefont {T.}~\bibnamefont {Liang}}, \bibinfo
  {author} {\bibfnamefont {M.}~\bibnamefont {Shi}},\ and\ \bibinfo {author}
  {\bibfnamefont {H.}~\bibnamefont {Chen}},\ }\bibfield  {title} {\bibinfo
  {title} {Graphene-like two-dimensional materials},\ }\href
  {https://doi.org/10.1021/cr300263a} {\bibfield  {journal} {\bibinfo
  {journal} {Chem. Rev.}\ }\textbf {\bibinfo {volume} {113}},\ \bibinfo {pages}
  {3766} (\bibinfo {year} {2013})}\BibitemShut {NoStop}%
\bibitem [{\citenamefont {Wang}\ \emph {et~al.}(2015)\citenamefont {Wang},
  \citenamefont {Deng}, \citenamefont {Liu},\ and\ \citenamefont
  {Liu}}]{rare2DmaterialsDiracCones_Wang_2015}%
  \BibitemOpen
  \bibfield  {author} {\bibinfo {author} {\bibfnamefont {J.}~\bibnamefont
  {Wang}}, \bibinfo {author} {\bibfnamefont {S.}~\bibnamefont {Deng}}, \bibinfo
  {author} {\bibfnamefont {Z.}~\bibnamefont {Liu}},\ and\ \bibinfo {author}
  {\bibfnamefont {Z.}~\bibnamefont {Liu}},\ }\bibfield  {title} {\bibinfo
  {title} {The rare two-dimensional materials with {Dirac} cones},\ }\href
  {https://doi.org/10.1093/nsr/nwu080} {\bibfield  {journal} {\bibinfo
  {journal} {National Science Review}\ }\textbf {\bibinfo {volume} {2}},\
  \bibinfo {pages} {22} (\bibinfo {year} {2015})}\BibitemShut {NoStop}%
\bibitem [{\citenamefont {Jariwala}\ \emph {et~al.}(2014)\citenamefont
  {Jariwala}, \citenamefont {Sangwan}, \citenamefont {Lauhon}, \citenamefont
  {Marks},\ and\ \citenamefont
  {Hersam}}]{EmergingDeviceApplications2DTMDs_Jariwala2014Feb}%
  \BibitemOpen
  \bibfield  {author} {\bibinfo {author} {\bibfnamefont {D.}~\bibnamefont
  {Jariwala}}, \bibinfo {author} {\bibfnamefont {V.~K.}\ \bibnamefont
  {Sangwan}}, \bibinfo {author} {\bibfnamefont {L.~J.}\ \bibnamefont {Lauhon}},
  \bibinfo {author} {\bibfnamefont {T.~J.}\ \bibnamefont {Marks}},\ and\
  \bibinfo {author} {\bibfnamefont {M.~C.}\ \bibnamefont {Hersam}},\ }\bibfield
   {title} {\bibinfo {title} {Emerging device applications for semiconducting
  two-dimensional transition metal dichalcogenides},\ }\href
  {https://doi.org/10.1021/nn500064s} {\bibfield  {journal} {\bibinfo
  {journal} {ACS Nano}\ }\textbf {\bibinfo {volume} {8}},\ \bibinfo {pages}
  {1102} (\bibinfo {year} {2014})}\BibitemShut {NoStop}%
\bibitem [{\citenamefont {Schaibley}\ \emph {et~al.}(2016)\citenamefont
  {Schaibley}, \citenamefont {Yu}, \citenamefont {Clark}, \citenamefont
  {Rivera}, \citenamefont {Ross}, \citenamefont {Seyler}, \citenamefont {Yao},\
  and\ \citenamefont {Xu}}]{Schaibley2016Aug}%
  \BibitemOpen
  \bibfield  {author} {\bibinfo {author} {\bibfnamefont {J.~R.}\ \bibnamefont
  {Schaibley}}, \bibinfo {author} {\bibfnamefont {H.}~\bibnamefont {Yu}},
  \bibinfo {author} {\bibfnamefont {G.}~\bibnamefont {Clark}}, \bibinfo
  {author} {\bibfnamefont {P.}~\bibnamefont {Rivera}}, \bibinfo {author}
  {\bibfnamefont {J.~S.}\ \bibnamefont {Ross}}, \bibinfo {author}
  {\bibfnamefont {K.~L.}\ \bibnamefont {Seyler}}, \bibinfo {author}
  {\bibfnamefont {W.}~\bibnamefont {Yao}},\ and\ \bibinfo {author}
  {\bibfnamefont {X.}~\bibnamefont {Xu}},\ }\bibfield  {title} {\bibinfo
  {title} {{Valleytronics in 2D materials}},\ }\href
  {https://doi.org/10.1038/natrevmats.2016.55} {\bibfield  {journal} {\bibinfo
  {journal} {Nat. Rev. Mater.}\ }\textbf {\bibinfo {volume} {1}},\ \bibinfo
  {pages} {1} (\bibinfo {year} {2016})}\BibitemShut {NoStop}%
\bibitem [{\citenamefont {Feng}\ \emph {et~al.}(2017)\citenamefont {Feng},
  \citenamefont {Shen}, \citenamefont {Yang}, \citenamefont {Wang},
  \citenamefont {Zeng}, \citenamefont {Wu}, \citenamefont {Chintalapati},\ and\
  \citenamefont {Chang}}]{Feng2017Sep}%
  \BibitemOpen
  \bibfield  {author} {\bibinfo {author} {\bibfnamefont {Y.~P.}\ \bibnamefont
  {Feng}}, \bibinfo {author} {\bibfnamefont {L.}~\bibnamefont {Shen}}, \bibinfo
  {author} {\bibfnamefont {M.}~\bibnamefont {Yang}}, \bibinfo {author}
  {\bibfnamefont {A.}~\bibnamefont {Wang}}, \bibinfo {author} {\bibfnamefont
  {M.}~\bibnamefont {Zeng}}, \bibinfo {author} {\bibfnamefont {Q.}~\bibnamefont
  {Wu}}, \bibinfo {author} {\bibfnamefont {S.}~\bibnamefont {Chintalapati}},\
  and\ \bibinfo {author} {\bibfnamefont {C.-R.}\ \bibnamefont {Chang}},\
  }\bibfield  {title} {\bibinfo {title} {{Prospects of spintronics based on 2D
  materials}},\ }\href {https://doi.org/10.1002/wcms.1313} {\bibfield
  {journal} {\bibinfo  {journal} {WIREs Comput. Mol. Sci.}\ }\textbf {\bibinfo
  {volume} {7}},\ \bibinfo {pages} {e1313} (\bibinfo {year}
  {2017})}\BibitemShut {NoStop}%
\bibitem [{\citenamefont {Zhu}\ \emph {et~al.}(2017)\citenamefont {Zhu},
  \citenamefont {Du},\ and\ \citenamefont {Lin}}]{Zhu2017Mar}%
  \BibitemOpen
  \bibfield  {author} {\bibinfo {author} {\bibfnamefont {C.}~\bibnamefont
  {Zhu}}, \bibinfo {author} {\bibfnamefont {D.}~\bibnamefont {Du}},\ and\
  \bibinfo {author} {\bibfnamefont {Y.}~\bibnamefont {Lin}},\ }\bibfield
  {title} {\bibinfo {title} {{Graphene-like 2D nanomaterial-based biointerfaces
  for biosensing applications}},\ }\href
  {https://doi.org/10.1016/j.bios.2016.06.045} {\bibfield  {journal} {\bibinfo
  {journal} {Biosens. Bioelectron.}\ }\textbf {\bibinfo {volume} {89}},\
  \bibinfo {pages} {43} (\bibinfo {year} {2017})}\BibitemShut {NoStop}%
\bibitem [{\citenamefont {Ahn}(2020)}]{Ahn2020Jun}%
  \BibitemOpen
  \bibfield  {author} {\bibinfo {author} {\bibfnamefont {E.~C.}\ \bibnamefont
  {Ahn}},\ }\bibfield  {title} {\bibinfo {title} {{2D materials for spintronic
  devices}},\ }\href {https://doi.org/10.1038/s41699-020-0152-0} {\bibfield
  {journal} {\bibinfo  {journal} {npj 2D Mater. Appl.}\ }\textbf {\bibinfo
  {volume} {4}},\ \bibinfo {pages} {1} (\bibinfo {year} {2020})}\BibitemShut
  {NoStop}%
\bibitem [{\citenamefont {Wehling}\ \emph {et~al.}(2014)\citenamefont
  {Wehling}, \citenamefont {Black‐Schaffer},\ and\ \citenamefont
  {Balatsky}}]{DiracMaterials_Oliver_2014}%
  \BibitemOpen
  \bibfield  {author} {\bibinfo {author} {\bibfnamefont {T.~O.}\ \bibnamefont
  {Wehling}}, \bibinfo {author} {\bibfnamefont {A.~M.}\ \bibnamefont
  {Black‐Schaffer}},\ and\ \bibinfo {author} {\bibfnamefont {A.~V.}\
  \bibnamefont {Balatsky}},\ }\bibfield  {title} {\bibinfo {title} {Dirac
  materials},\ }\bibfield  {journal} {\bibinfo  {journal} {Advances in
  Physics}\ }\textbf {\bibinfo {volume} {63}},\ \href
  {https://doi.org/10.1080/00018732.2014.927109} {10.1080/00018732.2014.927109}
  (\bibinfo {year} {2014})\BibitemShut {NoStop}%
\bibitem [{\citenamefont {Cayssol}(2013)}]{Cayssol2013}%
  \BibitemOpen
  \bibfield  {author} {\bibinfo {author} {\bibfnamefont {J.}~\bibnamefont
  {Cayssol}},\ }\bibfield  {title} {\bibinfo {title} {{Introduction to Dirac
  materials and topological insulators}},\ }\href
  {https://doi.org/10.1016/j.crhy.2013.09.012} {\bibfield  {journal} {\bibinfo
  {journal} {C. R. Phys.}\ }\textbf {\bibinfo {volume} {14}},\ \bibinfo {pages}
  {760} (\bibinfo {year} {2013})}\BibitemShut {NoStop}%
\bibitem [{\citenamefont {K{\"o}nig}\ \emph {et~al.}(2007)\citenamefont
  {K{\"o}nig}, \citenamefont {Wiedmann}, \citenamefont {Br{\"u}ne},
  \citenamefont {Roth}, \citenamefont {Buhmann}, \citenamefont {Molenkamp},
  \citenamefont {Qi},\ and\ \citenamefont {Zhang}}]{Konig2007}%
  \BibitemOpen
  \bibfield  {author} {\bibinfo {author} {\bibfnamefont {M.}~\bibnamefont
  {K{\"o}nig}}, \bibinfo {author} {\bibfnamefont {S.}~\bibnamefont {Wiedmann}},
  \bibinfo {author} {\bibfnamefont {C.}~\bibnamefont {Br{\"u}ne}}, \bibinfo
  {author} {\bibfnamefont {A.}~\bibnamefont {Roth}}, \bibinfo {author}
  {\bibfnamefont {H.}~\bibnamefont {Buhmann}}, \bibinfo {author} {\bibfnamefont
  {L.~W.}\ \bibnamefont {Molenkamp}}, \bibinfo {author} {\bibfnamefont {X.-L.}\
  \bibnamefont {Qi}},\ and\ \bibinfo {author} {\bibfnamefont {S.-C.}\
  \bibnamefont {Zhang}},\ }\bibfield  {title} {\bibinfo {title} {Quantum spin
  {Hall} insulator state in {HgTe} quantum wells},\ }\href
  {https://doi.org/10.1126/science.1148047} {\bibfield  {journal} {\bibinfo
  {journal} {Science}\ }\textbf {\bibinfo {volume} {318}},\ \bibinfo {pages}
  {766} (\bibinfo {year} {2007})}\BibitemShut {NoStop}%
\bibitem [{\citenamefont {Knez}\ \emph {et~al.}(2011)\citenamefont {Knez},
  \citenamefont {Du},\ and\ \citenamefont {Sullivan}}]{Sullivan2011}%
  \BibitemOpen
  \bibfield  {author} {\bibinfo {author} {\bibfnamefont {I.}~\bibnamefont
  {Knez}}, \bibinfo {author} {\bibfnamefont {R.-R.}\ \bibnamefont {Du}},\ and\
  \bibinfo {author} {\bibfnamefont {G.}~\bibnamefont {Sullivan}},\ }\bibfield
  {title} {\bibinfo {title} {Evidence for helical edge modes in inverted
  $\mathrm{InAs}/\mathrm{GaSb}$ quantum wells},\ }\href
  {https://doi.org/10.1103/PhysRevLett.107.136603} {\bibfield  {journal}
  {\bibinfo  {journal} {Phys. Rev. Lett.}\ }\textbf {\bibinfo {volume} {107}},\
  \bibinfo {pages} {136603} (\bibinfo {year} {2011})}\BibitemShut {NoStop}%
\bibitem [{\citenamefont {Hasan}\ and\ \citenamefont {Kane}(2010)}]{Hasan2010}%
  \BibitemOpen
  \bibfield  {author} {\bibinfo {author} {\bibfnamefont {M.~Z.}\ \bibnamefont
  {Hasan}}\ and\ \bibinfo {author} {\bibfnamefont {C.~L.}\ \bibnamefont
  {Kane}},\ }\bibfield  {title} {\bibinfo {title} {Colloquium: Topological
  insulators},\ }\href {https://doi.org/10.1103/RevModPhys.82.3045} {\bibfield
  {journal} {\bibinfo  {journal} {Rev. Mod. Phys.}\ }\textbf {\bibinfo {volume}
  {82}},\ \bibinfo {pages} {3045} (\bibinfo {year} {2010})}\BibitemShut
  {NoStop}%
\bibitem [{\citenamefont {Kou}\ \emph {et~al.}(2017)\citenamefont {Kou},
  \citenamefont {Ma}, \citenamefont {Sun}, \citenamefont {Heine},\ and\
  \citenamefont {Chen}}]{Kou2017Apr}%
  \BibitemOpen
  \bibfield  {author} {\bibinfo {author} {\bibfnamefont {L.}~\bibnamefont
  {Kou}}, \bibinfo {author} {\bibfnamefont {Y.}~\bibnamefont {Ma}}, \bibinfo
  {author} {\bibfnamefont {Z.}~\bibnamefont {Sun}}, \bibinfo {author}
  {\bibfnamefont {T.}~\bibnamefont {Heine}},\ and\ \bibinfo {author}
  {\bibfnamefont {C.}~\bibnamefont {Chen}},\ }\bibfield  {title} {\bibinfo
  {title} {Two-dimensional topological insulators: Progress and prospects},\
  }\href {https://doi.org/10.1021/acs.jpclett.7b00222} {\bibfield  {journal}
  {\bibinfo  {journal} {J. Phys. Chem. Lett.}\ }\textbf {\bibinfo {volume}
  {8}},\ \bibinfo {pages} {1905} (\bibinfo {year} {2017})}\BibitemShut
  {NoStop}%
\bibitem [{\citenamefont {Xia}\ \emph {et~al.}(2009)\citenamefont {Xia},
  \citenamefont {Qian}, \citenamefont {Hsieh}, \citenamefont {Wray},
  \citenamefont {Pal}, \citenamefont {Lin}, \citenamefont {Bansil},
  \citenamefont {Grauer}, \citenamefont {Hor}, \citenamefont {Cava},\ and\
  \citenamefont {Hasan}}]{Xia2009Jun}%
  \BibitemOpen
  \bibfield  {author} {\bibinfo {author} {\bibfnamefont {Y.}~\bibnamefont
  {Xia}}, \bibinfo {author} {\bibfnamefont {D.}~\bibnamefont {Qian}}, \bibinfo
  {author} {\bibfnamefont {D.}~\bibnamefont {Hsieh}}, \bibinfo {author}
  {\bibfnamefont {L.}~\bibnamefont {Wray}}, \bibinfo {author} {\bibfnamefont
  {A.}~\bibnamefont {Pal}}, \bibinfo {author} {\bibfnamefont {H.}~\bibnamefont
  {Lin}}, \bibinfo {author} {\bibfnamefont {A.}~\bibnamefont {Bansil}},
  \bibinfo {author} {\bibfnamefont {D.}~\bibnamefont {Grauer}}, \bibinfo
  {author} {\bibfnamefont {Y.~S.}\ \bibnamefont {Hor}}, \bibinfo {author}
  {\bibfnamefont {R.~J.}\ \bibnamefont {Cava}},\ and\ \bibinfo {author}
  {\bibfnamefont {M.~Z.}\ \bibnamefont {Hasan}},\ }\bibfield  {title} {\bibinfo
  {title} {{Observation of a large-gap topological-insulator class with a
  single Dirac cone on the surface}},\ }\href
  {https://doi.org/10.1038/nphys1274} {\bibfield  {journal} {\bibinfo
  {journal} {Nat. Phys.}\ }\textbf {\bibinfo {volume} {5}},\ \bibinfo {pages}
  {398} (\bibinfo {year} {2009})}\BibitemShut {NoStop}%
\bibitem [{\citenamefont {Hsieh}\ \emph {et~al.}(2008)\citenamefont {Hsieh},
  \citenamefont {Qian}, \citenamefont {Wray}, \citenamefont {Xia},
  \citenamefont {Hor}, \citenamefont {Cava},\ and\ \citenamefont
  {Hasan}}]{TIQSHP_Hsieh2008Apr}%
  \BibitemOpen
  \bibfield  {author} {\bibinfo {author} {\bibfnamefont {D.}~\bibnamefont
  {Hsieh}}, \bibinfo {author} {\bibfnamefont {D.}~\bibnamefont {Qian}},
  \bibinfo {author} {\bibfnamefont {L.}~\bibnamefont {Wray}}, \bibinfo {author}
  {\bibfnamefont {Y.}~\bibnamefont {Xia}}, \bibinfo {author} {\bibfnamefont
  {Y.~S.}\ \bibnamefont {Hor}}, \bibinfo {author} {\bibfnamefont {R.~J.}\
  \bibnamefont {Cava}},\ and\ \bibinfo {author} {\bibfnamefont {M.~Z.}\
  \bibnamefont {Hasan}},\ }\bibfield  {title} {\bibinfo {title} {{A topological
  Dirac insulator in a quantum spin Hall phase}},\ }\href
  {https://doi.org/10.1038/nature06843} {\bibfield  {journal} {\bibinfo
  {journal} {Nature}\ }\textbf {\bibinfo {volume} {452}},\ \bibinfo {pages}
  {970} (\bibinfo {year} {2008})}\BibitemShut {NoStop}%
\bibitem [{\citenamefont {Chen}\ \emph {et~al.}(2009)\citenamefont {Chen},
  \citenamefont {Analytis}, \citenamefont {Chu}, \citenamefont {Liu},
  \citenamefont {Mo}, \citenamefont {Qi}, \citenamefont {Zhang}, \citenamefont
  {Lu}, \citenamefont {Dai}, \citenamefont {Fang}, \citenamefont {Zhang},
  \citenamefont {Fisher}, \citenamefont {Hussain},\ and\ \citenamefont
  {Shen}}]{Chen2009Jul}%
  \BibitemOpen
  \bibfield  {author} {\bibinfo {author} {\bibfnamefont {Y.~L.}\ \bibnamefont
  {Chen}}, \bibinfo {author} {\bibfnamefont {J.~G.}\ \bibnamefont {Analytis}},
  \bibinfo {author} {\bibfnamefont {J.-H.}\ \bibnamefont {Chu}}, \bibinfo
  {author} {\bibfnamefont {Z.~K.}\ \bibnamefont {Liu}}, \bibinfo {author}
  {\bibfnamefont {S.-K.}\ \bibnamefont {Mo}}, \bibinfo {author} {\bibfnamefont
  {X.~L.}\ \bibnamefont {Qi}}, \bibinfo {author} {\bibfnamefont {H.~J.}\
  \bibnamefont {Zhang}}, \bibinfo {author} {\bibfnamefont {D.~H.}\ \bibnamefont
  {Lu}}, \bibinfo {author} {\bibfnamefont {X.}~\bibnamefont {Dai}}, \bibinfo
  {author} {\bibfnamefont {Z.}~\bibnamefont {Fang}}, \bibinfo {author}
  {\bibfnamefont {S.~C.}\ \bibnamefont {Zhang}}, \bibinfo {author}
  {\bibfnamefont {I.~R.}\ \bibnamefont {Fisher}}, \bibinfo {author}
  {\bibfnamefont {Z.}~\bibnamefont {Hussain}},\ and\ \bibinfo {author}
  {\bibfnamefont {Z.-X.}\ \bibnamefont {Shen}},\ }\bibfield  {title} {\bibinfo
  {title} {Experimental realization of a three-dimensional topological
  insulator, {Bi2Te3}},\ }\href {https://doi.org/10.1126/science.1173034}
  {\bibfield  {journal} {\bibinfo  {journal} {Science}\ }\textbf {\bibinfo
  {volume} {325}},\ \bibinfo {pages} {178} (\bibinfo {year}
  {2009})}\BibitemShut {NoStop}%
\bibitem [{\citenamefont {Bradlyn}\ \emph {et~al.}(2017)\citenamefont
  {Bradlyn}, \citenamefont {Elcoro}, \citenamefont {Cano}, \citenamefont
  {Vergniory}, \citenamefont {Wang}, \citenamefont {Felser}, \citenamefont
  {Aroyo},\ and\ \citenamefont {Bernevig}}]{Bradlyn2017}%
  \BibitemOpen
  \bibfield  {author} {\bibinfo {author} {\bibfnamefont {B.}~\bibnamefont
  {Bradlyn}}, \bibinfo {author} {\bibfnamefont {L.}~\bibnamefont {Elcoro}},
  \bibinfo {author} {\bibfnamefont {J.}~\bibnamefont {Cano}}, \bibinfo {author}
  {\bibfnamefont {M.~G.}\ \bibnamefont {Vergniory}}, \bibinfo {author}
  {\bibfnamefont {Z.}~\bibnamefont {Wang}}, \bibinfo {author} {\bibfnamefont
  {C.}~\bibnamefont {Felser}}, \bibinfo {author} {\bibfnamefont {M.~I.}\
  \bibnamefont {Aroyo}},\ and\ \bibinfo {author} {\bibfnamefont {B.~A.}\
  \bibnamefont {Bernevig}},\ }\bibfield  {title} {\bibinfo {title}
  {{Topological quantum chemistry}},\ }\href
  {https://doi.org/10.1038/nature23268} {\bibfield  {journal} {\bibinfo
  {journal} {Nature}\ }\textbf {\bibinfo {volume} {547}},\ \bibinfo {pages}
  {298} (\bibinfo {year} {2017})}\BibitemShut {NoStop}%
\bibitem [{\citenamefont {Liu}\ \emph {et~al.}(2011)\citenamefont {Liu},
  \citenamefont {Feng},\ and\ \citenamefont
  {Yao}}]{QSHEinSiliceneAnd2DGermaniun_Cheng_2011}%
  \BibitemOpen
  \bibfield  {author} {\bibinfo {author} {\bibfnamefont {C.~C.}\ \bibnamefont
  {Liu}}, \bibinfo {author} {\bibfnamefont {W.}~\bibnamefont {Feng}},\ and\
  \bibinfo {author} {\bibfnamefont {Y.}~\bibnamefont {Yao}},\ }\bibfield
  {title} {\bibinfo {title} {Quantum spin {Hall} effect in silicene and
  two-dimensional germanium},\ }\href
  {https://doi.org/10.1103/PhysRevLett.107.076802} {\bibfield  {journal}
  {\bibinfo  {journal} {Phys. Rev. Lett.}\ }\textbf {\bibinfo {volume} {107}},\
  \bibinfo {pages} {076802} (\bibinfo {year} {2011})}\BibitemShut {NoStop}%
\bibitem [{\citenamefont
  {Ezawa}(2012{\natexlab{a}})}]{TopologicalInsulatorHelicalModeSiliceneInhoumogeneusElectricField_Ezawa_2012}%
  \BibitemOpen
  \bibfield  {author} {\bibinfo {author} {\bibfnamefont {M.}~\bibnamefont
  {Ezawa}},\ }\bibfield  {title} {\bibinfo {title} {A topological insulator and
  helical zero mode in silicene under an inhomogeneous electric field},\ }\href
  {https://doi.org/10.1088/1367-2630/14/3/033003} {\bibfield  {journal}
  {\bibinfo  {journal} {New Journal of Physics}\ }\textbf {\bibinfo {volume}
  {14}},\ \bibinfo {pages} {033003} (\bibinfo {year}
  {2012}{\natexlab{a}})}\BibitemShut {NoStop}%
\bibitem [{\citenamefont
  {Ezawa}(2012{\natexlab{b}})}]{Valley-PolarizedMetalandQAHEinSilicene_Ezawa_2012}%
  \BibitemOpen
  \bibfield  {author} {\bibinfo {author} {\bibfnamefont {M.}~\bibnamefont
  {Ezawa}},\ }\bibfield  {title} {\bibinfo {title} {Valley-polarized metals and
  quantum anomalous {Hall} effect in silicene},\ }\href
  {https://doi.org/10.1103/PhysRevLett.109.055502} {\bibfield  {journal}
  {\bibinfo  {journal} {Phys. Rev. Lett.}\ }\textbf {\bibinfo {volume} {109}},\
  \bibinfo {pages} {055502} (\bibinfo {year} {2012}{\natexlab{b}})}\BibitemShut
  {NoStop}%
\bibitem [{\citenamefont {Lewenkopf}\ and\ \citenamefont
  {Mucciolo}(2013)}]{Lewenkopf2013Jun}%
  \BibitemOpen
  \bibfield  {author} {\bibinfo {author} {\bibfnamefont {C.~H.}\ \bibnamefont
  {Lewenkopf}}\ and\ \bibinfo {author} {\bibfnamefont {E.~R.}\ \bibnamefont
  {Mucciolo}},\ }\bibfield  {title} {\bibinfo {title} {{The recursive Green{'}s
  function method for graphene}},\ }\href
  {https://doi.org/10.1007/s10825-013-0458-7} {\bibfield  {journal} {\bibinfo
  {journal} {J. Comput. Electron.}\ }\textbf {\bibinfo {volume} {12}},\
  \bibinfo {pages} {203} (\bibinfo {year} {2013})}\BibitemShut {NoStop}%
\bibitem [{\citenamefont {Thorgilsson}\ \emph {et~al.}(2014)\citenamefont
  {Thorgilsson}, \citenamefont {Viktorsson},\ and\ \citenamefont
  {Erlingsson}}]{Thorgilsson2014Mar}%
  \BibitemOpen
  \bibfield  {author} {\bibinfo {author} {\bibfnamefont {G.}~\bibnamefont
  {Thorgilsson}}, \bibinfo {author} {\bibfnamefont {G.}~\bibnamefont
  {Viktorsson}},\ and\ \bibinfo {author} {\bibfnamefont {S.~I.}\ \bibnamefont
  {Erlingsson}},\ }\bibfield  {title} {\bibinfo {title} {Recursive {Green}{'}s
  function method for multi-terminal nanostructures},\ }\href
  {https://doi.org/10.1016/j.jcp.2013.12.054} {\bibfield  {journal} {\bibinfo
  {journal} {J. Comput. Phys.}\ }\textbf {\bibinfo {volume} {261}},\ \bibinfo
  {pages} {256} (\bibinfo {year} {2014})}\BibitemShut {NoStop}%
\bibitem [{\citenamefont {Ezawa}(2015)}]{Ezawa2015Oct}%
  \BibitemOpen
  \bibfield  {author} {\bibinfo {author} {\bibfnamefont {M.}~\bibnamefont
  {Ezawa}},\ }\bibfield  {title} {\bibinfo {title} {Monolayer topological
  insulators: Silicene, germanene, and stanene},\ }\href
  {https://doi.org/10.7566/JPSJ.84.121003} {\bibfield  {journal} {\bibinfo
  {journal} {J. Phys. Soc. Jpn.}\ }\textbf {\bibinfo {volume} {84}},\ \bibinfo
  {pages} {121003} (\bibinfo {year} {2015})}\BibitemShut {NoStop}%
\bibitem [{\citenamefont {Gerivani}\ and\ \citenamefont
  {Milani~Moghaddam}(2022)}]{Gerivani2022}%
  \BibitemOpen
  \bibfield  {author} {\bibinfo {author} {\bibfnamefont {S.}~\bibnamefont
  {Gerivani}}\ and\ \bibinfo {author} {\bibfnamefont {H.}~\bibnamefont
  {Milani~Moghaddam}},\ }\bibfield  {title} {\bibinfo {title} {{Intrinsic
  half-metallic properties of MnHm (M: Fe, V, Co, and Cr) in various space
  groups: A first-principles study}},\ }\href
  {https://doi.org/10.1016/j.jmmm.2021.168758} {\bibfield  {journal} {\bibinfo
  {journal} {J. Magn. Magn. Mater.}\ }\textbf {\bibinfo {volume} {547}},\
  \bibinfo {pages} {168758} (\bibinfo {year} {2022})}\BibitemShut {NoStop}%
\bibitem [{\citenamefont {Marmolejo-Tejada}\ and\ \citenamefont
  {Velasco-Medina}(2016)}]{Marmolejo-Tejada2016Feb}%
  \BibitemOpen
  \bibfield  {author} {\bibinfo {author} {\bibfnamefont {J.~M.}\ \bibnamefont
  {Marmolejo-Tejada}}\ and\ \bibinfo {author} {\bibfnamefont {J.}~\bibnamefont
  {Velasco-Medina}},\ }\bibfield  {title} {\bibinfo {title} {{Review on
  graphene nanoribbon devices for logic applications}},\ }\href
  {https://doi.org/10.1016/j.mejo.2015.11.006} {\bibfield  {journal} {\bibinfo
  {journal} {Microelectron. J.}\ }\textbf {\bibinfo {volume} {48}},\ \bibinfo
  {pages} {18} (\bibinfo {year} {2016})}\BibitemShut {NoStop}%
\bibitem [{\citenamefont {Wang}\ \emph {et~al.}(2021)\citenamefont {Wang},
  \citenamefont {Wang}, \citenamefont {Ma}, \citenamefont {Chen}, \citenamefont
  {Jiang}, \citenamefont {Chen}, \citenamefont {Xie}, \citenamefont {Li},\ and\
  \citenamefont {Wang}}]{Wang2021Dec}%
  \BibitemOpen
  \bibfield  {author} {\bibinfo {author} {\bibfnamefont {H.}~\bibnamefont
  {Wang}}, \bibinfo {author} {\bibfnamefont {H.~S.}\ \bibnamefont {Wang}},
  \bibinfo {author} {\bibfnamefont {C.}~\bibnamefont {Ma}}, \bibinfo {author}
  {\bibfnamefont {L.}~\bibnamefont {Chen}}, \bibinfo {author} {\bibfnamefont
  {C.}~\bibnamefont {Jiang}}, \bibinfo {author} {\bibfnamefont
  {C.}~\bibnamefont {Chen}}, \bibinfo {author} {\bibfnamefont {X.}~\bibnamefont
  {Xie}}, \bibinfo {author} {\bibfnamefont {A.-P.}\ \bibnamefont {Li}},\ and\
  \bibinfo {author} {\bibfnamefont {X.}~\bibnamefont {Wang}},\ }\bibfield
  {title} {\bibinfo {title} {{Graphene nanoribbons for quantum electronics}},\
  }\href {https://doi.org/10.1038/s42254-021-00370-x} {\bibfield  {journal}
  {\bibinfo  {journal} {Nat. Rev. Phys.}\ }\textbf {\bibinfo {volume} {3}},\
  \bibinfo {pages} {791} (\bibinfo {year} {2021})}\BibitemShut {NoStop}%
\bibitem [{\citenamefont {Brey}\ and\ \citenamefont
  {Fertig}(2006)}]{ElectronicStateGrapheneNanoDiracEquation_Brey_2006}%
  \BibitemOpen
  \bibfield  {author} {\bibinfo {author} {\bibfnamefont {L.}~\bibnamefont
  {Brey}}\ and\ \bibinfo {author} {\bibfnamefont {H.~A.}\ \bibnamefont
  {Fertig}},\ }\bibfield  {title} {\bibinfo {title} {Electronic states of
  graphene nanoribbons studied with the {Dirac} equation},\ }\href
  {https://doi.org/10.1103/PhysRevB.73.235411} {\bibfield  {journal} {\bibinfo
  {journal} {Phys. Rev. B}\ }\textbf {\bibinfo {volume} {73}},\ \bibinfo
  {pages} {235411} (\bibinfo {year} {2006})}\BibitemShut {NoStop}%
\bibitem [{\citenamefont {Wurm}\ \emph {et~al.}(2011)\citenamefont {Wurm},
  \citenamefont {Richter},\ and\ \citenamefont
  {Adagideli}}]{EdgeEffectsGrapheneNanostructures_Jurgen_2011}%
  \BibitemOpen
  \bibfield  {author} {\bibinfo {author} {\bibfnamefont {J.}~\bibnamefont
  {Wurm}}, \bibinfo {author} {\bibfnamefont {K.}~\bibnamefont {Richter}},\ and\
  \bibinfo {author} {\bibfnamefont {i.~d. I. m.~c.}\ \bibnamefont
  {Adagideli}},\ }\bibfield  {title} {\bibinfo {title} {Edge effects in
  graphene nanostructures: {From} multiple reflection expansion to density of
  states},\ }\href {https://doi.org/10.1103/PhysRevB.84.075468} {\bibfield
  {journal} {\bibinfo  {journal} {Phys. Rev. B}\ }\textbf {\bibinfo {volume}
  {84}},\ \bibinfo {pages} {075468} (\bibinfo {year} {2011})}\BibitemShut
  {NoStop}%
\bibitem [{\citenamefont {Aidelsburger}\ \emph {et~al.}(2018)\citenamefont
  {Aidelsburger}, \citenamefont {Nascimbene},\ and\ \citenamefont
  {Goldman}}]{Aidelsburger2018Sep}%
  \BibitemOpen
  \bibfield  {author} {\bibinfo {author} {\bibfnamefont {M.}~\bibnamefont
  {Aidelsburger}}, \bibinfo {author} {\bibfnamefont {S.}~\bibnamefont
  {Nascimbene}},\ and\ \bibinfo {author} {\bibfnamefont {N.}~\bibnamefont
  {Goldman}},\ }\bibfield  {title} {\bibinfo {title} {{Artificial gauge fields
  in materials and engineered systems}},\ }\href
  {https://doi.org/10.1016/j.crhy.2018.03.002} {\bibfield  {journal} {\bibinfo
  {journal} {C. R. Phys.}\ }\textbf {\bibinfo {volume} {19}},\ \bibinfo {pages}
  {394} (\bibinfo {year} {2018})}\BibitemShut {NoStop}%
\bibitem [{\citenamefont {Manjarrés}\ \emph {et~al.}(2009)\citenamefont
  {Manjarrés}, \citenamefont {Herrera},\ and\ \citenamefont
  {Gómez}}]{AndreevLevelsGrapheneSuperSurface_Manjarres_2009}%
  \BibitemOpen
  \bibfield  {author} {\bibinfo {author} {\bibfnamefont {D.~A.}\ \bibnamefont
  {Manjarrés}}, \bibinfo {author} {\bibfnamefont {W.~J.}\ \bibnamefont
  {Herrera}},\ and\ \bibinfo {author} {\bibfnamefont {S.}~\bibnamefont
  {Gómez}},\ }\bibfield  {title} {\bibinfo {title} {Andreev levels in a
  graphene–superconductor surface},\ }\href
  {https://doi.org/https://doi.org/10.1016/j.physb.2009.06.087} {\bibfield
  {journal} {\bibinfo  {journal} {Physica B: Condensed Matter}\ }\textbf
  {\bibinfo {volume} {404}},\ \bibinfo {pages} {2799} (\bibinfo {year}
  {2009})}\BibitemShut {NoStop}%
\bibitem [{\citenamefont {Herrera}\ \emph {et~al.}(2010)\citenamefont
  {Herrera}, \citenamefont {Burset},\ and\ \citenamefont
  {Yeyati}}]{GreenFunctionApproachWellDefinedEdges_Herrera_2010}%
  \BibitemOpen
  \bibfield  {author} {\bibinfo {author} {\bibfnamefont {W.~J.}\ \bibnamefont
  {Herrera}}, \bibinfo {author} {\bibfnamefont {P.}~\bibnamefont {Burset}},\
  and\ \bibinfo {author} {\bibfnamefont {A.~L.}\ \bibnamefont {Yeyati}},\
  }\bibfield  {title} {\bibinfo {title} {A {Green} function approach to
  graphene–superconductor junctions with well-defined edges},\ }\href
  {https://doi.org/10.1088/0953-8984/22/27/275304} {\bibfield  {journal}
  {\bibinfo  {journal} {Journal of Physics: Condensed Matter}\ }\textbf
  {\bibinfo {volume} {22}},\ \bibinfo {pages} {275304} (\bibinfo {year}
  {2010})}\BibitemShut {NoStop}%
\bibitem [{\citenamefont {Burset}\ \emph {et~al.}(2009)\citenamefont {Burset},
  \citenamefont {Herrera},\ and\ \citenamefont
  {Levy~Yeyati}}]{ProximityInterfaceBoundStateSuperGraphJunction_burset_2009}%
  \BibitemOpen
  \bibfield  {author} {\bibinfo {author} {\bibfnamefont {P.}~\bibnamefont
  {Burset}}, \bibinfo {author} {\bibfnamefont {W.}~\bibnamefont {Herrera}},\
  and\ \bibinfo {author} {\bibfnamefont {A.}~\bibnamefont {Levy~Yeyati}},\
  }\bibfield  {title} {\bibinfo {title} {Proximity-induced interface bound
  states in superconductor-graphene junctions},\ }\href
  {https://doi.org/10.1103/PhysRevB.80.041402} {\bibfield  {journal} {\bibinfo
  {journal} {Phys. Rev. B}\ }\textbf {\bibinfo {volume} {80}},\ \bibinfo
  {pages} {041402} (\bibinfo {year} {2009})}\BibitemShut {NoStop}%
\bibitem [{\citenamefont {G\'omez~P\'aez}\ \emph {et~al.}(2019)\citenamefont
  {G\'omez~P\'aez}, \citenamefont {Mart\'{\i}nez}, \citenamefont {Herrera},
  \citenamefont {Levy~Yeyati},\ and\ \citenamefont
  {Burset}}]{DiracPointAndreevTunnelingSuperlatticeGraphene-SuperconductorJunctions_Paez_2019}%
  \BibitemOpen
  \bibfield  {author} {\bibinfo {author} {\bibfnamefont {S.}~\bibnamefont
  {G\'omez~P\'aez}}, \bibinfo {author} {\bibfnamefont {C.}~\bibnamefont
  {Mart\'{\i}nez}}, \bibinfo {author} {\bibfnamefont {W.~J.}\ \bibnamefont
  {Herrera}}, \bibinfo {author} {\bibfnamefont {A.}~\bibnamefont
  {Levy~Yeyati}},\ and\ \bibinfo {author} {\bibfnamefont {P.}~\bibnamefont
  {Burset}},\ }\bibfield  {title} {\bibinfo {title} {Dirac point formation
  revealed by {Andreev} tunneling in superlattice-graphene/superconductor
  junctions},\ }\href {https://doi.org/10.1103/PhysRevB.100.205429} {\bibfield
  {journal} {\bibinfo  {journal} {Phys. Rev. B}\ }\textbf {\bibinfo {volume}
  {100}},\ \bibinfo {pages} {205429} (\bibinfo {year} {2019})}\BibitemShut
  {NoStop}%
\bibitem [{\citenamefont {Casas}\ \emph {et~al.}(2020)\citenamefont {Casas},
  \citenamefont {Páez},\ and\ \citenamefont
  {Herrera}}]{GreenFunctionTIjunctionwithMagneticandSuperRegion_Casas_2020}%
  \BibitemOpen
  \bibfield  {author} {\bibinfo {author} {\bibfnamefont {O.~E.}\ \bibnamefont
  {Casas}}, \bibinfo {author} {\bibfnamefont {S.~G.}\ \bibnamefont {Páez}},\
  and\ \bibinfo {author} {\bibfnamefont {W.~J.}\ \bibnamefont {Herrera}},\
  }\bibfield  {title} {\bibinfo {title} {A {Green}’s function approach to
  topological insulator junctions with magnetic and superconducting regions},\
  }\href {https://doi.org/10.1088/1361-648X/abafc9} {\bibfield  {journal}
  {\bibinfo  {journal} {Journal of Physics: Condensed Matter}\ }\textbf
  {\bibinfo {volume} {32}},\ \bibinfo {pages} {485302} (\bibinfo {year}
  {2020})}\BibitemShut {NoStop}%
\bibitem [{\citenamefont {Andelkovic}\ \emph {et~al.}(2023)\citenamefont
  {Andelkovic}, \citenamefont {Rakhimov}, \citenamefont {Chaves}, \citenamefont
  {Berdiyorov},\ and\ \citenamefont
  {Milo{\ifmmode\check{s}\else\v{s}\fi}evi{\ifmmode\acute{c}\else\'{c}\fi}}}]{Andelkovic2023}%
  \BibitemOpen
  \bibfield  {author} {\bibinfo {author} {\bibfnamefont {M.}~\bibnamefont
  {Andelkovic}}, \bibinfo {author} {\bibfnamefont {{\relax Kh}.~{\relax Yu}.}\
  \bibnamefont {Rakhimov}}, \bibinfo {author} {\bibfnamefont {A.}~\bibnamefont
  {Chaves}}, \bibinfo {author} {\bibfnamefont {G.~R.}\ \bibnamefont
  {Berdiyorov}},\ and\ \bibinfo {author} {\bibfnamefont {M.~V.}\ \bibnamefont
  {Milo{\ifmmode\check{s}\else\v{s}\fi}evi{\ifmmode\acute{c}\else\'{c}\fi}}},\
  }\bibfield  {title} {\bibinfo {title} {{Wave-packet propagation in a graphene
  geometric diode}},\ }\href {https://doi.org/10.1016/j.physe.2022.115607}
  {\bibfield  {journal} {\bibinfo  {journal} {Physica E}\ }\textbf {\bibinfo
  {volume} {147}},\ \bibinfo {pages} {115607} (\bibinfo {year}
  {2023})}\BibitemShut {NoStop}%
\bibitem [{\citenamefont {Linard}\ \emph {et~al.}(2023)\citenamefont {Linard},
  \citenamefont {Moura}, \citenamefont {Covaci}, \citenamefont {Milo\ifmmode
  \check{s}\else \v{s}\fi{}evi\ifmmode~\acute{c}\else \'{c}\fi{}},\ and\
  \citenamefont {Chaves}}]{Chaves2023}%
  \BibitemOpen
  \bibfield  {author} {\bibinfo {author} {\bibfnamefont {F.~J.~A.}\
  \bibnamefont {Linard}}, \bibinfo {author} {\bibfnamefont {V.~N.}\
  \bibnamefont {Moura}}, \bibinfo {author} {\bibfnamefont {L.}~\bibnamefont
  {Covaci}}, \bibinfo {author} {\bibfnamefont {M.~V.}\ \bibnamefont
  {Milo\ifmmode \check{s}\else \v{s}\fi{}evi\ifmmode~\acute{c}\else
  \'{c}\fi{}}},\ and\ \bibinfo {author} {\bibfnamefont {A.}~\bibnamefont
  {Chaves}},\ }\bibfield  {title} {\bibinfo {title} {Wave-packet scattering at
  a normal-superconductor interface in two-dimensional materials: A generalized
  theoretical approach},\ }\href {https://doi.org/10.1103/PhysRevB.107.165306}
  {\bibfield  {journal} {\bibinfo  {journal} {Phys. Rev. B}\ }\textbf {\bibinfo
  {volume} {107}},\ \bibinfo {pages} {165306} (\bibinfo {year}
  {2023})}\BibitemShut {NoStop}%
\bibitem [{\citenamefont {Lu}\ and\ \citenamefont
  {Tanaka}(2018)}]{StudyGreenFunctionTISurface_Lu_2018}%
  \BibitemOpen
  \bibfield  {author} {\bibinfo {author} {\bibfnamefont {B.}~\bibnamefont
  {Lu}}\ and\ \bibinfo {author} {\bibfnamefont {Y.}~\bibnamefont {Tanaka}},\
  }\bibfield  {title} {\bibinfo {title} {Study on {Green}’s function on
  topological insulator surface},\ }\bibfield  {journal} {\bibinfo  {journal}
  {Philosophical Transactions of the Royal Society A: Mathematical, Physical
  and Engineering Sciences}\ }\textbf {\bibinfo {volume} {376}},\ \href
  {https://doi.org/10.1098/rsta.2015.0246} {10.1098/rsta.2015.0246} (\bibinfo
  {year} {2018})\BibitemShut {NoStop}%
\bibitem [{\citenamefont {Burset}\ \emph {et~al.}(2008)\citenamefont {Burset},
  \citenamefont {Yeyati},\ and\ \citenamefont
  {Mart\'{\i}n-Rodero}}]{Burset2008}%
  \BibitemOpen
  \bibfield  {author} {\bibinfo {author} {\bibfnamefont {P.}~\bibnamefont
  {Burset}}, \bibinfo {author} {\bibfnamefont {A.~L.}\ \bibnamefont {Yeyati}},\
  and\ \bibinfo {author} {\bibfnamefont {A.}~\bibnamefont
  {Mart\'{\i}n-Rodero}},\ }\bibfield  {title} {\bibinfo {title} {Microscopic
  theory of the proximity effect in superconductor-graphene nanostructures},\
  }\href {https://doi.org/10.1103/PhysRevB.77.205425} {\bibfield  {journal}
  {\bibinfo  {journal} {Phys. Rev. B}\ }\textbf {\bibinfo {volume} {77}},\
  \bibinfo {pages} {205425} (\bibinfo {year} {2008})}\BibitemShut {NoStop}%
\bibitem [{\citenamefont {Qian}\ \emph {et~al.}(2014)\citenamefont {Qian},
  \citenamefont {Liu}, \citenamefont {Fu},\ and\ \citenamefont
  {Li}}]{Qian2014Dec}%
  \BibitemOpen
  \bibfield  {author} {\bibinfo {author} {\bibfnamefont {X.}~\bibnamefont
  {Qian}}, \bibinfo {author} {\bibfnamefont {J.}~\bibnamefont {Liu}}, \bibinfo
  {author} {\bibfnamefont {L.}~\bibnamefont {Fu}},\ and\ \bibinfo {author}
  {\bibfnamefont {J.}~\bibnamefont {Li}},\ }\bibfield  {title} {\bibinfo
  {title} {{Quantum spin Hall effect in two-dimensional transition metal
  dichalcogenides}},\ }\href {https://doi.org/10.1126/science.1256815}
  {\bibfield  {journal} {\bibinfo  {journal} {Science}\ }\textbf {\bibinfo
  {volume} {346}},\ \bibinfo {pages} {1344} (\bibinfo {year}
  {2014})}\BibitemShut {NoStop}%
\bibitem [{\citenamefont {Xiao}\ \emph {et~al.}(2012)\citenamefont {Xiao},
  \citenamefont {Liu}, \citenamefont {Feng}, \citenamefont {Xu},\ and\
  \citenamefont {Yao}}]{Xiao2012May}%
  \BibitemOpen
  \bibfield  {author} {\bibinfo {author} {\bibfnamefont {D.}~\bibnamefont
  {Xiao}}, \bibinfo {author} {\bibfnamefont {G.-B.}\ \bibnamefont {Liu}},
  \bibinfo {author} {\bibfnamefont {W.}~\bibnamefont {Feng}}, \bibinfo {author}
  {\bibfnamefont {X.}~\bibnamefont {Xu}},\ and\ \bibinfo {author}
  {\bibfnamefont {W.}~\bibnamefont {Yao}},\ }\bibfield  {title} {\bibinfo
  {title} {Coupled spin and valley physics in monolayers of
  ${\mathrm{mos}}_{2}$ and other group-{VI} dichalcogenides},\ }\href
  {https://doi.org/10.1103/PhysRevLett.108.196802} {\bibfield  {journal}
  {\bibinfo  {journal} {Phys. Rev. Lett.}\ }\textbf {\bibinfo {volume} {108}},\
  \bibinfo {pages} {196802} (\bibinfo {year} {2012})}\BibitemShut {NoStop}%
\bibitem [{\citenamefont {Fang}\ \emph {et~al.}(2015)\citenamefont {Fang},
  \citenamefont {Kuate~Defo}, \citenamefont {Shirodkar}, \citenamefont {Lieu},
  \citenamefont {Tritsaris},\ and\ \citenamefont {Kaxiras}}]{Fang2015Nov}%
  \BibitemOpen
  \bibfield  {author} {\bibinfo {author} {\bibfnamefont {S.}~\bibnamefont
  {Fang}}, \bibinfo {author} {\bibfnamefont {R.}~\bibnamefont {Kuate~Defo}},
  \bibinfo {author} {\bibfnamefont {S.~N.}\ \bibnamefont {Shirodkar}}, \bibinfo
  {author} {\bibfnamefont {S.}~\bibnamefont {Lieu}}, \bibinfo {author}
  {\bibfnamefont {G.~A.}\ \bibnamefont {Tritsaris}},\ and\ \bibinfo {author}
  {\bibfnamefont {E.}~\bibnamefont {Kaxiras}},\ }\bibfield  {title} {\bibinfo
  {title} {{Ab initio tight-binding Hamiltonian for transition metal
  dichalcogenides}},\ }\href {https://doi.org/10.1103/PhysRevB.92.205108}
  {\bibfield  {journal} {\bibinfo  {journal} {Phys. Rev. B}\ }\textbf {\bibinfo
  {volume} {92}},\ \bibinfo {pages} {205108} (\bibinfo {year}
  {2015})}\BibitemShut {NoStop}%
\bibitem [{\citenamefont {Korm{\ifmmode\acute{a}\else\'{a}\fi}nyos}\ \emph
  {et~al.}(2015)\citenamefont {Korm{\ifmmode\acute{a}\else\'{a}\fi}nyos},
  \citenamefont {Burkard}, \citenamefont {Gmitra}, \citenamefont {Fabian},
  \citenamefont {Z{\ifmmode\acute{o}\else\'{o}\fi}lyomi}, \citenamefont
  {Drummond},\ and\ \citenamefont {Fal{'}ko}}]{Kormanyos2015Apr}%
  \BibitemOpen
  \bibfield  {author} {\bibinfo {author} {\bibfnamefont {A.}~\bibnamefont
  {Korm{\ifmmode\acute{a}\else\'{a}\fi}nyos}}, \bibinfo {author} {\bibfnamefont
  {G.}~\bibnamefont {Burkard}}, \bibinfo {author} {\bibfnamefont
  {M.}~\bibnamefont {Gmitra}}, \bibinfo {author} {\bibfnamefont
  {J.}~\bibnamefont {Fabian}}, \bibinfo {author} {\bibfnamefont
  {V.}~\bibnamefont {Z{\ifmmode\acute{o}\else\'{o}\fi}lyomi}}, \bibinfo
  {author} {\bibfnamefont {N.~D.}\ \bibnamefont {Drummond}},\ and\ \bibinfo
  {author} {\bibfnamefont {V.}~\bibnamefont {Fal{'}ko}},\ }\bibfield  {title}
  {\bibinfo {title} {{k{$\cdot$}p theory for two-dimensional transition metal
  dichalcogenide semiconductors}},\ }\href
  {https://doi.org/10.1088/2053-1583/2/2/022001} {\bibfield  {journal}
  {\bibinfo  {journal} {2D Mater.}\ }\textbf {\bibinfo {volume} {2}},\ \bibinfo
  {pages} {022001} (\bibinfo {year} {2015})}\BibitemShut {NoStop}%
\bibitem [{\citenamefont {Ma}\ \emph {et~al.}(2013)\citenamefont {Ma},
  \citenamefont {Wang},\ and\ \citenamefont {Ding}}]{Ma2013Jan}%
  \BibitemOpen
  \bibfield  {author} {\bibinfo {author} {\bibfnamefont {L.}~\bibnamefont
  {Ma}}, \bibinfo {author} {\bibfnamefont {J.}~\bibnamefont {Wang}},\ and\
  \bibinfo {author} {\bibfnamefont {F.}~\bibnamefont {Ding}},\ }\bibfield
  {title} {\bibinfo {title} {Recent progress and challenges in graphene
  nanoribbon synthesis},\ }\href {https://doi.org/10.1002/cphc.201200253}
  {\bibfield  {journal} {\bibinfo  {journal} {ChemPhysChem}\ }\textbf {\bibinfo
  {volume} {14}},\ \bibinfo {pages} {47} (\bibinfo {year} {2013})}\BibitemShut
  {NoStop}%
\bibitem [{\citenamefont {Si}\ \emph {et~al.}(2014)\citenamefont {Si},
  \citenamefont {Liu}, \citenamefont {Xu}, \citenamefont {Wu}, \citenamefont
  {Gu},\ and\ \citenamefont {Duan}}]{Si2014Mar}%
  \BibitemOpen
  \bibfield  {author} {\bibinfo {author} {\bibfnamefont {C.}~\bibnamefont
  {Si}}, \bibinfo {author} {\bibfnamefont {J.}~\bibnamefont {Liu}}, \bibinfo
  {author} {\bibfnamefont {Y.}~\bibnamefont {Xu}}, \bibinfo {author}
  {\bibfnamefont {J.}~\bibnamefont {Wu}}, \bibinfo {author} {\bibfnamefont
  {B.-L.}\ \bibnamefont {Gu}},\ and\ \bibinfo {author} {\bibfnamefont
  {W.}~\bibnamefont {Duan}},\ }\bibfield  {title} {\bibinfo {title}
  {{Functionalized germanene as a prototype of large-gap two-dimensional
  topological insulators}},\ }\href
  {https://doi.org/10.1103/PhysRevB.89.115429} {\bibfield  {journal} {\bibinfo
  {journal} {Phys. Rev. B}\ }\textbf {\bibinfo {volume} {89}},\ \bibinfo
  {pages} {115429} (\bibinfo {year} {2014})}\BibitemShut {NoStop}%
\bibitem [{\citenamefont {Ezawa}(2012{\natexlab{c}})}]{Ezawa2012Mar}%
  \BibitemOpen
  \bibfield  {author} {\bibinfo {author} {\bibfnamefont {M.}~\bibnamefont
  {Ezawa}},\ }\bibfield  {title} {\bibinfo {title} {{A topological insulator
  and helical zero mode in silicene under an inhomogeneous electric field}},\
  }\href {https://doi.org/10.1088/1367-2630/14/3/033003} {\bibfield  {journal}
  {\bibinfo  {journal} {New J. Phys.}\ }\textbf {\bibinfo {volume} {14}},\
  \bibinfo {pages} {033003} (\bibinfo {year} {2012}{\natexlab{c}})}\BibitemShut
  {NoStop}%
\bibitem [{\citenamefont {Molle}\ \emph {et~al.}(2017)\citenamefont {Molle},
  \citenamefont {Goldberger}, \citenamefont {Houssa}, \citenamefont {Xu},
  \citenamefont {Zhang},\ and\ \citenamefont {Akinwande}}]{Molle2017Feb}%
  \BibitemOpen
  \bibfield  {author} {\bibinfo {author} {\bibfnamefont {A.}~\bibnamefont
  {Molle}}, \bibinfo {author} {\bibfnamefont {J.}~\bibnamefont {Goldberger}},
  \bibinfo {author} {\bibfnamefont {M.}~\bibnamefont {Houssa}}, \bibinfo
  {author} {\bibfnamefont {Y.}~\bibnamefont {Xu}}, \bibinfo {author}
  {\bibfnamefont {S.-C.}\ \bibnamefont {Zhang}},\ and\ \bibinfo {author}
  {\bibfnamefont {D.}~\bibnamefont {Akinwande}},\ }\bibfield  {title} {\bibinfo
  {title} {{Buckled two-dimensional Xene sheets}},\ }\href
  {https://doi.org/10.1038/nmat4802} {\bibfield  {journal} {\bibinfo  {journal}
  {Nat. Mater.}\ }\textbf {\bibinfo {volume} {16}},\ \bibinfo {pages} {163}
  (\bibinfo {year} {2017})}\BibitemShut {NoStop}%
\bibitem [{\citenamefont {Rutter}\ \emph {et~al.}(2007)\citenamefont {Rutter},
  \citenamefont {Crain}, \citenamefont {Guisinger}, \citenamefont {Li},
  \citenamefont {First},\ and\ \citenamefont {Stroscio}}]{Rutter2007Jul}%
  \BibitemOpen
  \bibfield  {author} {\bibinfo {author} {\bibfnamefont {G.~M.}\ \bibnamefont
  {Rutter}}, \bibinfo {author} {\bibfnamefont {J.~N.}\ \bibnamefont {Crain}},
  \bibinfo {author} {\bibfnamefont {N.~P.}\ \bibnamefont {Guisinger}}, \bibinfo
  {author} {\bibfnamefont {T.}~\bibnamefont {Li}}, \bibinfo {author}
  {\bibfnamefont {P.~N.}\ \bibnamefont {First}},\ and\ \bibinfo {author}
  {\bibfnamefont {J.~A.}\ \bibnamefont {Stroscio}},\ }\bibfield  {title}
  {\bibinfo {title} {Scattering and interference in epitaxial graphene},\
  }\href {https://doi.org/10.1126/science.1142882} {\bibfield  {journal}
  {\bibinfo  {journal} {Science}\ }\textbf {\bibinfo {volume} {317}},\ \bibinfo
  {pages} {219} (\bibinfo {year} {2007})}\BibitemShut {NoStop}%
\bibitem [{\citenamefont {Rosales}\ and\ \citenamefont
  {Gonz{\ifmmode\acute{a}\else\'{a}\fi}lez}(2013)}]{Rosales2013Jan}%
  \BibitemOpen
  \bibfield  {author} {\bibinfo {author} {\bibfnamefont {L.}~\bibnamefont
  {Rosales}}\ and\ \bibinfo {author} {\bibfnamefont {J.~W.}\ \bibnamefont
  {Gonz{\ifmmode\acute{a}\else\'{a}\fi}lez}},\ }\bibfield  {title} {\bibinfo
  {title} {{Transport properties of two finite armchair graphene
  nanoribbons}},\ }\href {https://doi.org/10.1186/1556-276X-8-1} {\bibfield
  {journal} {\bibinfo  {journal} {Nanoscale Res. Lett.}\ }\textbf {\bibinfo
  {volume} {8}},\ \bibinfo {pages} {1} (\bibinfo {year} {2013})}\BibitemShut
  {NoStop}%
\bibitem [{\citenamefont {Raja}\ \emph {et~al.}(2017)\citenamefont {Raja},
  \citenamefont {Chaves}, \citenamefont {Yu}, \citenamefont {Arefe},
  \citenamefont {Hill}, \citenamefont {Rigosi}, \citenamefont {Berkelbach},
  \citenamefont {Nagler}, \citenamefont
  {Sch{\ifmmode\ddot{u}\else\"{u}\fi}ller}, \citenamefont {Korn}, \citenamefont
  {Nuckolls}, \citenamefont {Hone}, \citenamefont {Brus}, \citenamefont
  {Heinz}, \citenamefont {Reichman},\ and\ \citenamefont
  {Chernikov}}]{Raja2017May}%
  \BibitemOpen
  \bibfield  {author} {\bibinfo {author} {\bibfnamefont {A.}~\bibnamefont
  {Raja}}, \bibinfo {author} {\bibfnamefont {A.}~\bibnamefont {Chaves}},
  \bibinfo {author} {\bibfnamefont {J.}~\bibnamefont {Yu}}, \bibinfo {author}
  {\bibfnamefont {G.}~\bibnamefont {Arefe}}, \bibinfo {author} {\bibfnamefont
  {H.~M.}\ \bibnamefont {Hill}}, \bibinfo {author} {\bibfnamefont {A.~F.}\
  \bibnamefont {Rigosi}}, \bibinfo {author} {\bibfnamefont {T.~C.}\
  \bibnamefont {Berkelbach}}, \bibinfo {author} {\bibfnamefont
  {P.}~\bibnamefont {Nagler}}, \bibinfo {author} {\bibfnamefont
  {C.}~\bibnamefont {Sch{\ifmmode\ddot{u}\else\"{u}\fi}ller}}, \bibinfo
  {author} {\bibfnamefont {T.}~\bibnamefont {Korn}}, \bibinfo {author}
  {\bibfnamefont {C.}~\bibnamefont {Nuckolls}}, \bibinfo {author}
  {\bibfnamefont {J.}~\bibnamefont {Hone}}, \bibinfo {author} {\bibfnamefont
  {L.~E.}\ \bibnamefont {Brus}}, \bibinfo {author} {\bibfnamefont {T.~F.}\
  \bibnamefont {Heinz}}, \bibinfo {author} {\bibfnamefont {D.~R.}\ \bibnamefont
  {Reichman}},\ and\ \bibinfo {author} {\bibfnamefont {A.}~\bibnamefont
  {Chernikov}},\ }\bibfield  {title} {\bibinfo {title} {{Coulomb engineering of
  the bandgap and excitons in two-dimensional materials}},\ }\href
  {https://doi.org/10.1038/ncomms15251} {\bibfield  {journal} {\bibinfo
  {journal} {Nat. Commun.}\ }\textbf {\bibinfo {volume} {8}},\ \bibinfo {pages}
  {1} (\bibinfo {year} {2017})}\BibitemShut {NoStop}%
\bibitem [{\citenamefont {Chaves}\ \emph {et~al.}(2020)\citenamefont {Chaves},
  \citenamefont {Azadani}, \citenamefont {Alsalman}, \citenamefont {da~Costa},
  \citenamefont {Frisenda}, \citenamefont {Chaves}, \citenamefont {Song},
  \citenamefont {Kim}, \citenamefont {He}, \citenamefont {Zhou}, \citenamefont
  {Castellanos-Gomez}, \citenamefont {Peeters}, \citenamefont {Liu},
  \citenamefont {Hinkle}, \citenamefont {Oh}, \citenamefont {Ye}, \citenamefont
  {Koester}, \citenamefont {Lee}, \citenamefont {Avouris}, \citenamefont
  {Wang},\ and\ \citenamefont {Low}}]{Chaves2020Aug}%
  \BibitemOpen
  \bibfield  {author} {\bibinfo {author} {\bibfnamefont {A.}~\bibnamefont
  {Chaves}}, \bibinfo {author} {\bibfnamefont {J.~G.}\ \bibnamefont {Azadani}},
  \bibinfo {author} {\bibfnamefont {H.}~\bibnamefont {Alsalman}}, \bibinfo
  {author} {\bibfnamefont {D.~R.}\ \bibnamefont {da~Costa}}, \bibinfo {author}
  {\bibfnamefont {R.}~\bibnamefont {Frisenda}}, \bibinfo {author}
  {\bibfnamefont {A.~J.}\ \bibnamefont {Chaves}}, \bibinfo {author}
  {\bibfnamefont {S.~H.}\ \bibnamefont {Song}}, \bibinfo {author}
  {\bibfnamefont {Y.~D.}\ \bibnamefont {Kim}}, \bibinfo {author} {\bibfnamefont
  {D.}~\bibnamefont {He}}, \bibinfo {author} {\bibfnamefont {J.}~\bibnamefont
  {Zhou}}, \bibinfo {author} {\bibfnamefont {A.}~\bibnamefont
  {Castellanos-Gomez}}, \bibinfo {author} {\bibfnamefont {F.~M.}\ \bibnamefont
  {Peeters}}, \bibinfo {author} {\bibfnamefont {Z.}~\bibnamefont {Liu}},
  \bibinfo {author} {\bibfnamefont {C.~L.}\ \bibnamefont {Hinkle}}, \bibinfo
  {author} {\bibfnamefont {S.-H.}\ \bibnamefont {Oh}}, \bibinfo {author}
  {\bibfnamefont {P.~D.}\ \bibnamefont {Ye}}, \bibinfo {author} {\bibfnamefont
  {S.~J.}\ \bibnamefont {Koester}}, \bibinfo {author} {\bibfnamefont {Y.~H.}\
  \bibnamefont {Lee}}, \bibinfo {author} {\bibfnamefont {{\relax
  Ph}.}~\bibnamefont {Avouris}}, \bibinfo {author} {\bibfnamefont
  {X.}~\bibnamefont {Wang}},\ and\ \bibinfo {author} {\bibfnamefont
  {T.}~\bibnamefont {Low}},\ }\bibfield  {title} {\bibinfo {title} {{Bandgap
  engineering of two-dimensional semiconductor materials}},\ }\href
  {https://doi.org/10.1038/s41699-020-00162-4} {\bibfield  {journal} {\bibinfo
  {journal} {npj 2D Mater. Appl.}\ }\textbf {\bibinfo {volume} {4}},\ \bibinfo
  {pages} {1} (\bibinfo {year} {2020})}\BibitemShut {NoStop}%
\bibitem [{\citenamefont {Tao}\ \emph {et~al.}(2015)\citenamefont {Tao},
  \citenamefont {Cinquanta}, \citenamefont {Chiappe}, \citenamefont
  {Grazianetti}, \citenamefont {Fanciulli}, \citenamefont {Dubey},
  \citenamefont {Molle},\ and\ \citenamefont {Akinwande}}]{Tao2015Mar}%
  \BibitemOpen
  \bibfield  {author} {\bibinfo {author} {\bibfnamefont {L.}~\bibnamefont
  {Tao}}, \bibinfo {author} {\bibfnamefont {E.}~\bibnamefont {Cinquanta}},
  \bibinfo {author} {\bibfnamefont {D.}~\bibnamefont {Chiappe}}, \bibinfo
  {author} {\bibfnamefont {C.}~\bibnamefont {Grazianetti}}, \bibinfo {author}
  {\bibfnamefont {M.}~\bibnamefont {Fanciulli}}, \bibinfo {author}
  {\bibfnamefont {M.}~\bibnamefont {Dubey}}, \bibinfo {author} {\bibfnamefont
  {A.}~\bibnamefont {Molle}},\ and\ \bibinfo {author} {\bibfnamefont
  {D.}~\bibnamefont {Akinwande}},\ }\bibfield  {title} {\bibinfo {title}
  {{Silicene field-effect transistors operating at room temperature}},\ }\href
  {https://doi.org/10.1038/nnano.2014.325} {\bibfield  {journal} {\bibinfo
  {journal} {Nat. Nanotechnol.}\ }\textbf {\bibinfo {volume} {10}},\ \bibinfo
  {pages} {227} (\bibinfo {year} {2015})}\BibitemShut {NoStop}%
\bibitem [{\citenamefont {Roy}\ \emph {et~al.}(2014)\citenamefont {Roy},
  \citenamefont {Tosun}, \citenamefont {Kang}, \citenamefont {Sachid},
  \citenamefont {Desai}, \citenamefont {Hettick}, \citenamefont {Hu},\ and\
  \citenamefont {Javey}}]{Roy2014Jun}%
  \BibitemOpen
  \bibfield  {author} {\bibinfo {author} {\bibfnamefont {T.}~\bibnamefont
  {Roy}}, \bibinfo {author} {\bibfnamefont {M.}~\bibnamefont {Tosun}}, \bibinfo
  {author} {\bibfnamefont {J.~S.}\ \bibnamefont {Kang}}, \bibinfo {author}
  {\bibfnamefont {A.~B.}\ \bibnamefont {Sachid}}, \bibinfo {author}
  {\bibfnamefont {S.~B.}\ \bibnamefont {Desai}}, \bibinfo {author}
  {\bibfnamefont {M.}~\bibnamefont {Hettick}}, \bibinfo {author} {\bibfnamefont
  {C.~C.}\ \bibnamefont {Hu}},\ and\ \bibinfo {author} {\bibfnamefont
  {A.}~\bibnamefont {Javey}},\ }\bibfield  {title} {\bibinfo {title}
  {{Field-effect transistors built from all two-dimensional material
  components}},\ }\href {https://doi.org/10.1021/nn501723y} {\bibfield
  {journal} {\bibinfo  {journal} {ACS Nano}\ }\textbf {\bibinfo {volume} {8}},\
  \bibinfo {pages} {6259} (\bibinfo {year} {2014})}\BibitemShut {NoStop}%
\bibitem [{\citenamefont {Guimar{\ifmmode\tilde{a}\else\~{a}\fi}es}\ \emph
  {et~al.}(2016)\citenamefont {Guimar{\ifmmode\tilde{a}\else\~{a}\fi}es},
  \citenamefont {Gao}, \citenamefont {Han}, \citenamefont {Kang}, \citenamefont
  {Xie}, \citenamefont {Kim}, \citenamefont {Muller}, \citenamefont {Ralph},\
  and\ \citenamefont {Park}}]{Guimaraes2016Jun}%
  \BibitemOpen
  \bibfield  {author} {\bibinfo {author} {\bibfnamefont {M.~H.~D.}\
  \bibnamefont {Guimar{\ifmmode\tilde{a}\else\~{a}\fi}es}}, \bibinfo {author}
  {\bibfnamefont {H.}~\bibnamefont {Gao}}, \bibinfo {author} {\bibfnamefont
  {Y.}~\bibnamefont {Han}}, \bibinfo {author} {\bibfnamefont {K.}~\bibnamefont
  {Kang}}, \bibinfo {author} {\bibfnamefont {S.}~\bibnamefont {Xie}}, \bibinfo
  {author} {\bibfnamefont {C.-J.}\ \bibnamefont {Kim}}, \bibinfo {author}
  {\bibfnamefont {D.~A.}\ \bibnamefont {Muller}}, \bibinfo {author}
  {\bibfnamefont {D.~C.}\ \bibnamefont {Ralph}},\ and\ \bibinfo {author}
  {\bibfnamefont {J.}~\bibnamefont {Park}},\ }\bibfield  {title} {\bibinfo
  {title} {Atomically thin ohmic edge contacts between two-dimensional
  materials},\ }\href {https://doi.org/10.1021/acsnano.6b02879} {\bibfield
  {journal} {\bibinfo  {journal} {ACS Nano}\ }\textbf {\bibinfo {volume}
  {10}},\ \bibinfo {pages} {6392} (\bibinfo {year} {2016})}\BibitemShut
  {NoStop}%
\bibitem [{\citenamefont {L{\ifmmode\acute{e}\else\'{e}\fi}onard}\ and\
  \citenamefont {Talin}(2011)}]{Leonard2011Dec}%
  \BibitemOpen
  \bibfield  {author} {\bibinfo {author} {\bibfnamefont {F.}~\bibnamefont
  {L{\ifmmode\acute{e}\else\'{e}\fi}onard}}\ and\ \bibinfo {author}
  {\bibfnamefont {A.~A.}\ \bibnamefont {Talin}},\ }\bibfield  {title} {\bibinfo
  {title} {{Electrical contacts to one- and two-dimensional nanomaterials}},\
  }\href {https://doi.org/10.1038/nnano.2011.196} {\bibfield  {journal}
  {\bibinfo  {journal} {Nat. Nanotechnol.}\ }\textbf {\bibinfo {volume} {6}},\
  \bibinfo {pages} {773} (\bibinfo {year} {2011})}\BibitemShut {NoStop}%
\bibitem [{\citenamefont {Allain}\ \emph {et~al.}(2015)\citenamefont {Allain},
  \citenamefont {Kang}, \citenamefont {Banerjee},\ and\ \citenamefont
  {Kis}}]{Allain2015Dec}%
  \BibitemOpen
  \bibfield  {author} {\bibinfo {author} {\bibfnamefont {A.}~\bibnamefont
  {Allain}}, \bibinfo {author} {\bibfnamefont {J.}~\bibnamefont {Kang}},
  \bibinfo {author} {\bibfnamefont {K.}~\bibnamefont {Banerjee}},\ and\
  \bibinfo {author} {\bibfnamefont {A.}~\bibnamefont {Kis}},\ }\bibfield
  {title} {\bibinfo {title} {{Electrical contacts to two-dimensional
  semiconductors}},\ }\href {https://doi.org/10.1038/nmat4452} {\bibfield
  {journal} {\bibinfo  {journal} {Nat. Mater.}\ }\textbf {\bibinfo {volume}
  {14}},\ \bibinfo {pages} {1195} (\bibinfo {year} {2015})}\BibitemShut
  {NoStop}%
\bibitem [{\citenamefont {Wang}\ \emph {et~al.}(2013)\citenamefont {Wang},
  \citenamefont {Meric}, \citenamefont {Huang}, \citenamefont {Gao},
  \citenamefont {Gao}, \citenamefont {Tran}, \citenamefont {Taniguchi},
  \citenamefont {Watanabe}, \citenamefont {Campos}, \citenamefont {Muller},
  \citenamefont {Guo}, \citenamefont {Kim}, \citenamefont {Hone}, \citenamefont
  {Shepard},\ and\ \citenamefont
  {Dean}}]{OneDimensionalElectricalContactto2DMaterial_wang_2013}%
  \BibitemOpen
  \bibfield  {author} {\bibinfo {author} {\bibfnamefont {L.}~\bibnamefont
  {Wang}}, \bibinfo {author} {\bibfnamefont {I.}~\bibnamefont {Meric}},
  \bibinfo {author} {\bibfnamefont {P.~Y.}\ \bibnamefont {Huang}}, \bibinfo
  {author} {\bibfnamefont {Q.}~\bibnamefont {Gao}}, \bibinfo {author}
  {\bibfnamefont {Y.}~\bibnamefont {Gao}}, \bibinfo {author} {\bibfnamefont
  {H.}~\bibnamefont {Tran}}, \bibinfo {author} {\bibfnamefont {T.}~\bibnamefont
  {Taniguchi}}, \bibinfo {author} {\bibfnamefont {K.}~\bibnamefont {Watanabe}},
  \bibinfo {author} {\bibfnamefont {L.~M.}\ \bibnamefont {Campos}}, \bibinfo
  {author} {\bibfnamefont {D.~A.}\ \bibnamefont {Muller}}, \bibinfo {author}
  {\bibfnamefont {J.}~\bibnamefont {Guo}}, \bibinfo {author} {\bibfnamefont
  {P.}~\bibnamefont {Kim}}, \bibinfo {author} {\bibfnamefont {J.}~\bibnamefont
  {Hone}}, \bibinfo {author} {\bibfnamefont {K.~L.}\ \bibnamefont {Shepard}},\
  and\ \bibinfo {author} {\bibfnamefont {C.~R.}\ \bibnamefont {Dean}},\
  }\bibfield  {title} {\bibinfo {title} {One-dimensional electrical contact to
  a two-dimensional material},\ }\href
  {https://doi.org/10.1126/science.1244358} {\bibfield  {journal} {\bibinfo
  {journal} {Science}\ }\textbf {\bibinfo {volume} {342}},\ \bibinfo {pages}
  {614} (\bibinfo {year} {2013})}\BibitemShut {NoStop}%
\bibitem [{\citenamefont
  {Milo{\ifmmode\check{s}\else\v{s}\fi}evi{\ifmmode\acute{c}\else\'{c}\fi}}\
  and\ \citenamefont {Mandrus}(2021)}]{Milosevic2021}%
  \BibitemOpen
  \bibfield  {author} {\bibinfo {author} {\bibfnamefont {M.~V.}\ \bibnamefont
  {Milo{\ifmmode\check{s}\else\v{s}\fi}evi{\ifmmode\acute{c}\else\'{c}\fi}}}\
  and\ \bibinfo {author} {\bibfnamefont {D.}~\bibnamefont {Mandrus}},\
  }\bibfield  {title} {\bibinfo {title} {{2D Quantum materials: Magnetism and
  superconductivity}},\ }\bibfield  {journal} {\bibinfo  {journal} {J. Appl.
  Phys.}\ }\textbf {\bibinfo {volume} {130}},\ \href
  {https://doi.org/10.1063/5.0075774} {10.1063/5.0075774} (\bibinfo {year}
  {2021})\BibitemShut {NoStop}%
\bibitem [{\citenamefont {Lu}\ and\ \citenamefont
  {Sun}(2021)}]{AntiferroSuperJunctionCAR_WeiTao_2021}%
  \BibitemOpen
  \bibfield  {author} {\bibinfo {author} {\bibfnamefont {W.-T.}\ \bibnamefont
  {Lu}}\ and\ \bibinfo {author} {\bibfnamefont {Q.-F.}\ \bibnamefont {Sun}},\
  }\bibfield  {title} {\bibinfo {title} {Electrical control of crossed
  {Andreev} reflection and spin-valley switch in antiferromagnet/superconductor
  junctions},\ }\href {https://doi.org/10.1103/PhysRevB.104.045418} {\bibfield
  {journal} {\bibinfo  {journal} {Phys. Rev. B}\ }\textbf {\bibinfo {volume}
  {104}},\ \bibinfo {pages} {045418} (\bibinfo {year} {2021})}\BibitemShut
  {NoStop}%
\bibitem [{\citenamefont {Shafiei}\ \emph {et~al.}(2022)\citenamefont
  {Shafiei}, \citenamefont {Fazileh}, \citenamefont {Peeters},\ and\
  \citenamefont {Milo\ifmmode \check{s}\else
  \v{s}\fi{}evi\ifmmode~\acute{c}\else \'{c}\fi{}}}]{Milosevic2022}%
  \BibitemOpen
  \bibfield  {author} {\bibinfo {author} {\bibfnamefont {M.}~\bibnamefont
  {Shafiei}}, \bibinfo {author} {\bibfnamefont {F.}~\bibnamefont {Fazileh}},
  \bibinfo {author} {\bibfnamefont {F.~m. c.~M.}\ \bibnamefont {Peeters}},\
  and\ \bibinfo {author} {\bibfnamefont {M.~V.}\ \bibnamefont {Milo\ifmmode
  \check{s}\else \v{s}\fi{}evi\ifmmode~\acute{c}\else \'{c}\fi{}}},\ }\bibfield
   {title} {\bibinfo {title} {Controlling the hybridization gap and transport
  in a thin-film topological insulator: Effect of strain, and electric and
  magnetic field},\ }\href {https://doi.org/10.1103/PhysRevB.106.035119}
  {\bibfield  {journal} {\bibinfo  {journal} {Phys. Rev. B}\ }\textbf {\bibinfo
  {volume} {106}},\ \bibinfo {pages} {035119} (\bibinfo {year}
  {2022})}\BibitemShut {NoStop}%
\bibitem [{\citenamefont {Kheirabadi}\ and\ \citenamefont
  {Langari}(2022)}]{Kheirabadi2022}%
  \BibitemOpen
  \bibfield  {author} {\bibinfo {author} {\bibfnamefont {N.}~\bibnamefont
  {Kheirabadi}}\ and\ \bibinfo {author} {\bibfnamefont {A.}~\bibnamefont
  {Langari}},\ }\bibfield  {title} {\bibinfo {title} {Quantum nonlinear planar
  hall effect in bilayer graphene: An orbital effect of a steady in-plane
  magnetic field},\ }\href {https://doi.org/10.1103/PhysRevB.106.245143}
  {\bibfield  {journal} {\bibinfo  {journal} {Phys. Rev. B}\ }\textbf {\bibinfo
  {volume} {106}},\ \bibinfo {pages} {245143} (\bibinfo {year}
  {2022})}\BibitemShut {NoStop}%
\end{thebibliography}%

\end{document}